\documentclass[paper=a4, fontsize=11pt]{scrartcl}
\usepackage{amsmath,amsfonts,amsthm}
\usepackage[english]{babel}	
\usepackage[pdftex]{graphicx}
\usepackage{caption}
\usepackage[T1]{fontenc}
\usepackage[utf8]{inputenc}
\usepackage{authblk}
\usepackage[colorlinks=true,breaklinks=true]{hyperref}
\usepackage{float}
\usepackage{pdflscape}

\title{Ion counting efficiencies at the IGISOL facility}

\author{A. Al-Adili \footnote{ali.al-adili@physics.uu.se} }
\author{K. Jansson} 
\author{M. Lantz \footnote{mattias.lantz@physics.uu.se} }
\author{A. Solders} 

\author{C. Gustavsson}
\author{A. Mattera}
\author{A. V. Prokofiev}
\author{V. Rakopoulos}
\author{D. Tarr\'{i}o}
\author{S. Wiberg}
\author{M. \"Osterlund}

\author{S. Pomp \footnote{stephan.pomp@physics.uu.se}}

\affil{The Nuclear Reaction Group - Division of Applied Nuclear Physics  Uppsala University, Sweden}

\newcommand{\class}[1]{\texttt{#1}}

\begin{document}

\maketitle


\begin{abstract}

\section*{Abstract}
At the IGISOL-JYFLTRAP facility, fission mass yields can be studied at high precision. Fission fragments from a U target are passing through a Ni foil and entering a gas filled chamber. The collected fragments are guided through a mass separator to a Penning trap where their masses are identified. This simulation work focuses on how different fission fragment properties (mass, charge and energy) affect the stopping efficiency in the gas cell. In addition, different experimental parameters are varied (e. g. U and Ni thickness and He gas pressure) to study their impact on the stopping efficiency. The simulations were performed using the Geant4 package and the SRIM code.

The main results suggest a small variation in the stopping efficiency as a function of mass, charge and kinetic energy. It is predicted that heavy fragments are stopped about 9\% less efficiently than the light fragments. However it was found that the properties of the U, Ni and the He gas influences this behaviour. Hence it could be possible to optimize the efficiency.

\end{abstract}

 \pagebreak
\tableofcontents
\pagebreak

\begin{footnotesize}
\listoffigures
\end{footnotesize}
\pagebreak

\section*{List of abbreviations}

\begin{itemize}

\item[$\diamond$] ChargeAfterTarget = Ion charge when leaving the U target.

\item[$\diamond$] ChargeAfterFoil = Ion charge when leaving the Ni foil.

\item[$\diamond$] ChargeAfterEvent = Final Ion charge when stopping or leaving the simulation volume.

\item[$\diamond$] Geant4 = GEometry ANd Tracking (simulation package of the passage of particles through matter (developed at CERN \cite{GeantNIM03}).

\item[$\diamond$] GEF = GEneral Fission model (Fission code by Schmidt and Jurado \cite{GEF}). 

\item[$\diamond$] EnergyAfterTarget = Ion energy after leaving the U target.

\item[$\diamond$] EnergyAfterFoil = Ion energy after leaving the Ni foil.  

\item[$\diamond$] EnergyAfterEvent = Final ion energy when stopping or leaving the simulation volume. 

\item[$\diamond$] FF = Fission fragments.

\item[$\diamond$] IGISOL = Ion Guide Isotope Separator On-Line (IGISOL) at the Jyväskylä facility.

\item[$\diamond$] SRIM = The Stopping and Range of Ions in Matter. Stopping power calculation package by Ziegler \cite{Ziegler85}.

\item[$\diamond$] TKE = Total kinetic energy of both fission fragments.

\item[$\diamond$] (X/Y/Z)AfterTarget = Ion position when leaving the U target in (X/Y/Z).

\item[$\diamond$] (X/Y/Z)AfterFoil = Ion position when leaving the Ni foil in (X/Y/Z).

\item[$\diamond$] (X/Y/Z)AfterEvent = Final ion position after fully stop in (X/Y/Z).

\end{itemize}

\clearpage 

\section{\label{sectIntro}Introduction}

This report describes the simulations performed by the Uppsala group concerning the ion stopping efficiency at the IGISOL fission ion guide \cite{pentil,moore}. One application of the IGISOL technique is to study independent fission yields, i.e. to determine distribution of fission products over mass A and charge Z. \cite{IAEA}.

\begin{figure}[H]
\centering
    \includegraphics[width=0.8\textwidth]{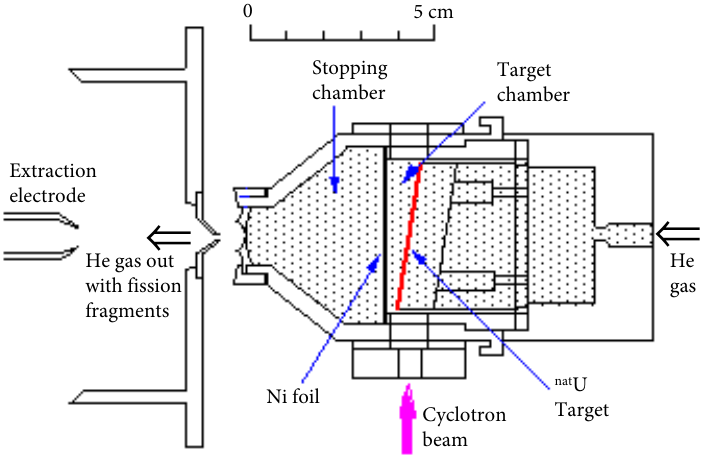}
\caption[The IGISOL target.]{The IGISOL target design \cite{IGISOLdesign}. The cyclotron beam irradiates a tilted U target and the fission fragments enter the target chamber which is separated from the stopping chamber by a Ni foil.}
\label{Fig1}
\end{figure}

The emitted fission products are identified using the Penning trap JYFLTRAP. Fission is induced e.g.~by a proton beam impinging on a natural Uranium target (15\,mg/cm$^2$ thick) and one of the two fragments is emitted in the direction of the stopping chamber (see Fig.~\ref{Fig1}). The cyclotron beam irradiates the U-sample tilted 7 degrees with respect to the beam direction, in order to increase the effective target thickness. In order to reach the stopping chamber, resulting fission products have to escape the Uranium target and to pass a separating Ni foil (1\,mg/cm$^2$). The purpose of the Ni foil is to prevent the plasma created by the beam from getting into the ion guide. Helium gas held at a pressure of 200\,torr ($\sim 250$\,mbar) is used to collect the ions in the stopping chamber. If the fission products are slowed down sufficiently, they can be collected by the He-gas flow and extracted to the mass separator.

This work investigates to what extent different ion properties affect the stopping efficiency. The fission process produces a wide range of nuclides with atomic masses ranging from $A =$ 70 to 160. In addition, fragments show a large spread of kinetic energies, $\sim$ 0.3 - 1.2\,MeV/u. Therefore it is important to understand the differences in stopping power as a function of mass, charge and energy. Eventually the purpose is to decrease the systematic uncertainty of fission measurements at the IGISOL facility. 

\subsection{\label{sectFF}Brief introduction to fission}
For the current simulations it is important to consider some main aspects on the properties of nuclear fission. Firstly, the large spread in different nuclides produced. Secondly, the wide variation in kinetic/excitation energy. Thirdly, the difference between the pre-neutron emission fragments and the fission products (hence the link to the prompt neutron emission). For an extensive survey on the fission process the reader is referred to Ref. \cite{bible} and references therein.

In this chapter, we consider fission of actinide nuclei. An actinide nucleus has relatively low binding energy and can undergo fission by adding relatively small activation energy. Fissile nuclei (e. g. $^{235}$U, $^{236}$Np) have odd number of neutrons and can fission with thermal neutrons, in contrast to fertile nuclei (e. g. $^{238}$U, $^{237}$Np), which need larger neutron kinetic energy. When a nucleus captures a neutron (or a charged particle or a photon) it gets excited above the ground state and may overcome the fission barrier. Nuclear deformation gives rise to a strong reduction in nuclear force as a function of elongation, in contrast to the repelling Coulomb interaction. The two fragments begin to form, and as soon as the nucleus reaches a critical state (saddle point), fission cannot be prevented. Classically the Liquid Drop Model (LDM) was employed to depict the nuclear potential energy. The LDM, however, has not allowed one to explain such phenomenon as asymmetric mass split, prompt neutron distribution and isomeric fission. All these were explained later by introduction of multi-humped barrier which was implied by nuclear shell corrections \cite{bible}. 

\subsubsection*{Fragment mass}
Due to the properties of the fission barrier, the formed fragments are typically not equally sized. It is commonly agreed that a higher outer fission barrier suppresses the formation of symmetric fission (relative to the asymmetric mass split). At higher excitation energies the probability of symmetric fission becomes higher. At low and intermediate energies, one observes clearly an asymmetric mass distribution with the most probable masses around A$\sim$100 and  A$\sim$140 amu, in case of fission of e.g. $^{238}$U. This stability as a function of fissioing system, is due to magic shells for instance around Z = 50 and N = 82 and deformed shells at N = 88, albeit recent studies have suggested a rather stable configuration due to Z = 54 \cite{GEF}. Figure~\ref{Figneutron} shows a typical double-humped mass yield distribution. 

The modelling of the mass yields has been improved considerably due to the development in computation power. Nowadays sophisticated models try to map the nuclear potential energy landscape in a multi-dimensional deformation space \cite{möller,brosa}. Fragment properties are modelled as a result of different fission mode selection, often together with a statistical population of states (for example via a random neck rupture or a random walk along the potential energy landscape).

\subsubsection*{Kinetic energy} 
The large energy release in fission is owed to the repelling electromagnetic force. About 200 MeV energy, released in each fission event is found mainly in kinetic energy of the fragments. The Q value is shared between the total kinetic energy (TKE) and the Total Excitation Energy (TXE).  The TKE is directly related to the pre-scission shape. Compact pre-scission shape results in a larger repelling force and hence a large kinetic energy. The opposite is true for large deformation in the fissioning nucleus which results instead in a smaller kinetic energy release. The average TKE is at maximum around masses A$\sim$110 and $\sim$130 and decreases as the fragment mass ratio increases. In addition the TKE release is anti-correlated to the number of emitted prompt neutrons. The single fragment kinetic energy is strongly mass-dependent. Due to momentum conservation, light fragments have larger kinetic energies than the heavy fragments. In addition the energy stays fairly constant in the light fragments, as will be seen later in sect. \ref{secions}.

\subsubsection*{Neutron emission}
\begin{figure}[H]
\centering
    \includegraphics[width=0.6\textwidth]{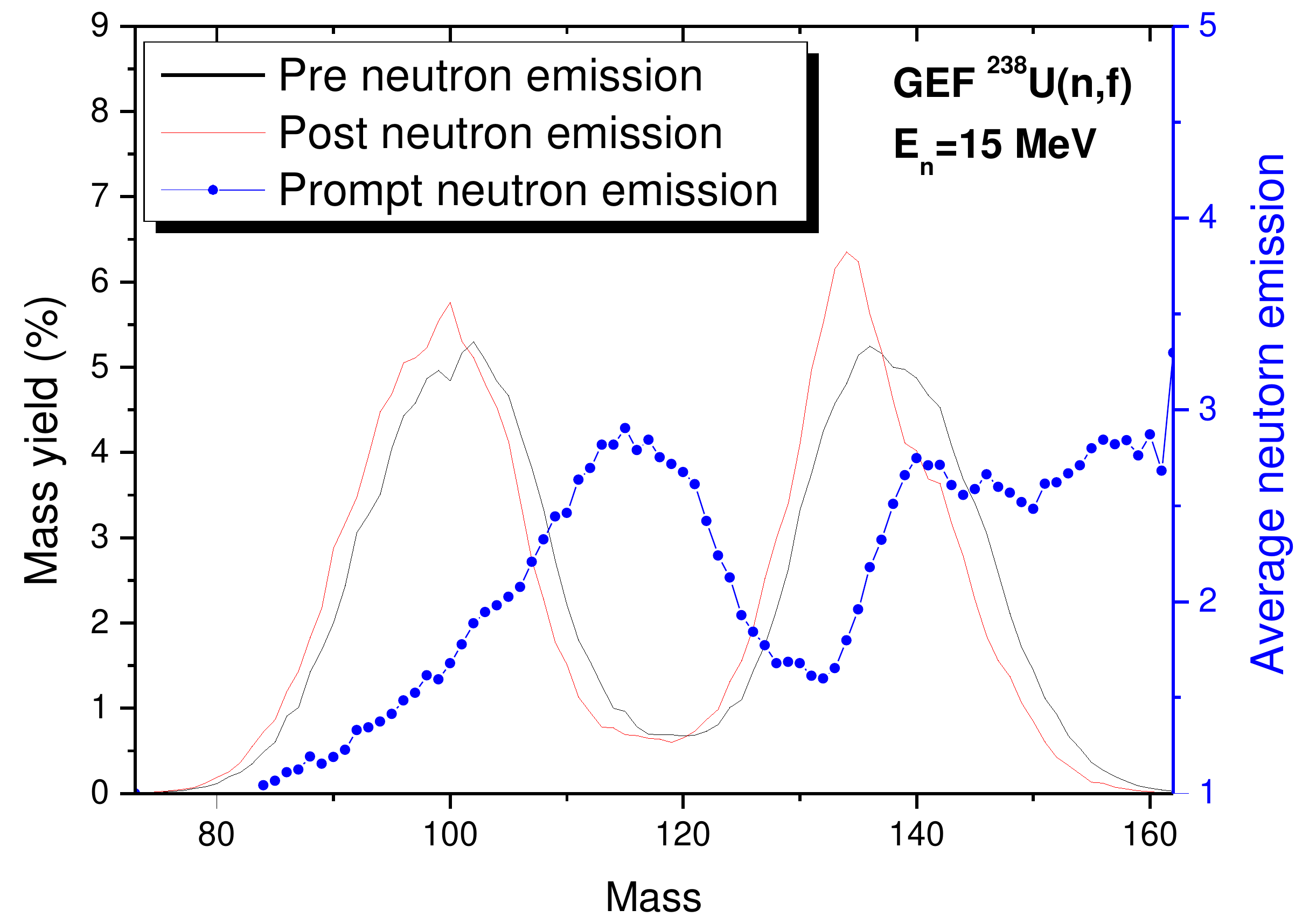}
\caption[The prompt neutron emission.]{The prompt neutron emission together with the fragment mass yield before and after neutron emission, in fission of $^{238}$U induced by 15-MeV neutrons. The data are simulated using the GEF code \cite{GEF}.}
\label{Figneutron}
\end{figure}

Most of fission fragments are neutron-rich, and they de-excite promptly after scission, by neutron and gamma emission. The neutron emission is mass-dependent with a substantial decrease due to magic shells around A = 132. Figure~\ref{Figneutron} shows the mass distribution before and after neutron emission together with the neutron emission curve as simulated using the GEF code for the $^{238}$U(n,f) reaction. The prompt neutron distribution is often referred to as the ``sawtooth shape''. The increased number of emitted neutrons is a direct measure of the deformation of the nascent fragments. A fragment with a fairly spherical shape (e.g. around A = 132) exhibits very low neutron emission. Since the neutron emission happens very fast (in the time scales of 10$^{-17}$ s), the post-neutron emission distribution is used in the simulations. 

\clearpage
\section{\label{secions}Ion selection for the simulations}
In order to select a representative set of fission products, the GEF code was used to simulate the reaction $^{238}$U(n,f) at incident neutron energy $E_n=15$ MeV \cite{GEF}. The GEF output has the advantage of giving both mass- and kinetic energy distributions for both the pre and post-neutron emission cases. The expected distribution of fission products from (n,f) is relatively similar to (p,f) as will be seen in sect. \ref{sect:comparisontoexpyields}. 

\subsection{GEF calculations}
\begin{figure}[H]
\centering
    \includegraphics[width=0.95\textwidth]{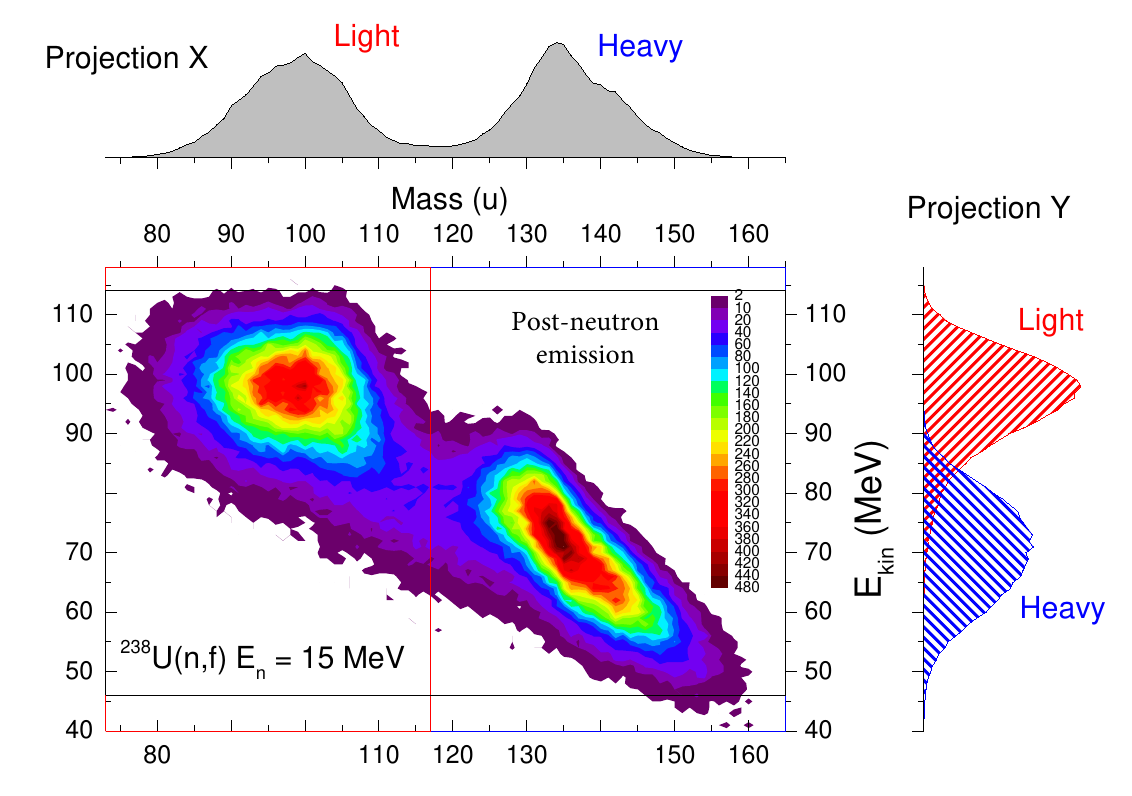}
\caption[The fission yield as a function of the mass and kinetic energy for $^{238}$U(n,f).]{The fission yield as a function of mass and kinetic energy for $^{238}$U(n,f). The data were simulated with the GEF fission code \cite{GEF}.}
\label{Fig2}
\end{figure}

\begin{figure}[b!]
\centering
    \includegraphics[width=0.95\textwidth]{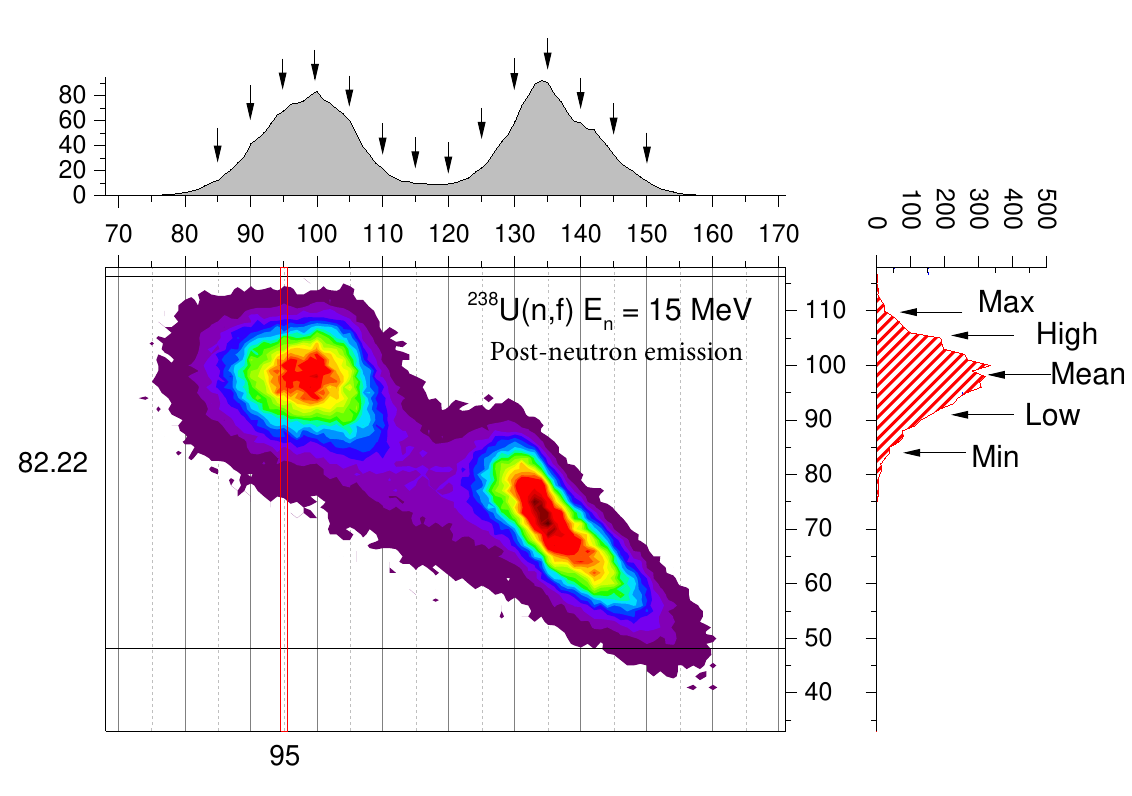}
\caption[The chosen isotopes.]{14 masses were chosen for the analysis; $A =$ 85, 90, 95, \ldots,145, 150. For each chosen mass 5 energies were chosen (min, low, mean, high and max energies).}
\label{Fig3}
\end{figure}

Figure~\ref{Fig2} shows the GEF simulated, two-dimensional scatter plot of the yield dependence on the mass and energy. The light fragments have energies around 1\,MeV/u whereas the heavy fragments have about half that energy per nucleon. The entire energy and mass distributions are projected on the sides in Fig.~\ref{Fig2}. Figure~\ref{Fig3} shows the selected ions ($A =$ 85, 90, 95, \ldots, 145, 150) and for each mass its characteristic kinetic energies (min, low, mean, high and max). These are chosen around 20\%, 50\% and full peak of the energy distribution height. From the GEF code the most probable $Z$ for each fragment mass was selected. Table \ref{table1} lists all the simulated ions. These ions are then used as input into the Geant4 simulation code.

\begin{table}[H] 
\caption{The ions selected for the Geant4 simulation (see Fig.~\ref{Fig3}). 1e7 events were simulated for each case.} 
\centering 
\begin{tabular}{c|c|c|ccccc} 
\hline
\hline
\bf{Element} & \bf{Mass}	& \bf{Z}	& \multicolumn{5}{c}{\bf{Energy (MeV)}} \\
&	& 	& Min  & Low  & Mean  &  High  & Max    \\
\hline
Se &85 & 34 & 85 & 92 & 100 & 106 & 110 \\ 
Kr &90	& 36	&88	&92	&97	&105	& 108	\\
Sr &95	& 38	& 86	&91	&98	&105	& 108	\\
Zr &100	& 40 &	85	&91	&98	&105&	108	\\
Mo &105	& 42 &	82	&87	&96	&103&	107	\\
Ru &110	& 44	&73	&81	&89	&96&	102	\\
Rh &115	& 45	&69	&74	&81	&88	&94	\\
Cd &120	& 48	&65	&70	&77	&87	&90	\\
In &125	& 49	&66	&72	&79	&85	&88	\\
Sn &130	& 50	&66	&71	&78	&84	&88	\\
I &135	& 53	&60	&64	&71	&79	&82	\\
Xe &140	& 54	&56	&58	&67	&72	&75	\\
Ba &145	& 56	&53	&56	&61	&66	&69	\\
Ce &150	& 58	&47	&51	&56	&62&	66	\\
\hline \hline
\end{tabular}
\label{table1}
\end{table} 

\subsection{\label{sect:comparisontoexpyields}Comparison to experimental fission yields}
\begin{figure}[b!]
\centering
    \includegraphics[width=0.9\textwidth]{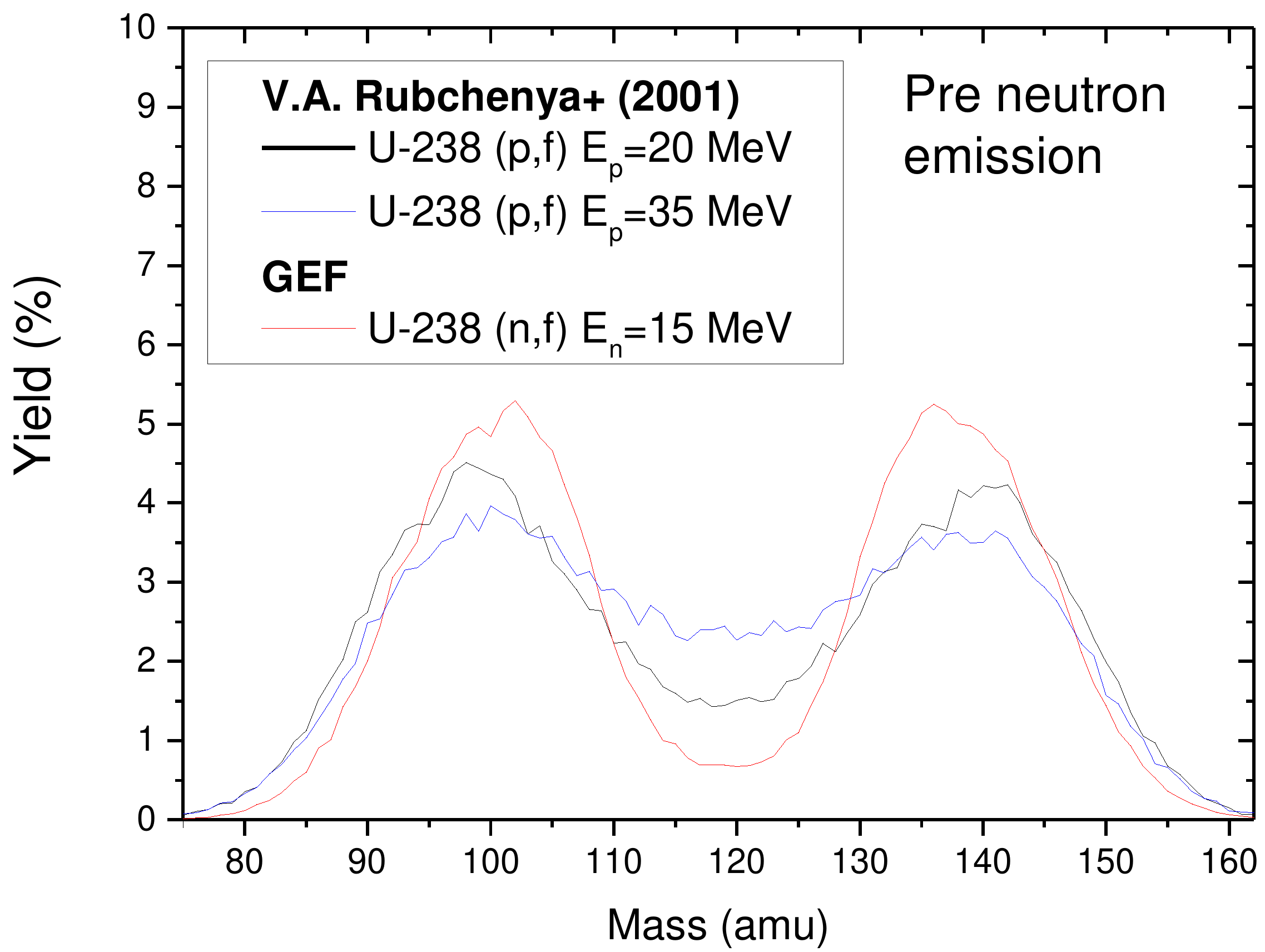}
\caption[The chosen isotopes compared to exp. data.]{Experimental data on $^{238}$U(p,f) from the HENDES setup, compared to the GEF simulation for $^{238}$U(n,f) \cite{Rubch}. The selected 14 masses (A = 85 - 150 amu) from the GEF simulation cover well the expected fission fragments from the (p,f) data. Note that the fragment distributions are before the neutron emission contrary to Fig.~\ref{Fig3} (available data).}
\label{FigRub}
\end{figure}

The $^{238}$U(n,f) reaction was chosen since it can be simulated using the GEF code. The wide selection range of fission products (A = 85 to 150) guaranties a decent coverage of the expected fission fragments as found in the comparison to experimental data \cite{Rubch}. Figure~\ref{FigRub} shows $^{238}$U(p,f) at proton energies of 20 MeV and 35 MeV compared to the GEF calculations for (n,f). The main differences emerge in the ratio between symmetric and asymmetric fission yields which is due to the different excitation energies. The range of possible fission fragments is however relatively similar. This is partly because of the same compound nuclear mass (A = 239) and partly because of fragment shell structure which contributes to the final mass selection. Note that the fragment distributions presented in Fig.~\ref{FigRub} are before emitting the prompt neutrons in contrast to the post-neutron emission distributions shown in Fig.~\ref{Fig3} (available compared data). In reality only post-neutron emission products will enter the chamber as the neutron emission happens almost immediately.

\clearpage
\section{Stopping powers}
\begin{figure}[b!]
\centering
    \includegraphics[width=0.49\textwidth]{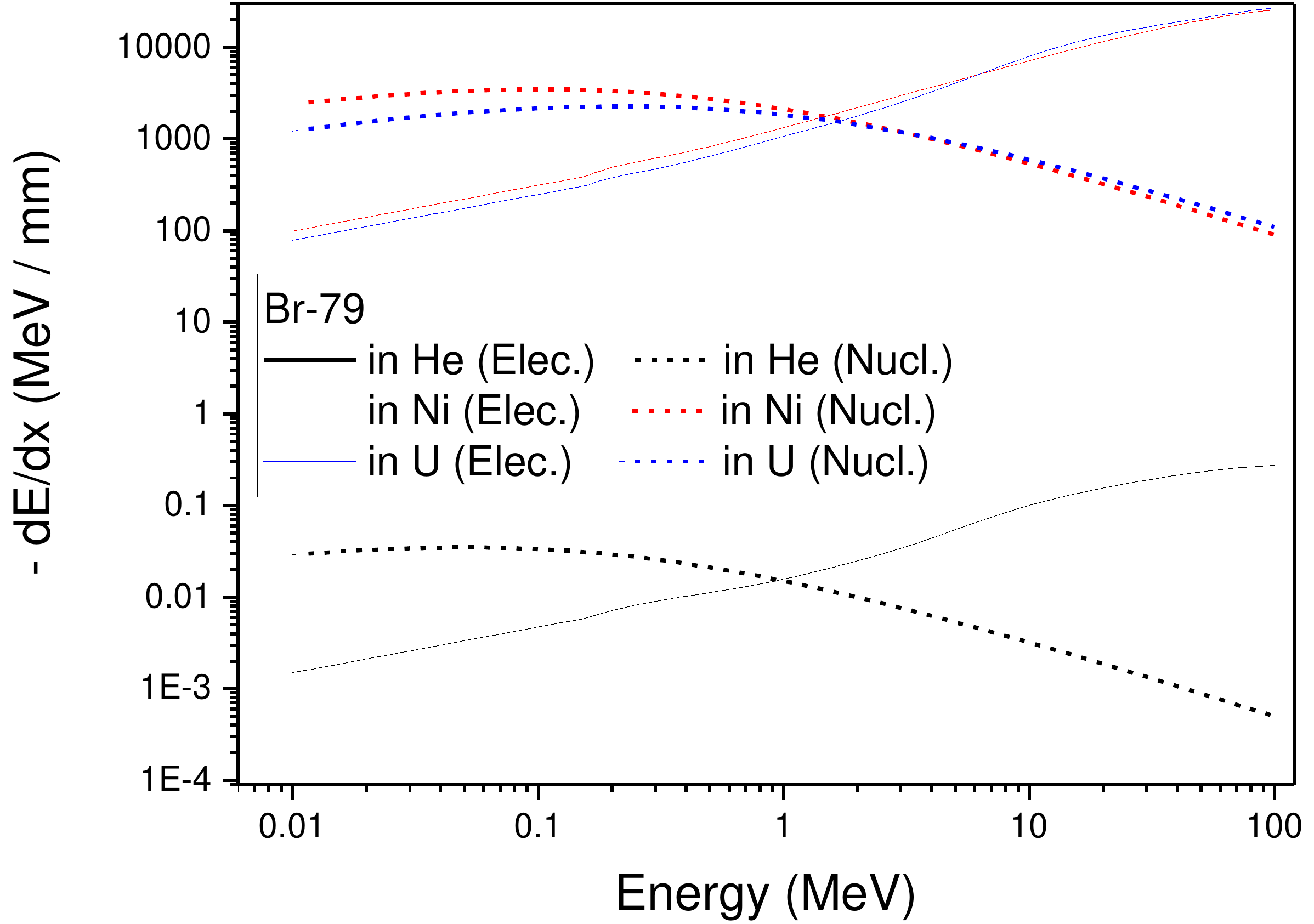}
    \includegraphics[width=0.49\textwidth]{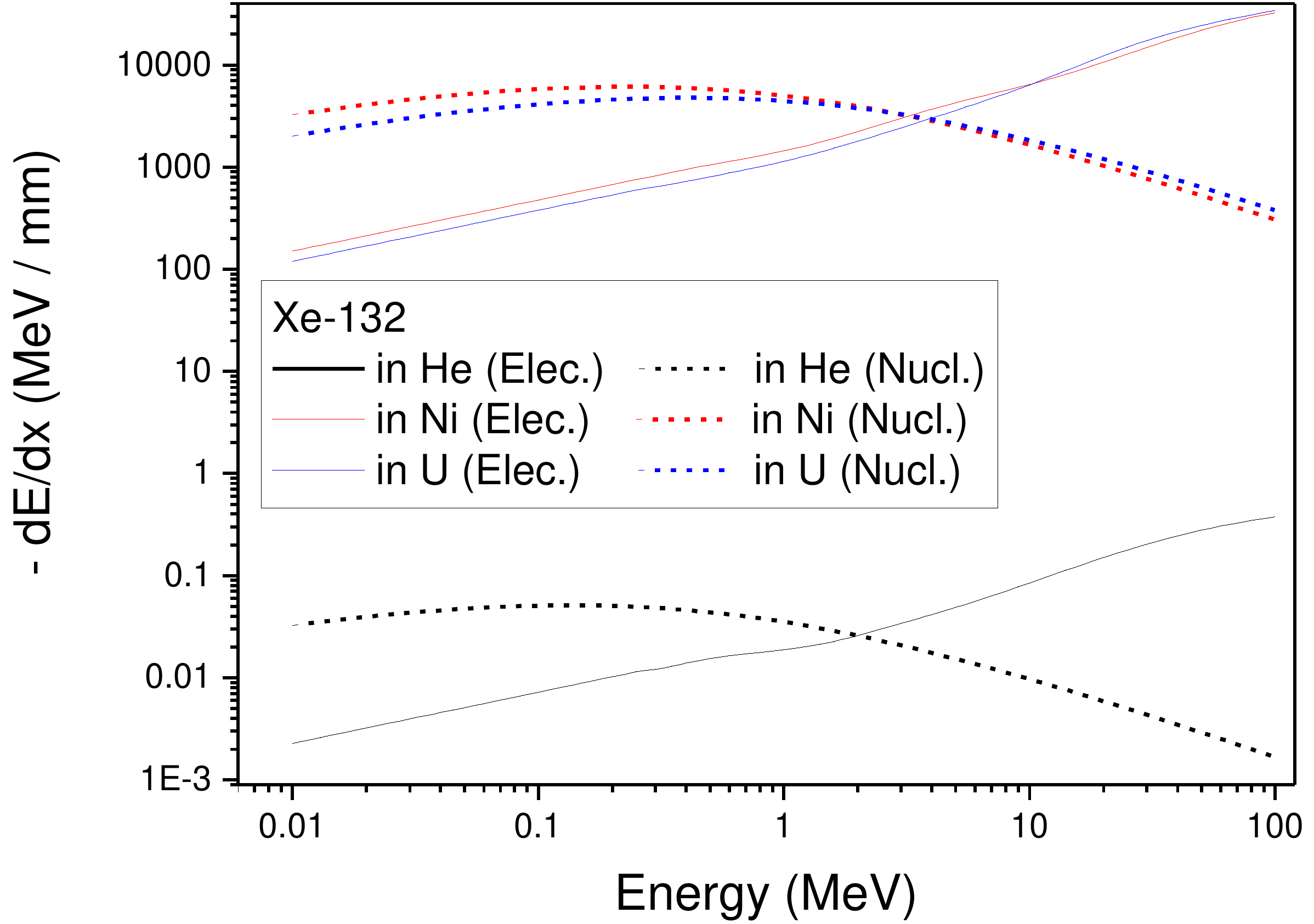}
    \includegraphics[width=0.49\textwidth]{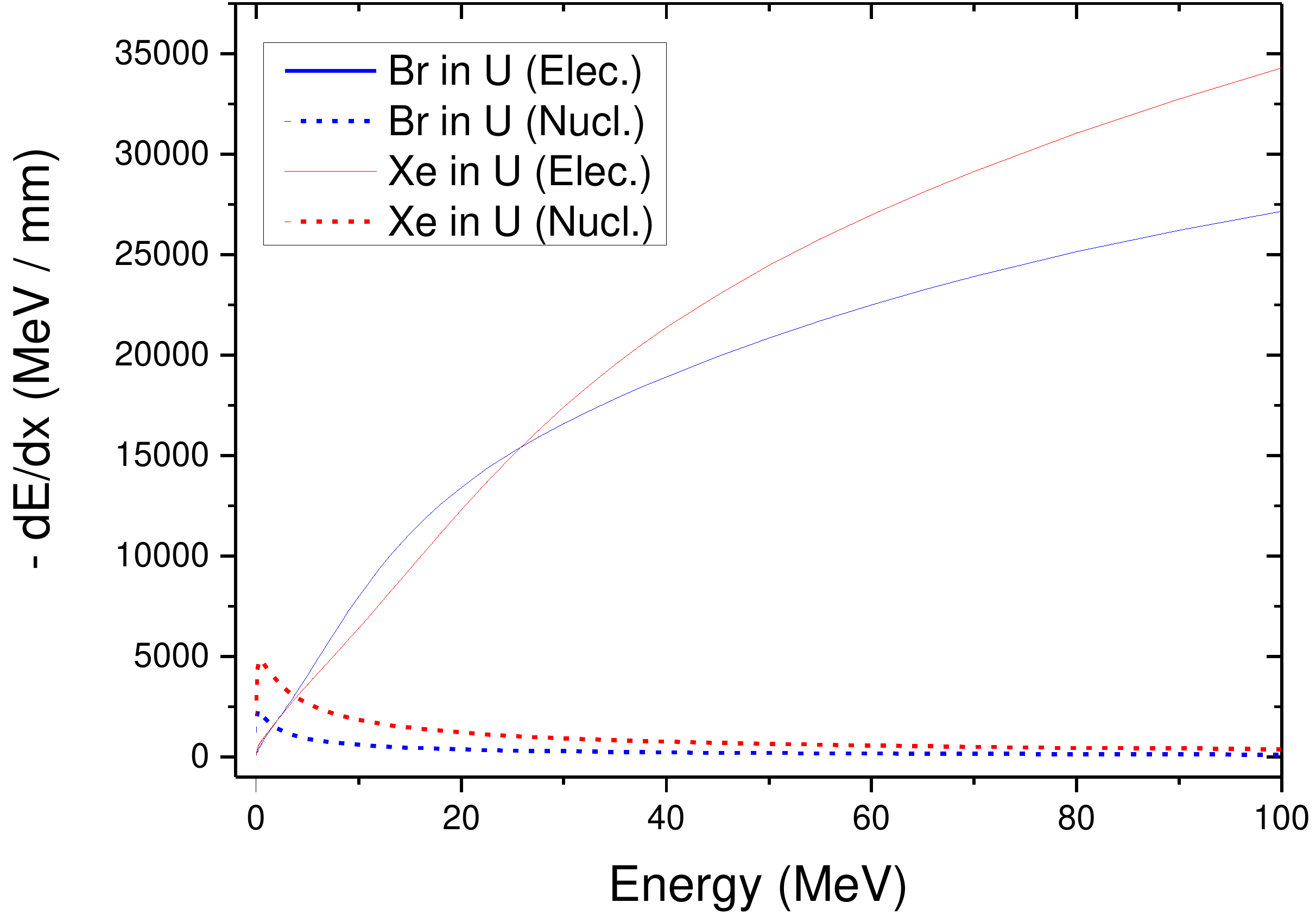} 
    \includegraphics[width=0.49\textwidth]{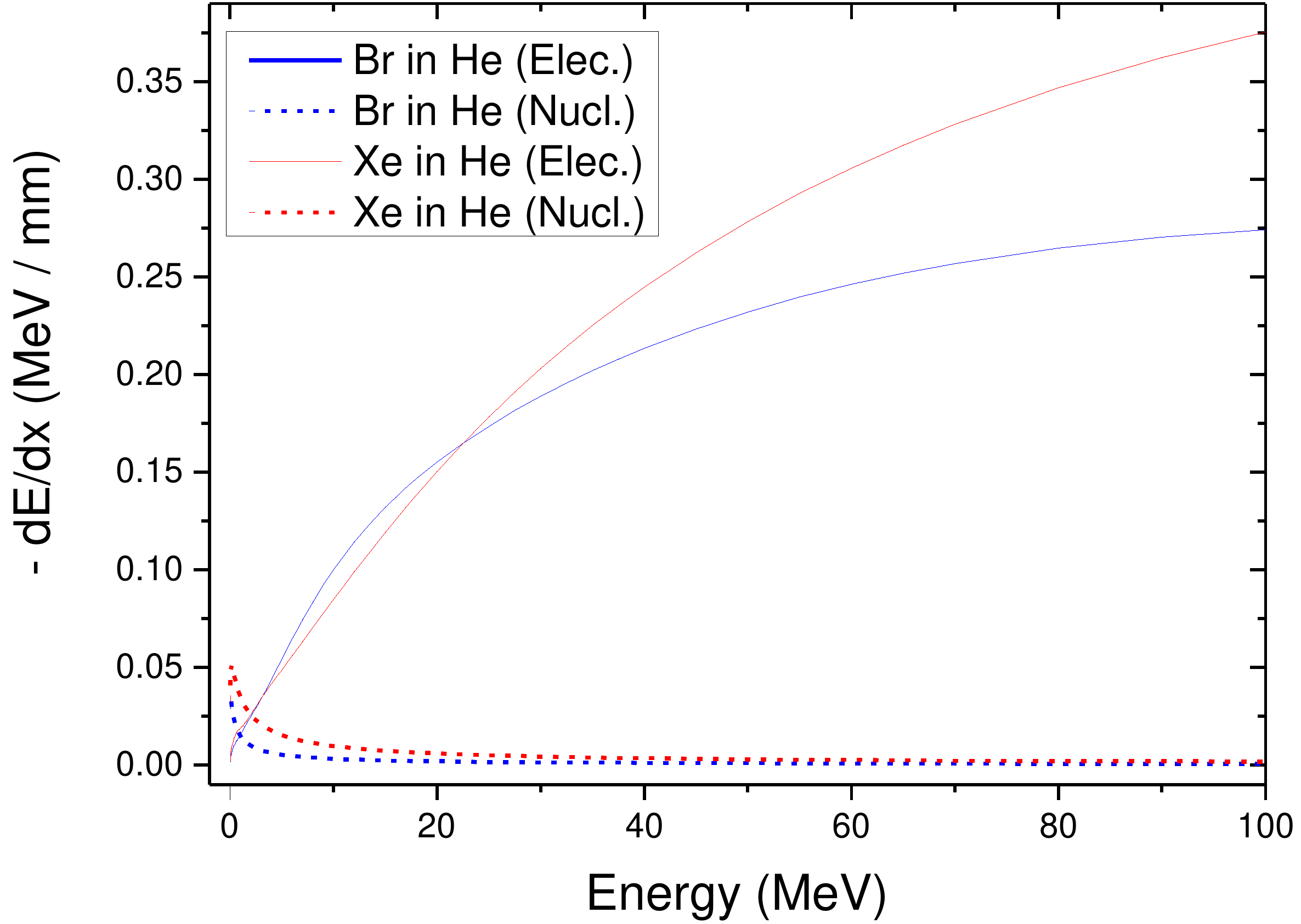}  
\caption[Stopping power of fission fragments.]{The SRIM stopping powers for Br-79 and Xe-132, in He, U and Ni. The stopping power is composed of interaction with electrons and with the target nuclei.}
\label{Fig4}
\end{figure}

The ion stopping power depends on both an interaction with the electrons $\lbrace{-dE/dx}\rbrace_e$ and with the nuclei $\lbrace{-dE/dx}\rbrace_N$ in the target. Figure~\ref{Fig4} shows calculated -dE/dx from SRIM, for Xe-132 and Br-79 in He, Ni and U. A combination of effects makes it difficult to predict the dependence on mass and charge of the ion stopping efficiency. At lower energies, the scattering against nuclei is dominant and is stronger for heavier nuclei. For energies between a few MeV and $\sim$25 MeV, scattering with electrons, $\lbrace{-dE/dx}\rbrace_e$, is larger. Note that below $\sim$25 MeV, $\lbrace{-dE/dx}\rbrace_e$ is larger for lighter ions whereas above $\sim$25 MeV it becomes larger for heavy ions.

Fission fragments are highly charged after formation (typically effective charge of 15-20+) but due to the large kinetic energy they tend to have about half the range of $\alpha$ particles \cite[p42,p521]{knoll}. $\alpha$ particles lose their energy mainly when they reach low energies, in the end of the track. Immediately after formation the fission fragments start to pick-up electrons from the surroundings. In contrast to $\alpha$ particles they lose their energy more evenly along the entire path due to a continuous reduction in the effective charge which compensates for the lower velocity \cite[p42,p521]{knoll}.

\clearpage
\section{Geant4 geometry}
\begin{figure}[b!]
\centering
    \includegraphics[width=0.7\textwidth]{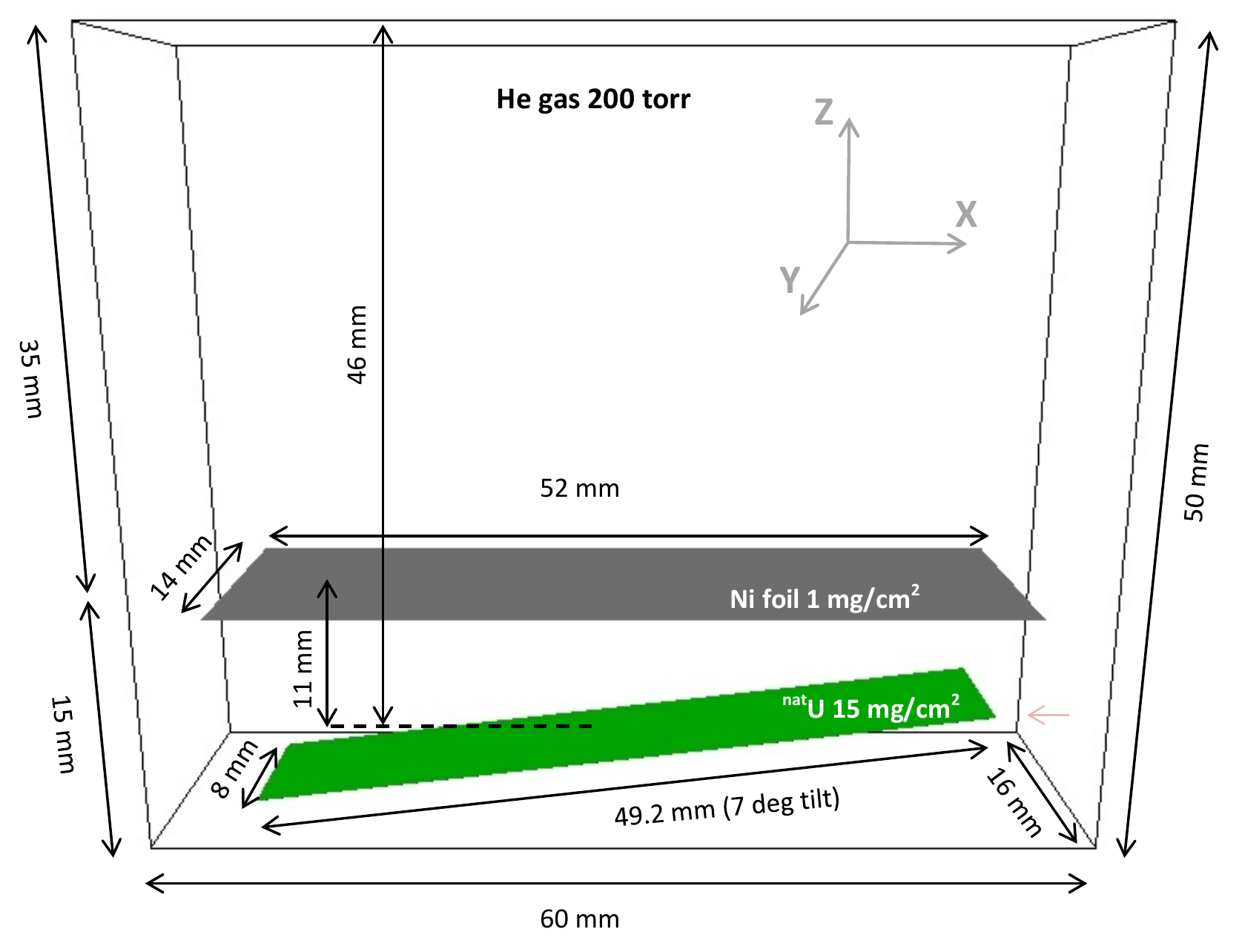}
    \includegraphics[width=0.44\textwidth]{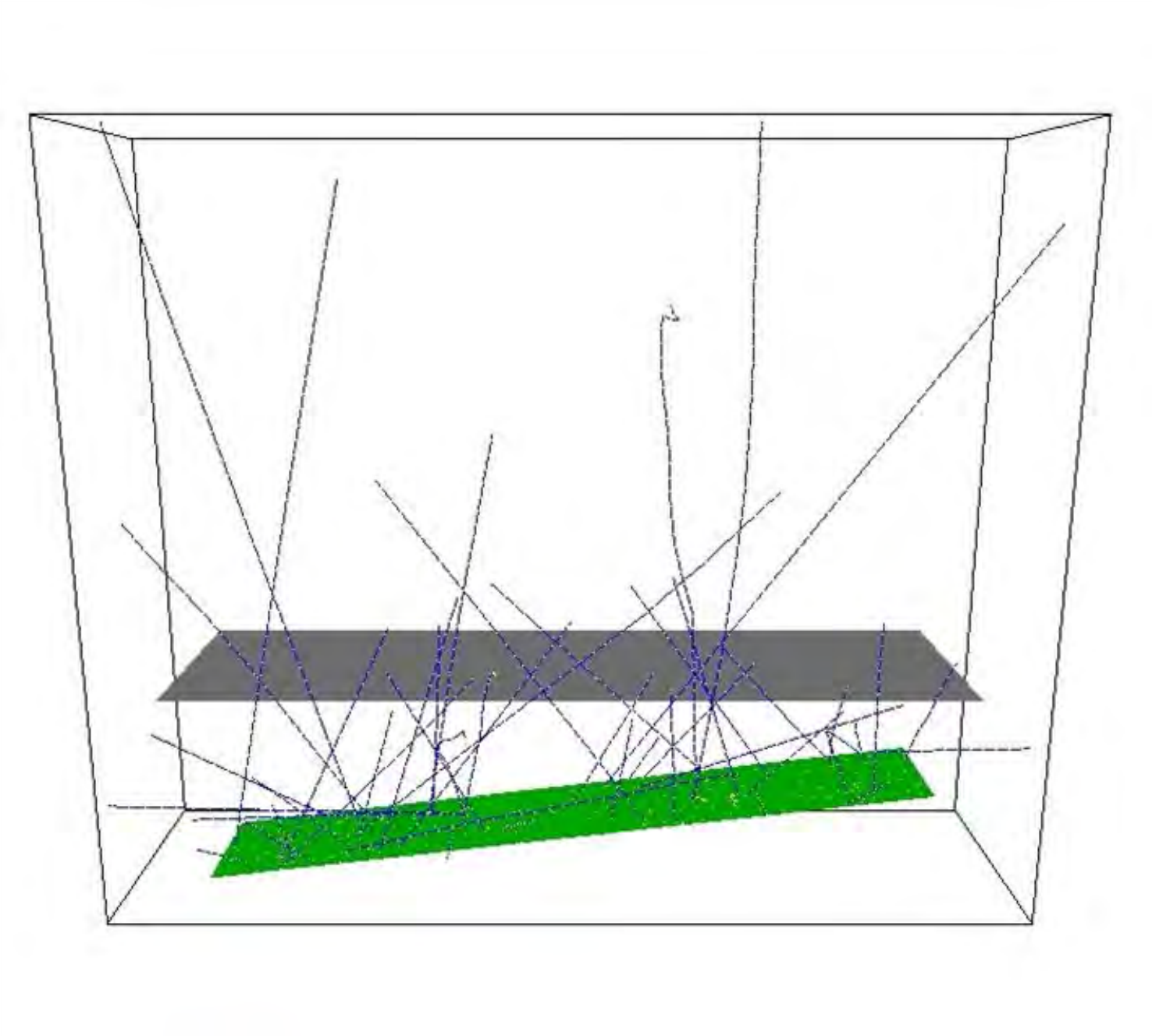}
    \includegraphics[width=0.13\textwidth]{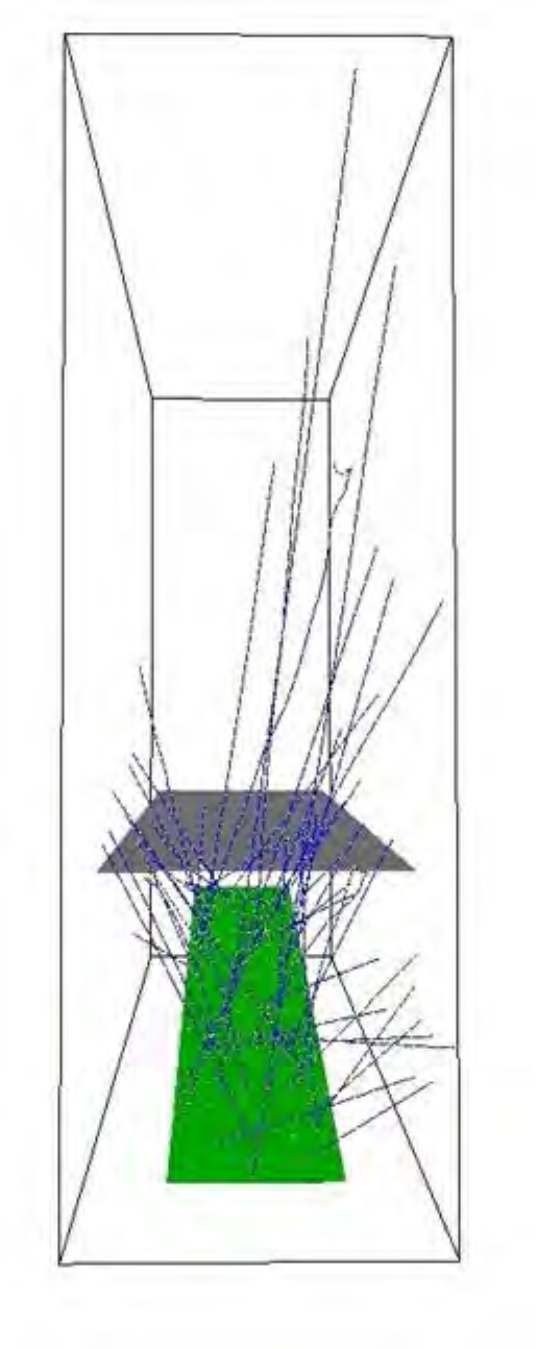} 
\caption[Geant4 target design.]{The chamber geometry in Geant4. The chamber is filled with He gas held at 200\,torr. The ions are started at random positions in all three dimensions of the target. The ion emission angle is also randomly selected in 2$\pi$. In the lower plots some typical ion tracks are shown in two projections.}
\label{Fig5}
\end{figure}

Geant4, a Monte Carlo simulation toolkit \cite{GeantNIM03}, was employed to simulate the ion stopping efficiency. The calculations were also compared to SRIM simulations \cite{Ziegler85}, in order to validate the results. The chamber geometry was adopted from Ref. \cite{sonoda}. The start positions of the ions are selected randomly in all three dimensions of the target. The ion emission angle is also randomly selected in 2$\pi$ (for a 4$\pi$ case see sect. \ref{secGeant4details}). Further details on the calculations can be found in sect. \ref{secGeant4details}. Figure~\ref{Fig5} shows the chamber geometry.  

\clearpage
\section{Simulations\label{sectsimulations}}
\begin{figure}[b!]
\centering
    \includegraphics[width=0.47\textwidth]{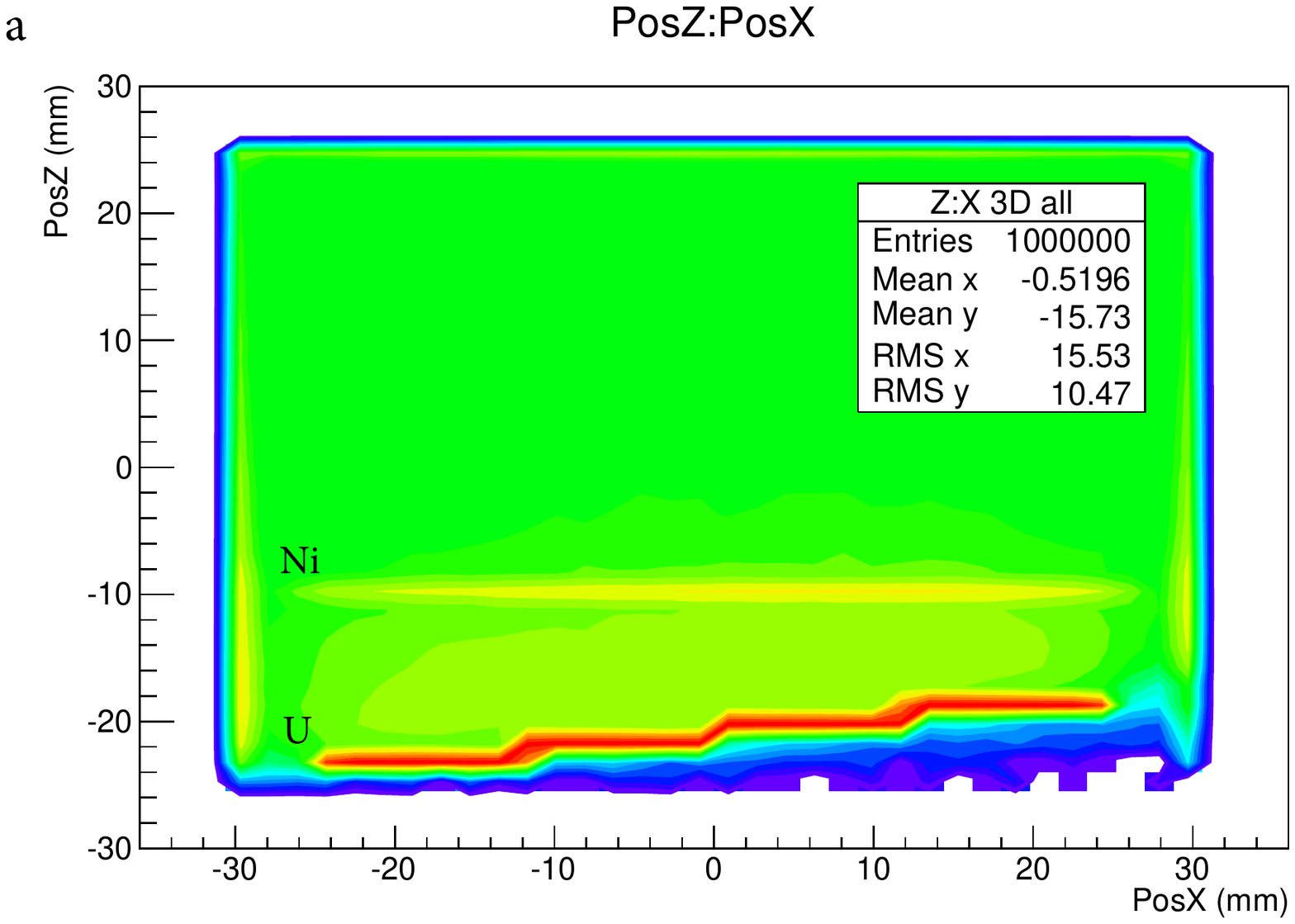}
    \includegraphics[width=0.51\textwidth]{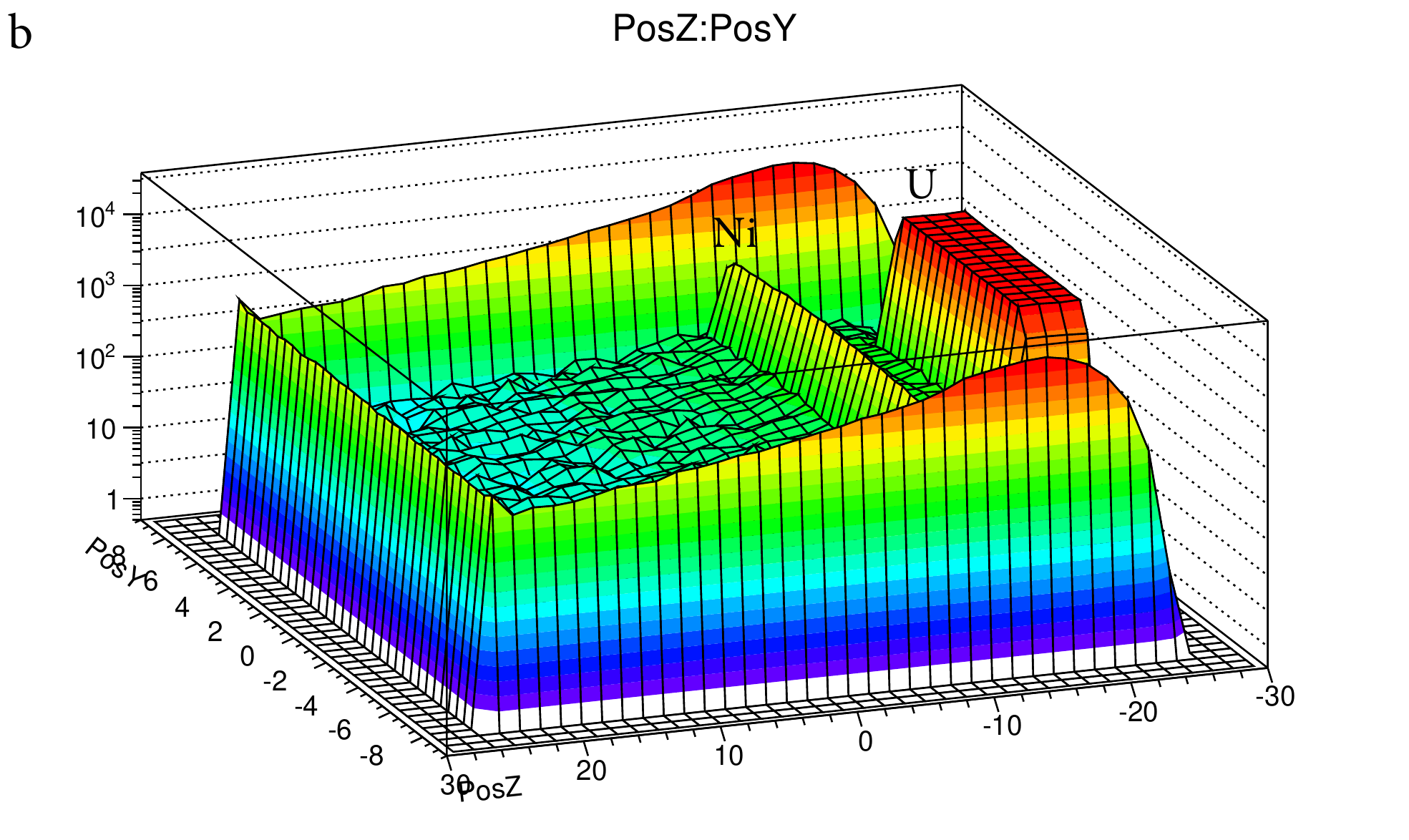}
\caption[Stopped ions in the simulation.]{Number of stopped ions in the $XZ$- (a) and $YZ$-plane (b). Most events are lost either in the target, in the Ni foil or in the walls of the chamber.}
\label{Fig6}
\end{figure}

\begin{figure}[!]
\centering
    \includegraphics[width=0.7\textwidth]{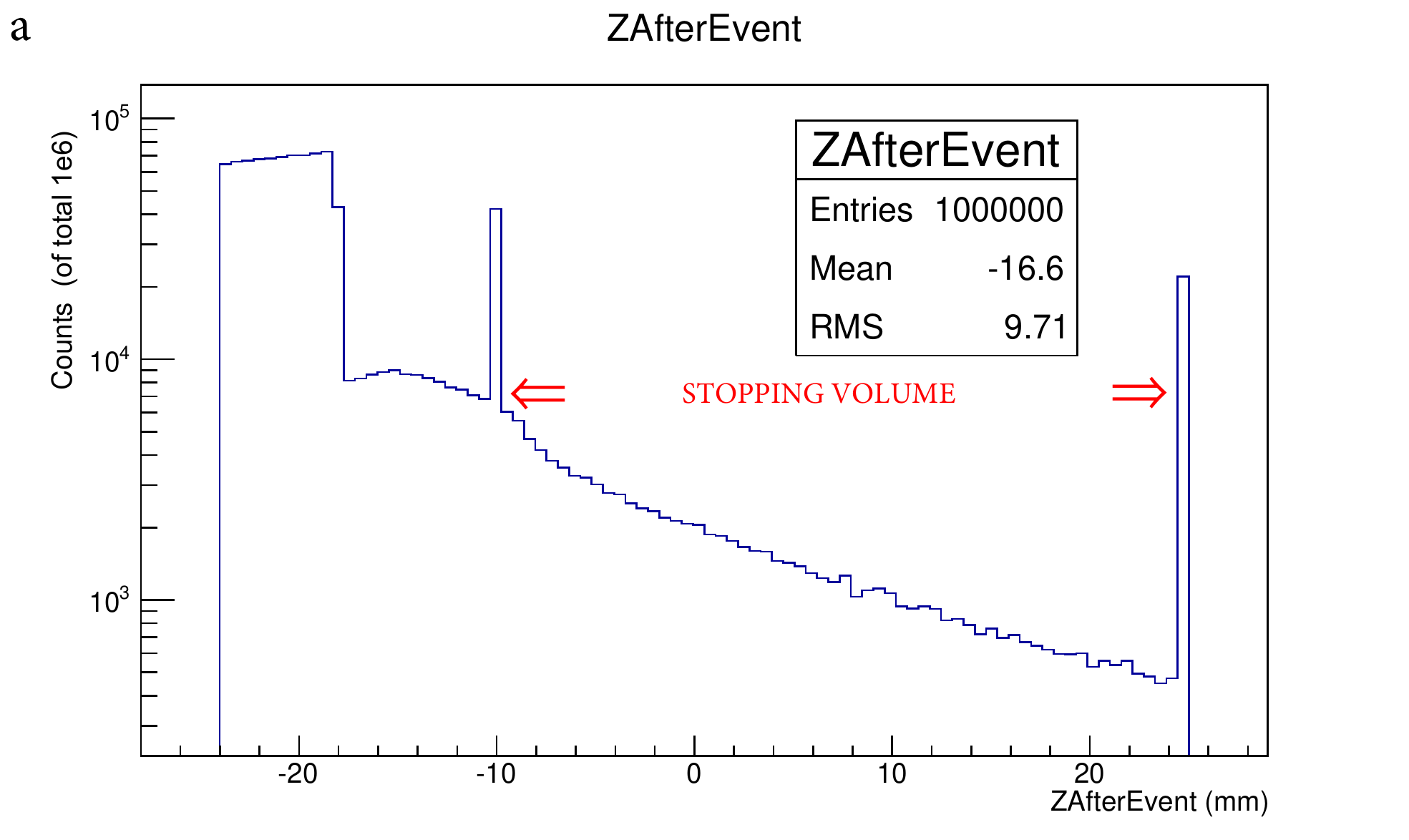}
    \includegraphics[width=0.7\textwidth]{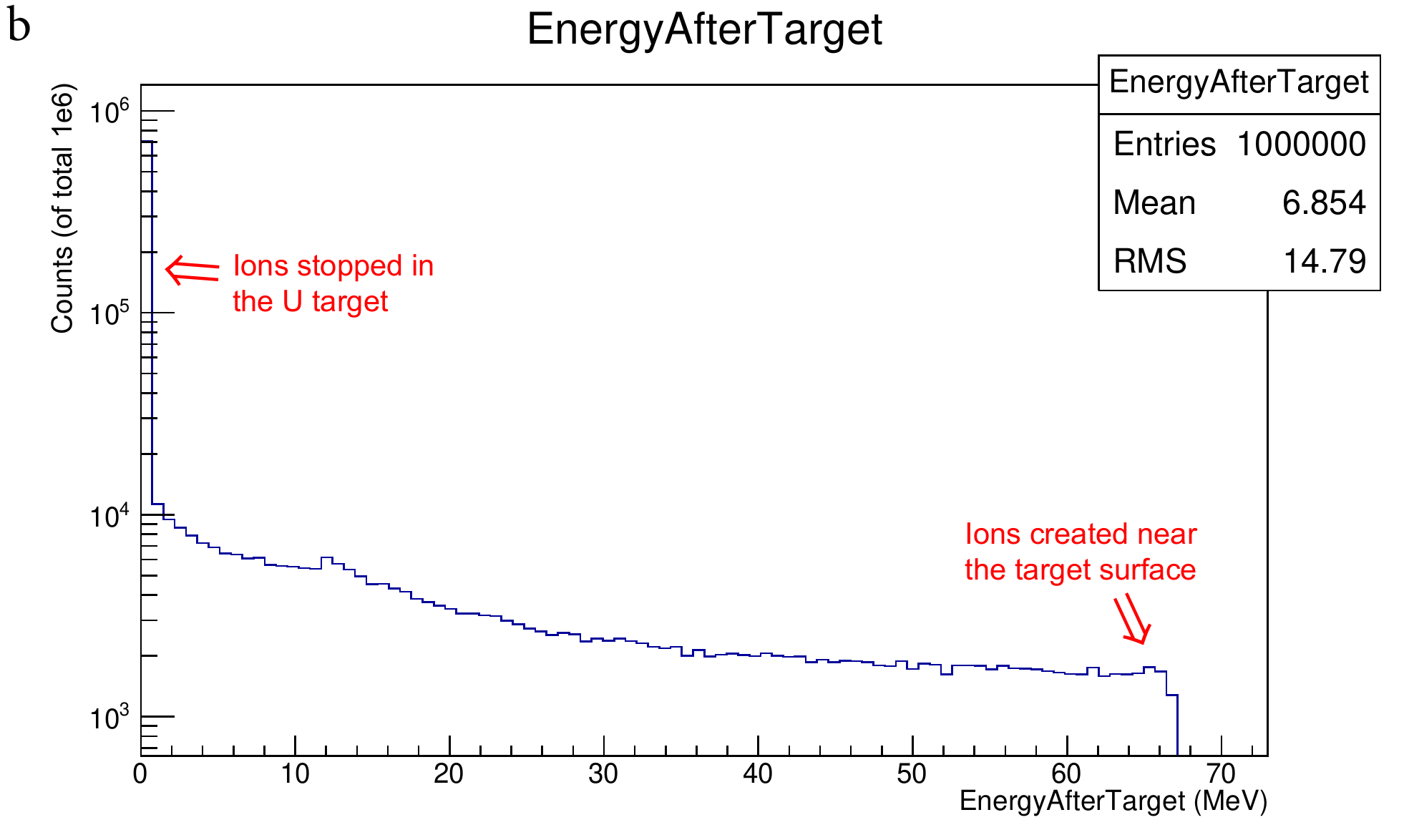}
    \includegraphics[width=0.7\textwidth]{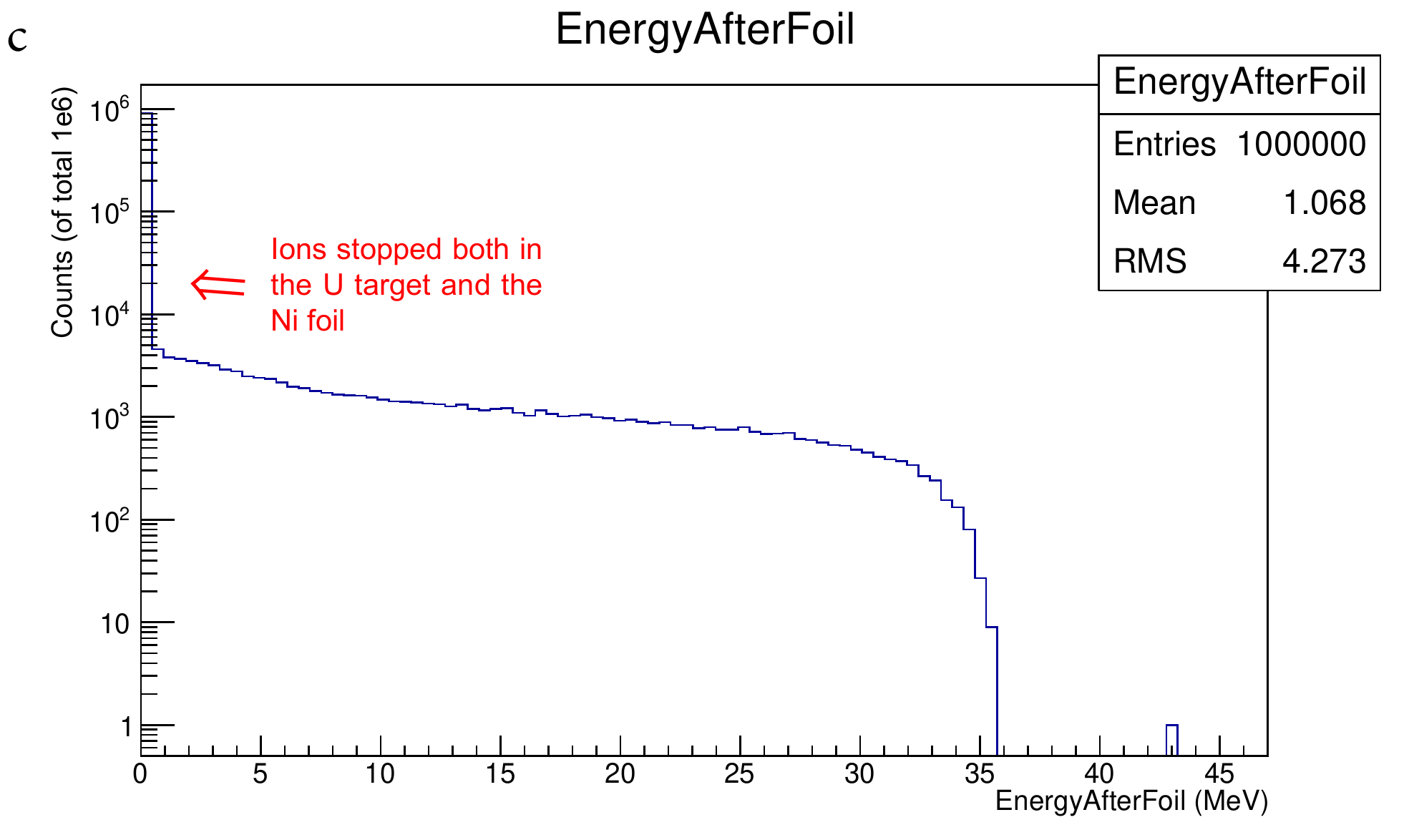}
\caption[The particle range and energies after U target and Ni foil.]{a) The range of the ions (in mm) along the ion guide. b) Energy After leaving the U target. c) Energy after ions leave the Ni foil.  }
\label{Fig7}
\end{figure}

Table \ref{table1} shows the simulated ions along with the statistics. In the following example $10^6$ ions from the run $A = 140$, $Z = 54$ and $E = 67$\,MeV is discussed. In Fig.~\ref{Fig6} the stopped ions are shown as a function of position in $X$ and $Z$ (a) and $Y$ and $Z$ (b). The vast majority of all ions are stopped in the Uranium target or hitting the walls of the chamber. A significant fraction is also stopped in the Ni foil. Below 0.5\% of all ions in 4$\pi$ are stopped in the stopping volume. The range of the fission products is shown in Fig.~\ref{Fig7}a where the Ni foil is at position -10 mm and the volume ends at position 25 mm. Also the energies after leaving the target and Ni foil are shown in Fig.~\ref{Fig7}b,c. 

\begin{figure}[b!]
\centering
    \includegraphics[width=0.495\textwidth]{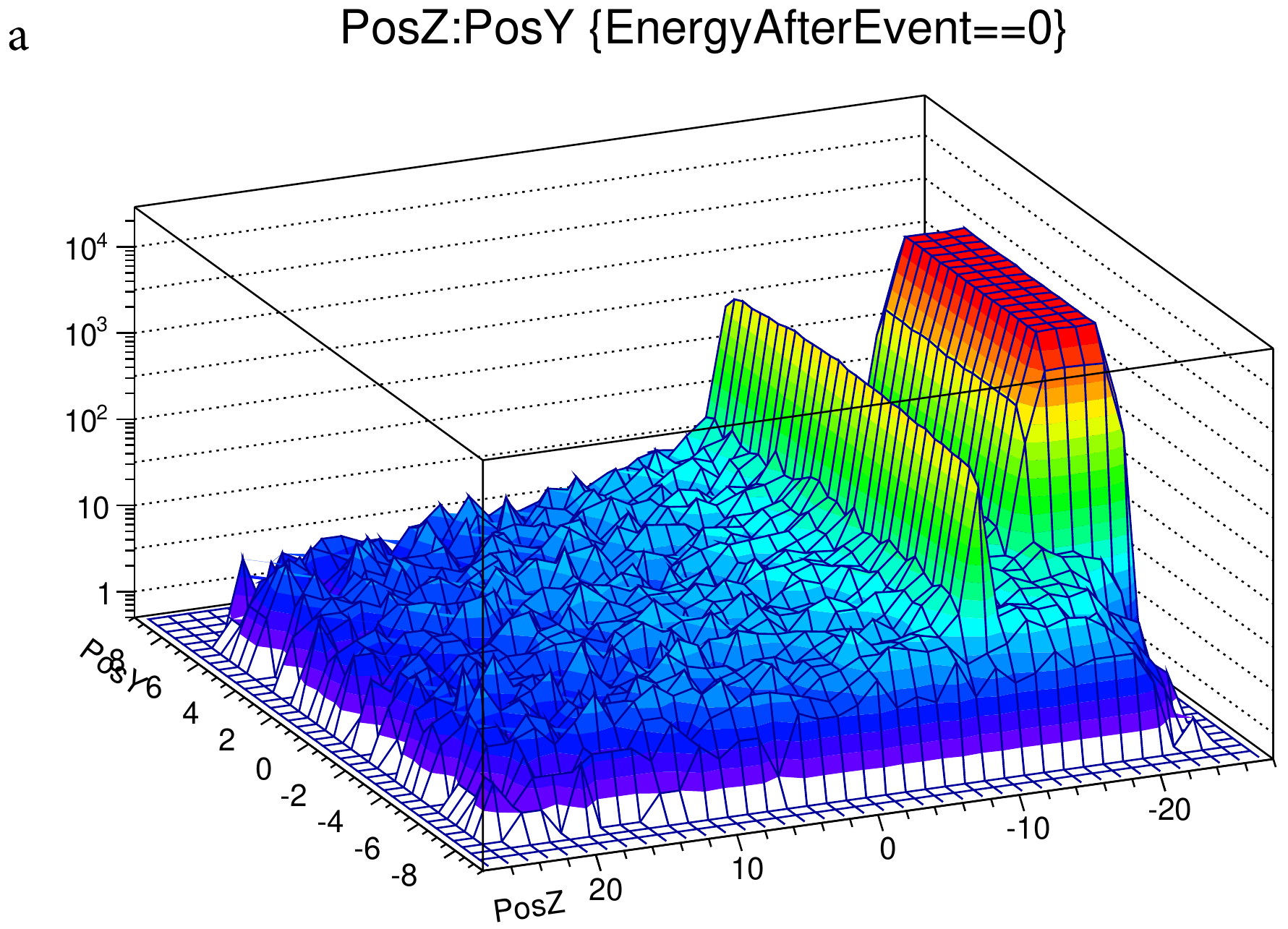}
    \includegraphics[width=0.495\textwidth]{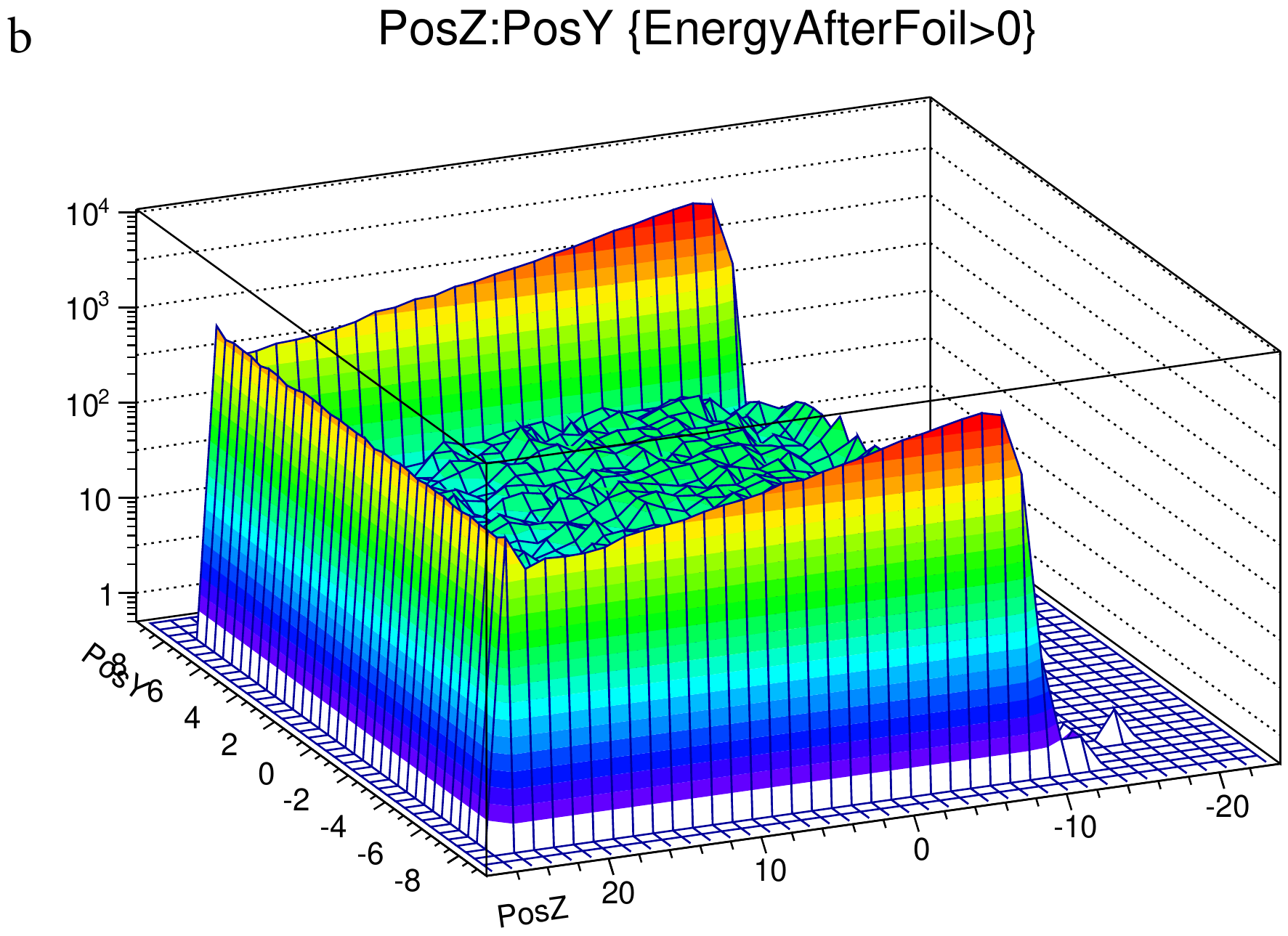}    
    \includegraphics[width=0.95\textwidth]{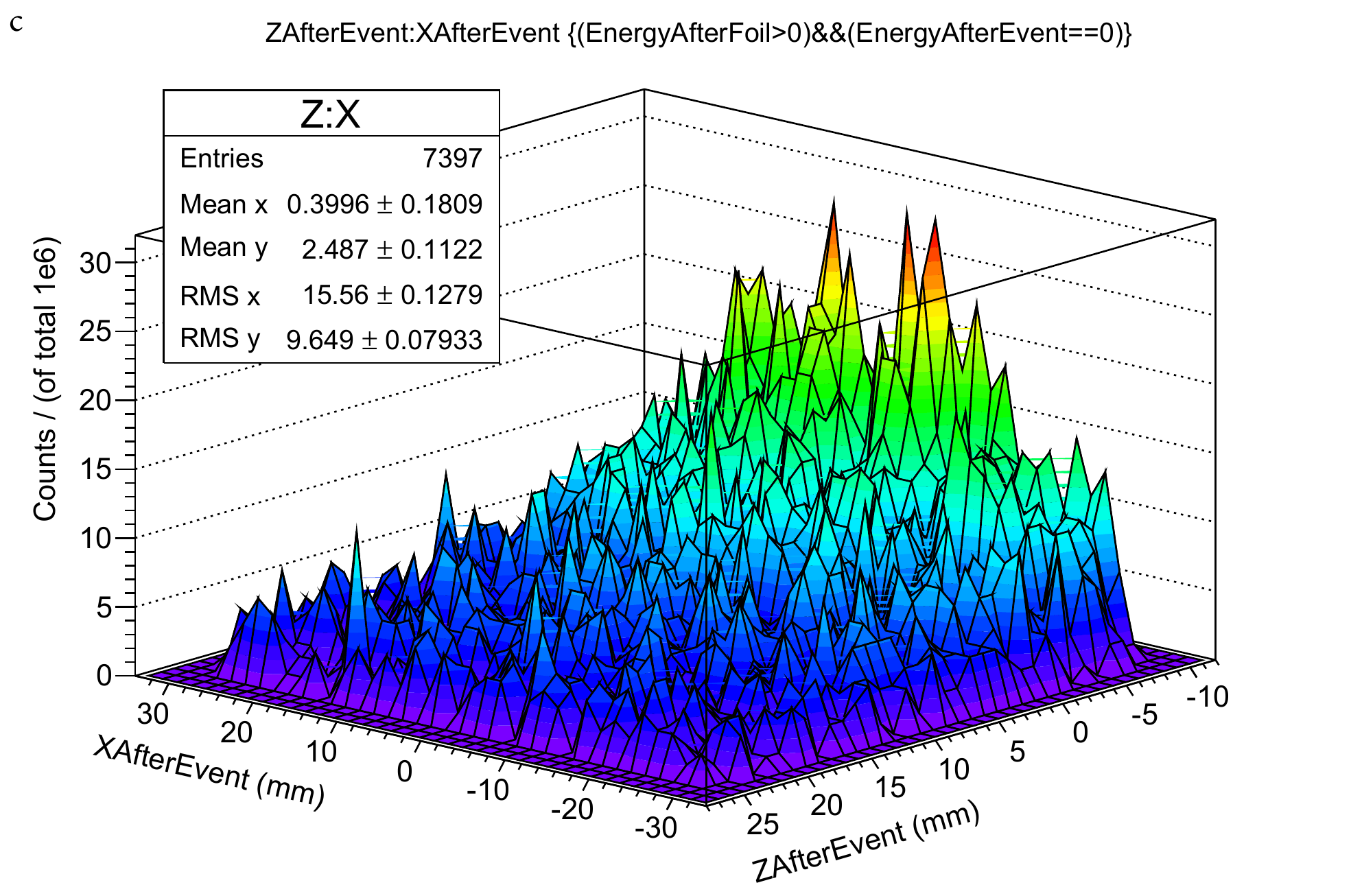}    
\caption[The conditions applied to cut the data.]{Visualization of the conditions applied to select the ions which are stopped in the stopping chamber. In (a) and (b) the individual cuts on the final ion energy and the energy after leaving the foil are shown. By combining both cuts in (c) one obtains the final number of ions stopped in the stopping volume. }
\label{Fig8}
\end{figure}

In order to select the ions stopped in the buffer gas from the total data sets, two constraints were put on the ion energies.  All presented data in this work are denoted with ``CUT'' if they fulfil both of the following criteria: 

\begin{itemize}
\item[$\diamond$] Final kinetic energy of ion = 0 : All ions that end up having $E = 0$ are stopped somewhere inside the reaction chamber. Figure~\ref{Fig8}a shows these ions ending up in the target, foil and in the stopping chamber. This condition rejects the ions hitting the walls.

\item[$\diamond$] Energy after the ions leave the Ni foil $>$ 0 : This ensures that all selected ions pass the foil. Hence, all ions stopped in the target and Ni foil are rejected (See Fig.~\ref{Fig8}b). 
\end{itemize}

\begin{figure}[H]
\centering
    \includegraphics[width=0.6\textwidth]{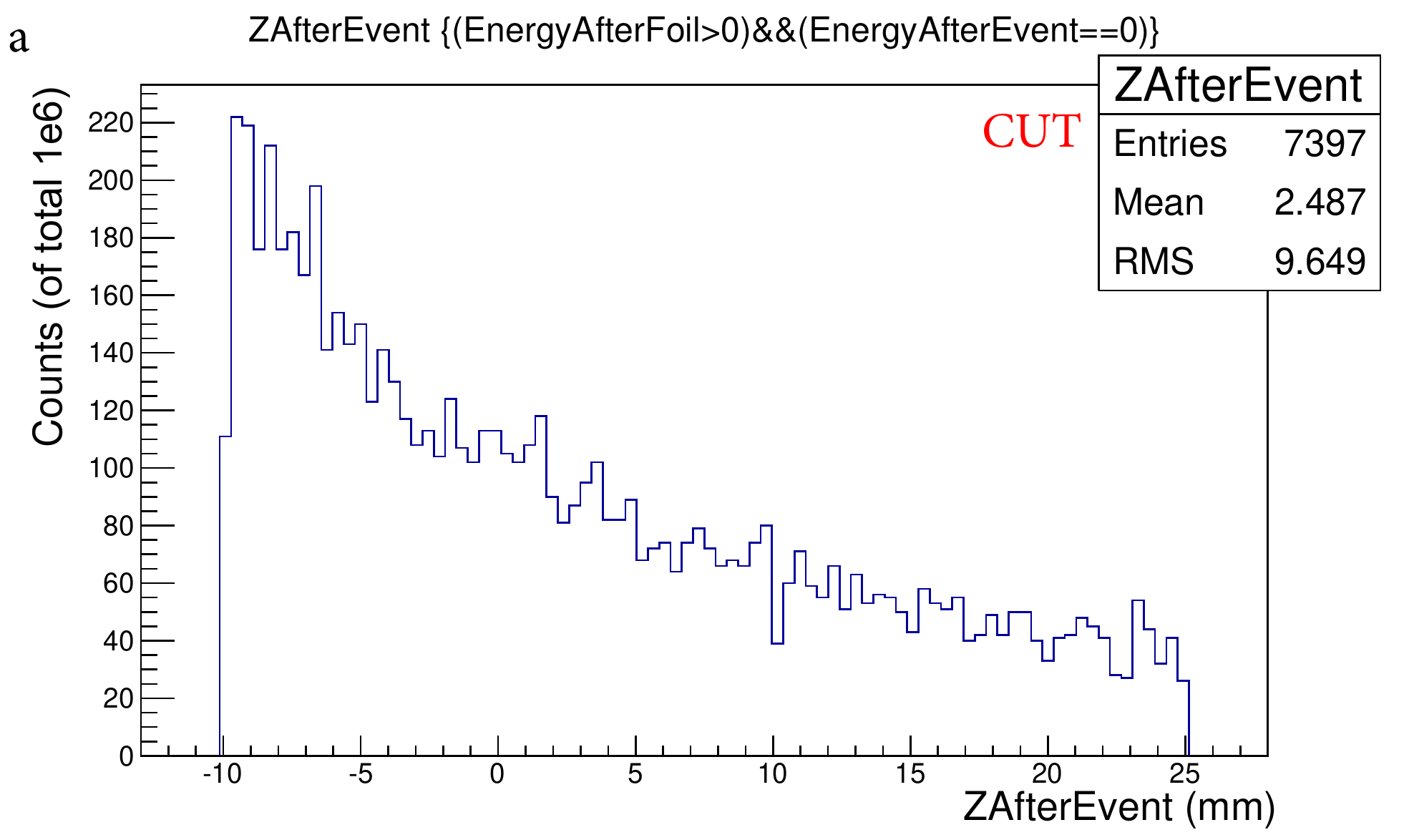}
    \includegraphics[width=0.6\textwidth]{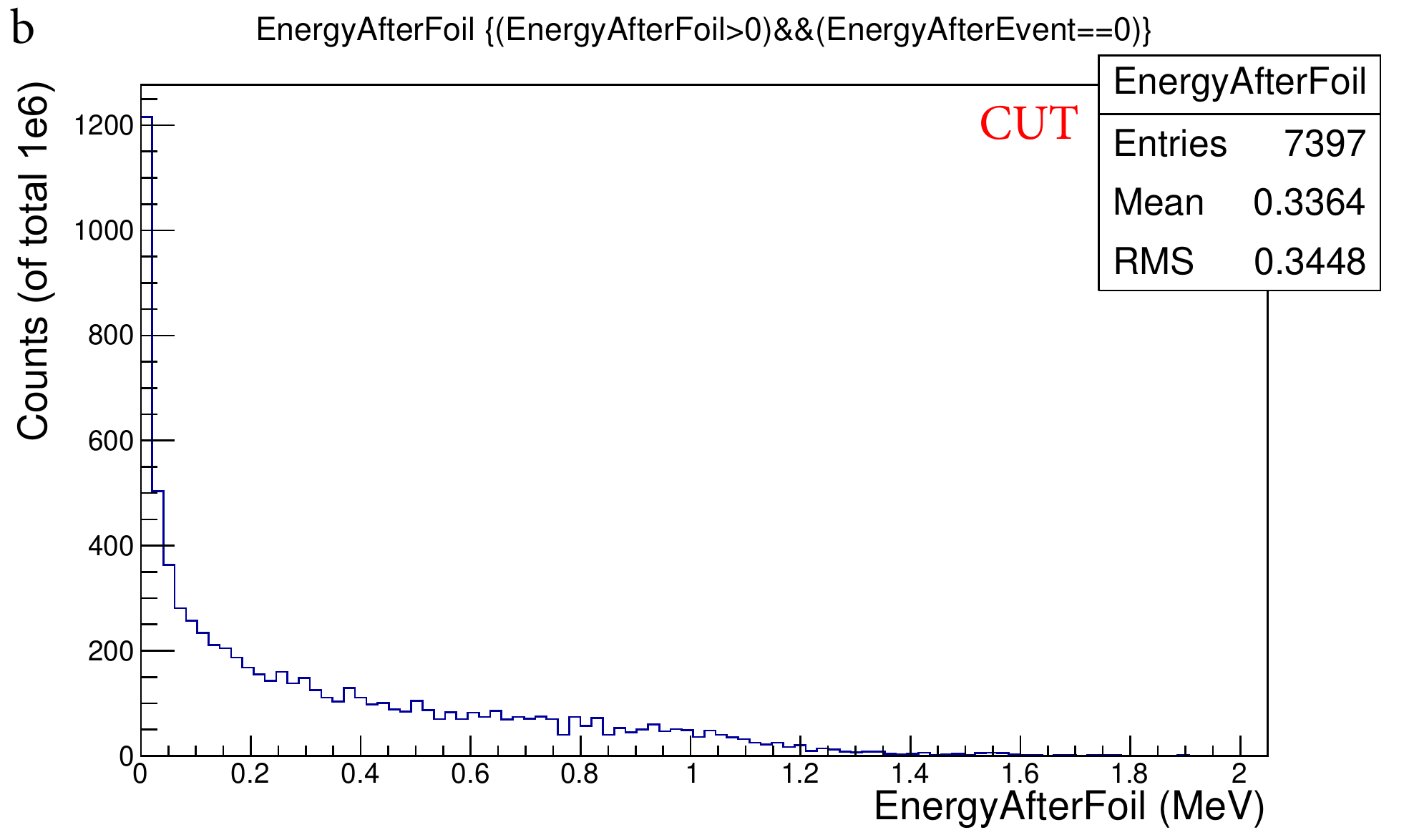}
    \includegraphics[width=0.6\textwidth]{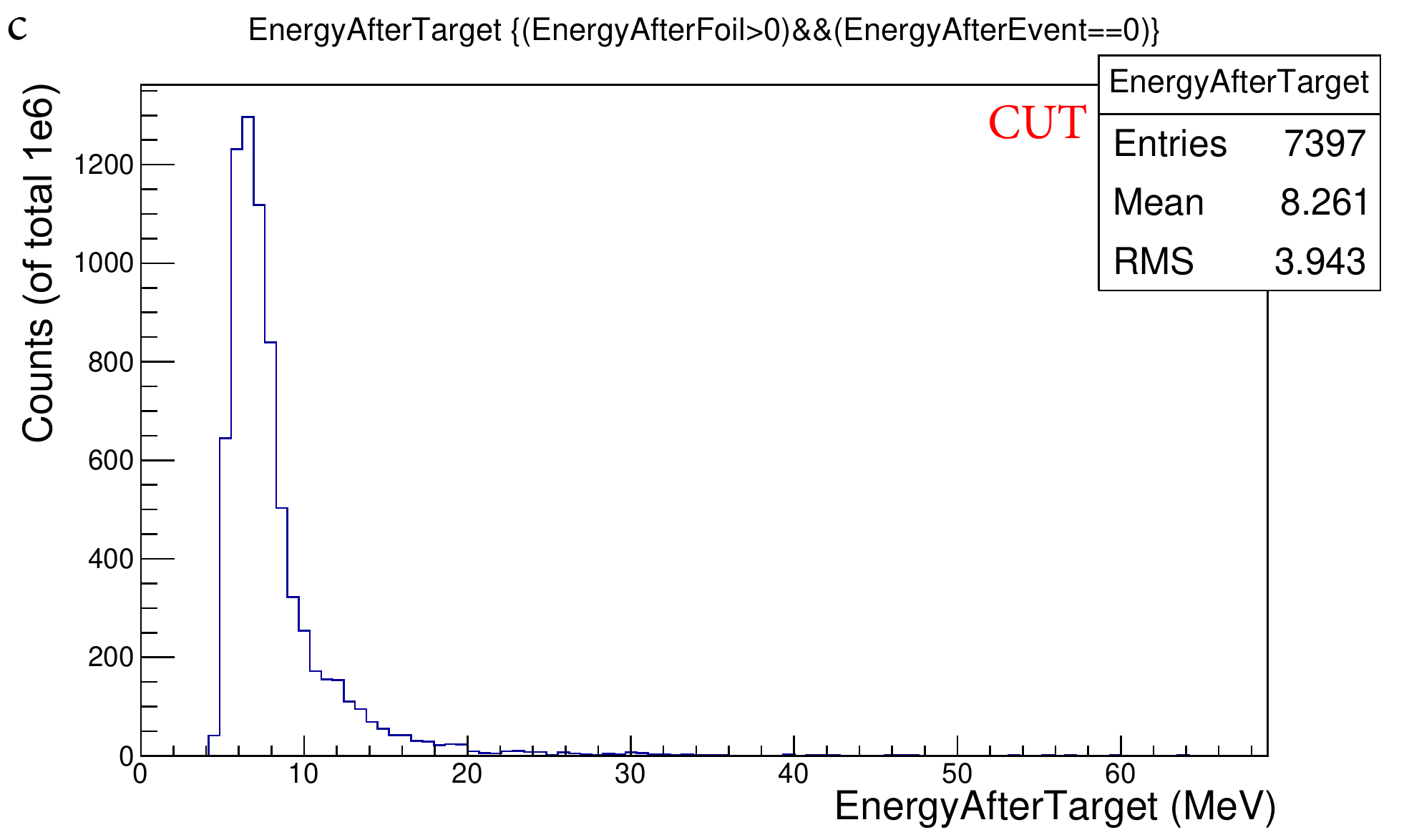}
\caption[(a) The range along the z direction. (b) Energy after foil. (b) Energy after target.]{(a) The range along the Z direction. (b) Energy after foil. (c) Energy after target, for the ions stopped in the gas cell.}
\label{Fig9}
\end{figure}

\begin{figure}[b!]
\centering
    \includegraphics[width=0.49\textwidth]{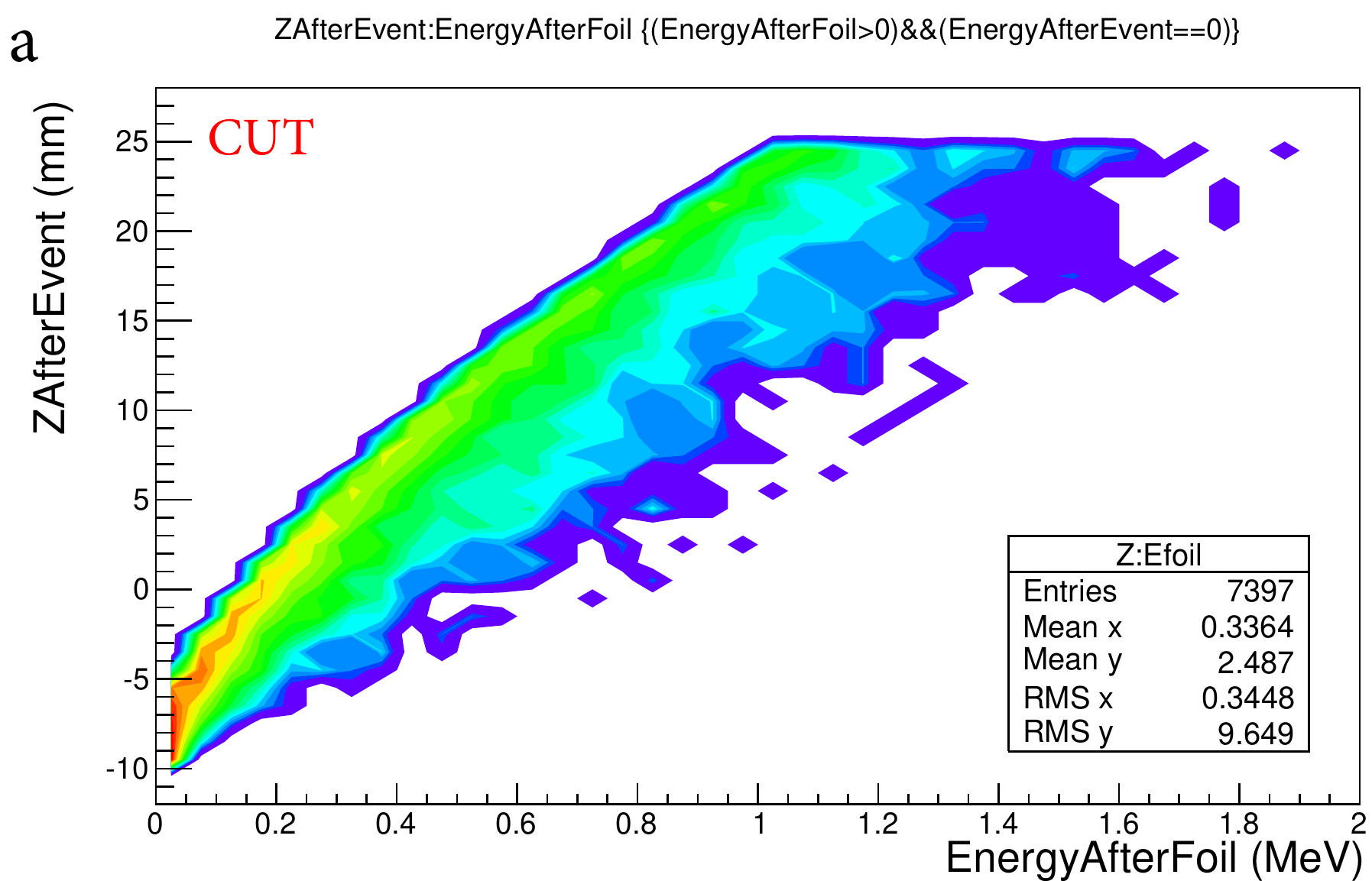}
    \includegraphics[width=0.49\textwidth]{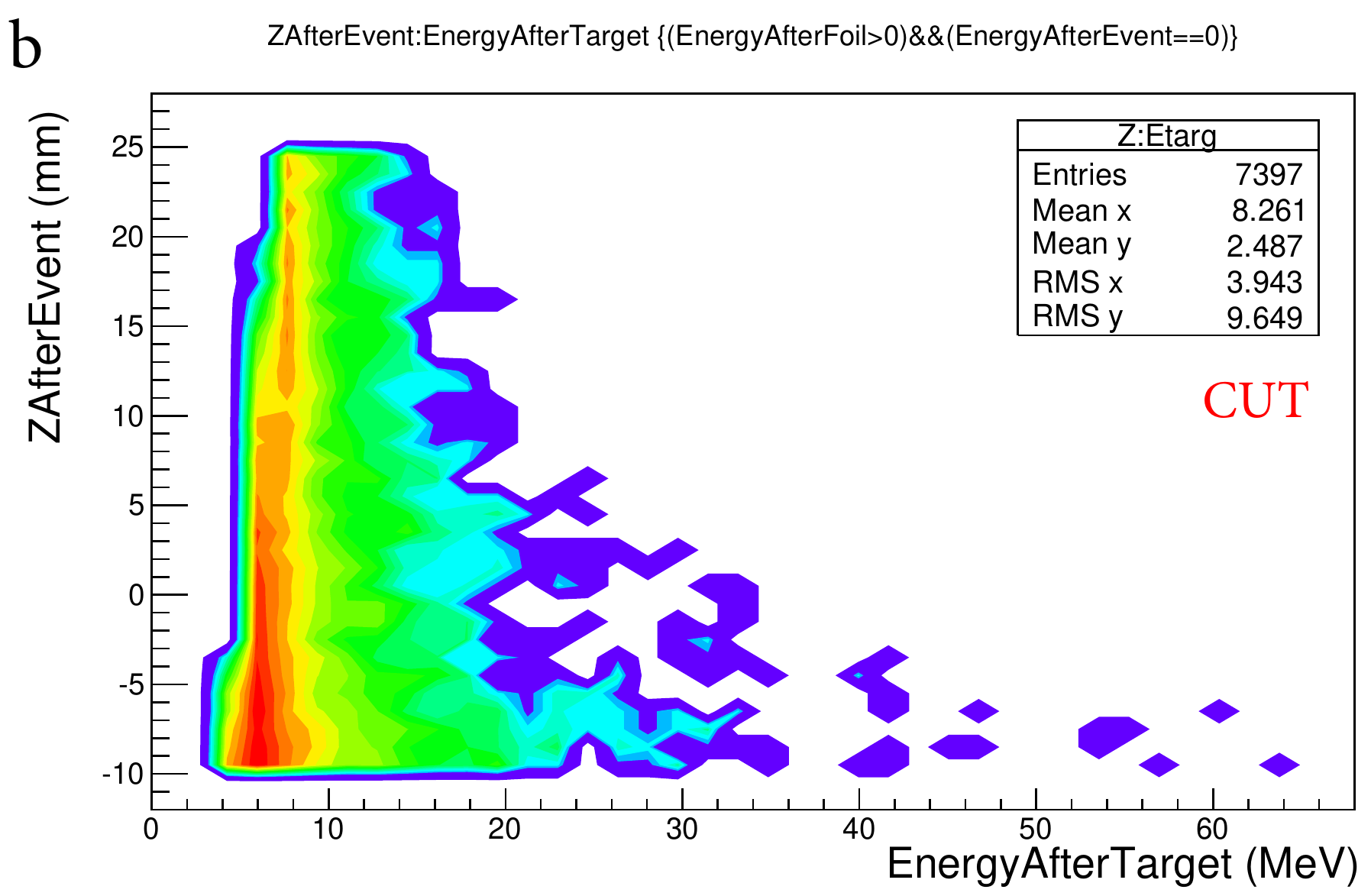}
    \includegraphics[width=0.49\textwidth]{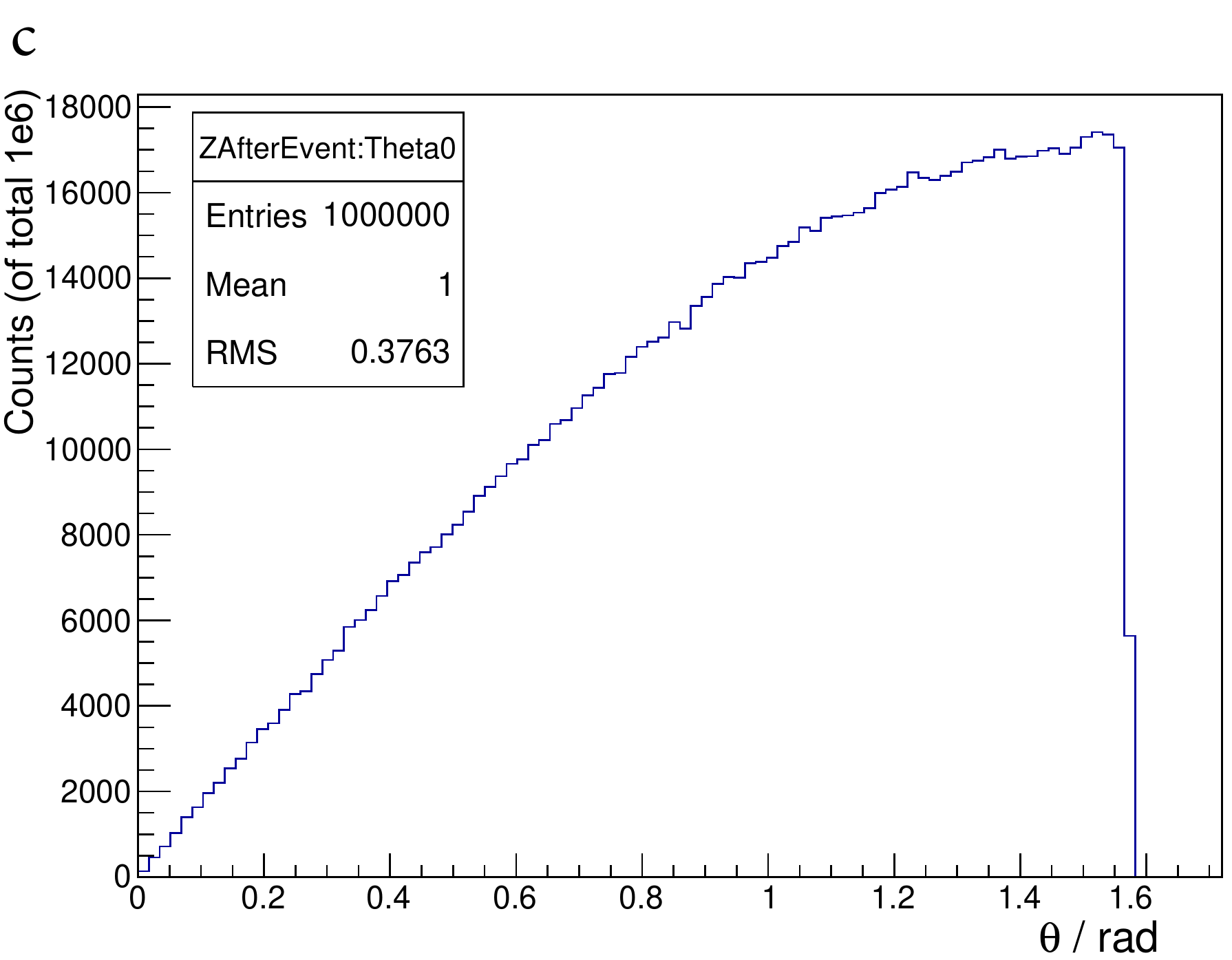}
    \includegraphics[width=0.49\textwidth]{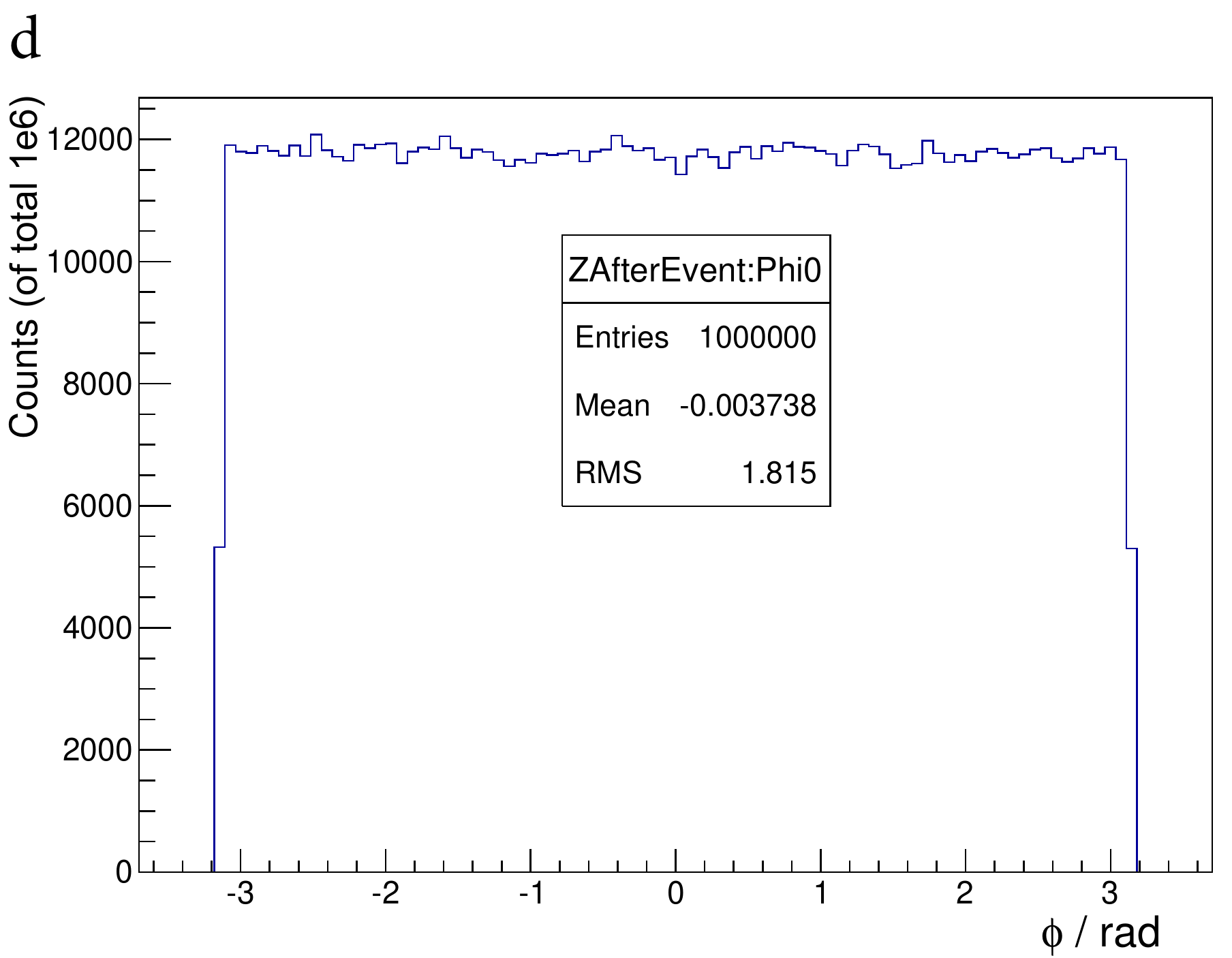}
    \includegraphics[width=0.49\textwidth]{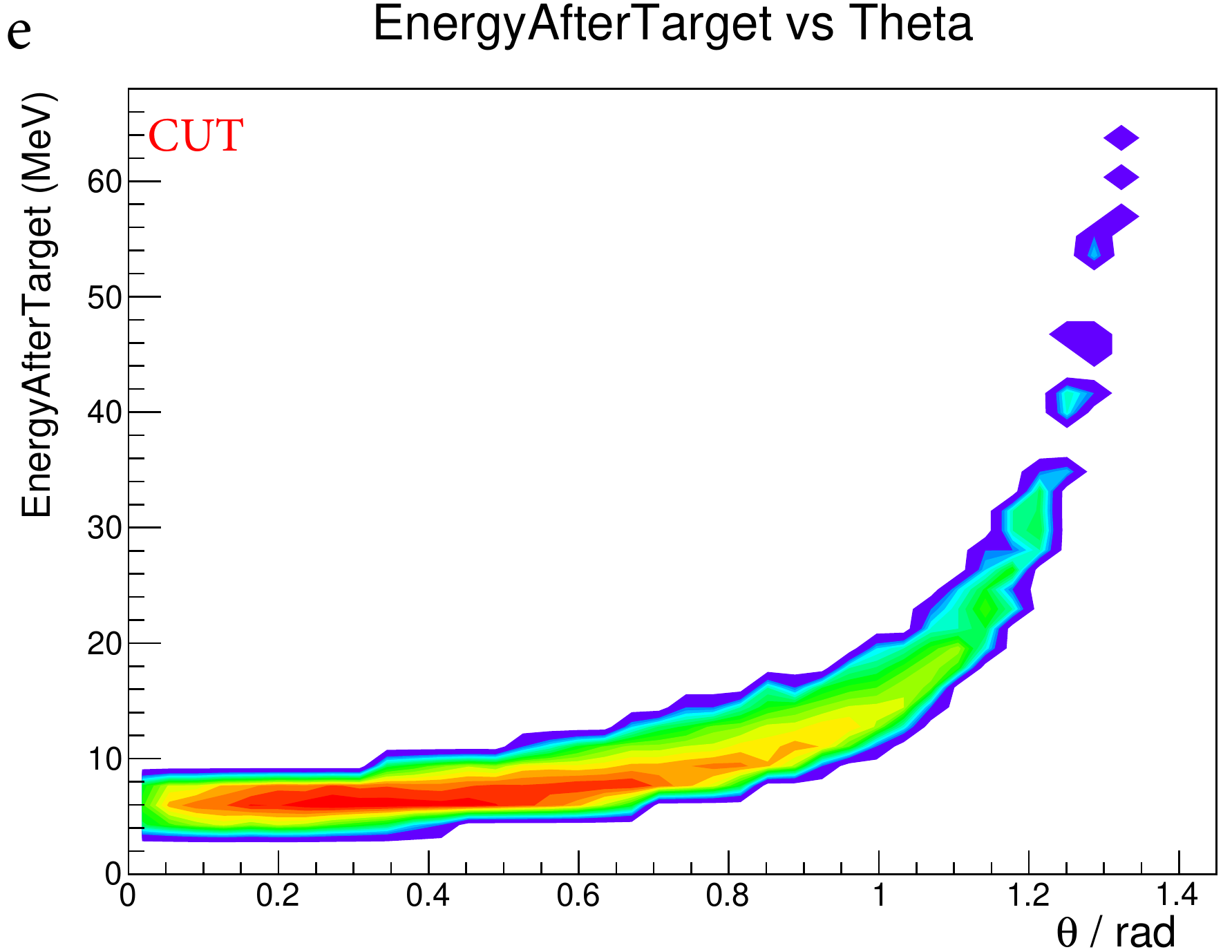}
    \includegraphics[width=0.49\textwidth]{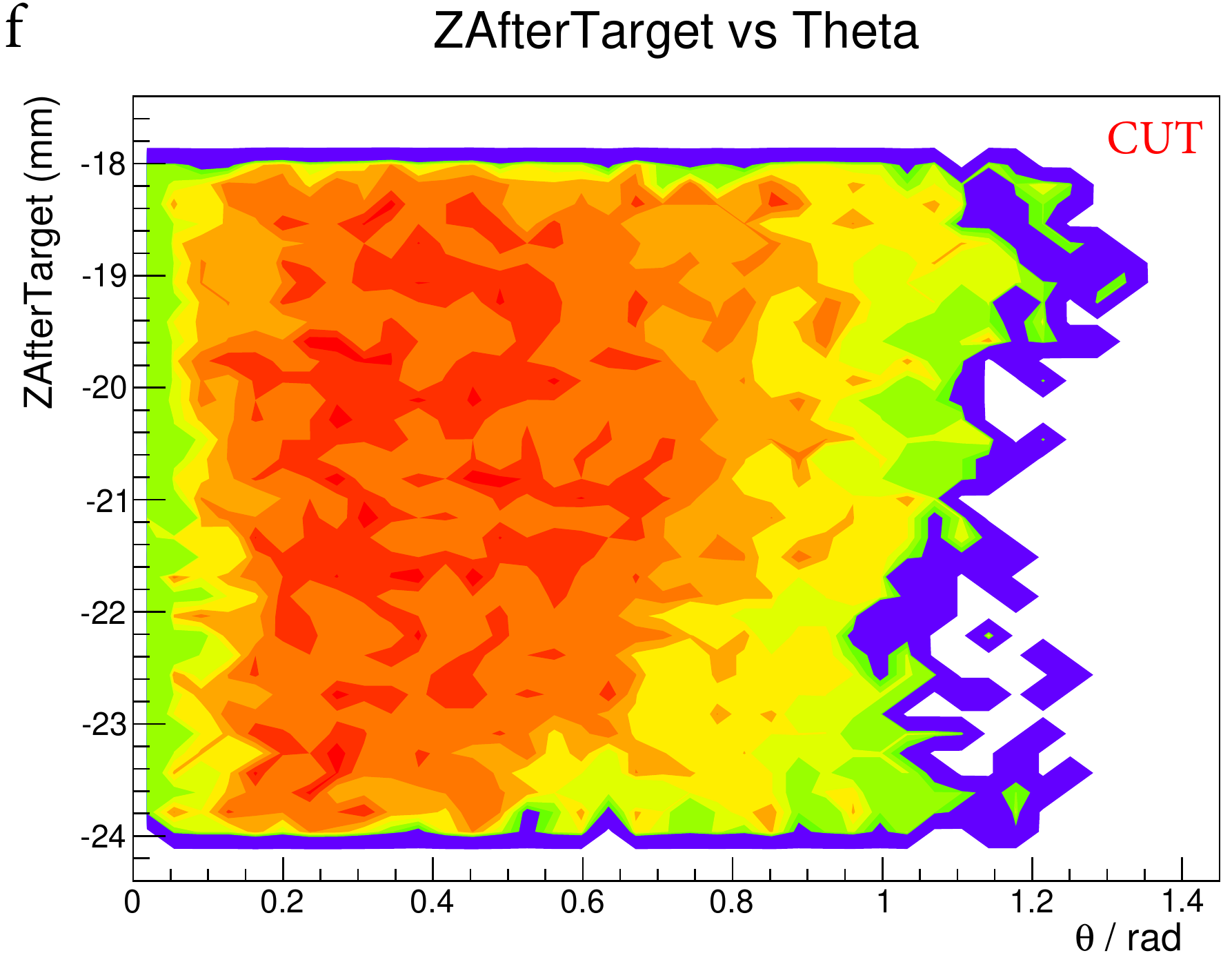}
\caption[Plots on various observables.]{(a) A two-dimensional scatter plot showing the fragment energy after leaving the Ni foil versus $Z$-range. Higher energies give longer range. In (b) the fragment energy after leaving the target is shown as a function of $Z$-range. (c and d) The random emission angles, $\theta$ and $\phi$, for all ions. In (e) the energy after target is shown as a function of $\theta$ for the stopped ions. The ions that are emitted in larger angles must have larger energies to pass the target and foil. (f) The $Z$ position of ions after leaving the target as a function of $\theta$.}
\label{Fig10}	
\end{figure}

The combined effect of both conditions is shown in Fig.~\ref{Fig8}c. Out of $10^6$ simulated ions in 2$\pi$ about 1\% were stopped in the stopping volume. In all simulations most ions were stopped right after the Ni foil and the number gradually declines as a function of the $Z$-range. This was observed for all ions independently of mass and energy. The one-dimensional plot of the range is seen in Fig.~\ref{Fig9}a. The energy distribution after leaving the Ni foil is shown in Fig.~\ref{Fig9}b. It was noted that nearly all stopped ions leave the foil with kinetic energies below $\sim 2$\,MeV, again more or less regardless of mass and energy. Figure~\ref{Fig9}c shows the ion energy after leaving the target for the cut selection applied. Ions with a mean energy around 9\,MeV after leaving the Uranium target are collected in the stopping volume. For lower kinetic energies the ions have an increased risk of being stopped in the target or foil, for higher energies the ions risk hitting the walls. Figure~\ref{Fig10} shows the relation between the ion energy after foil (a) and target (b) and the final ion range. Plots (c) and (d) show the simulated emission angles both in $\theta$ and $\phi$. In addition, plots (e) and (f) show the correlation between the emission angle and the energy and position after ions leave the U target. 

\clearpage
\section{Mass dependence of stopping efficiency} 
\begin{figure}[b!]
\centering
    \includegraphics[width=0.95\textwidth]{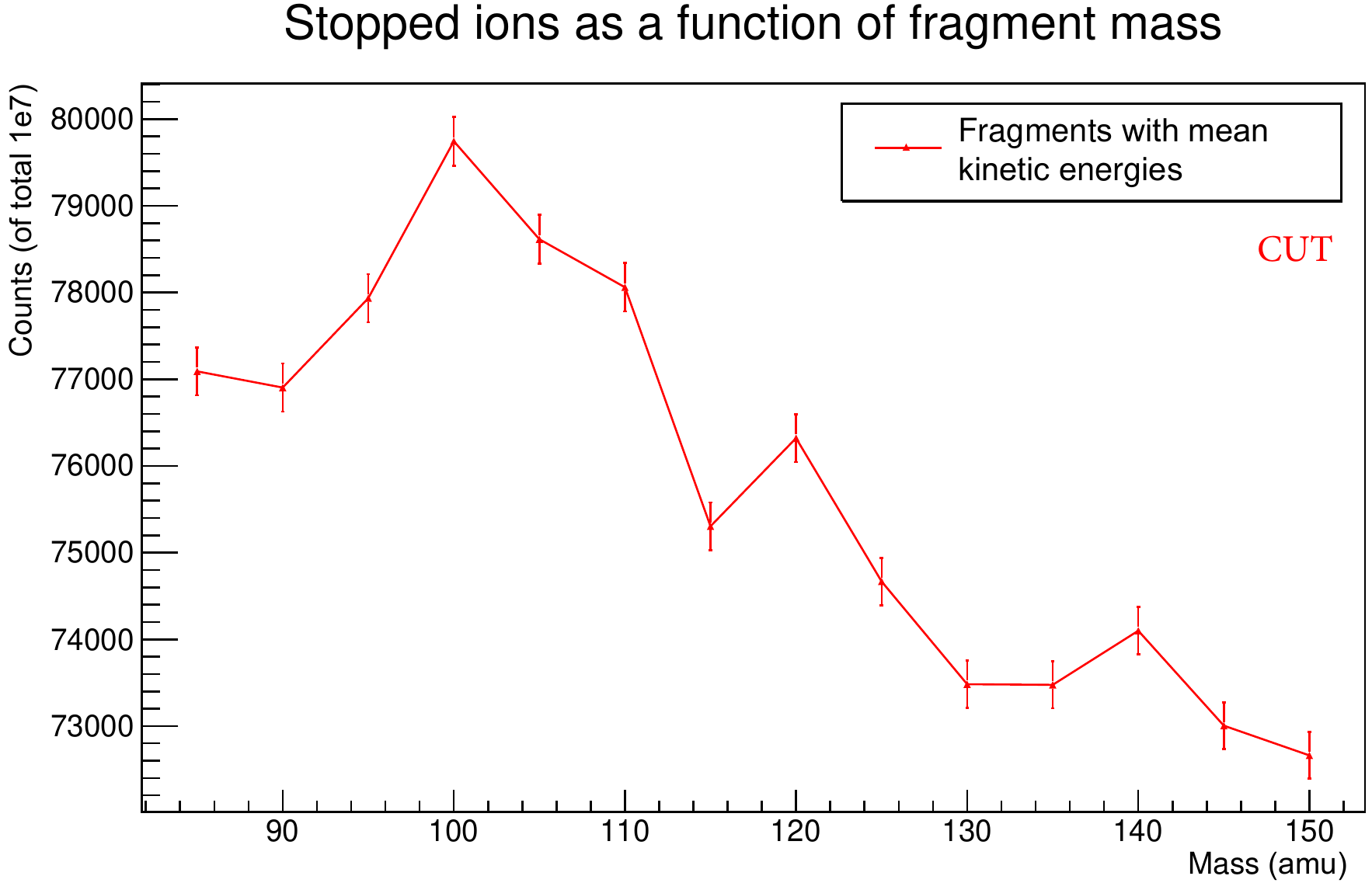}
\caption[The total stopped ions in the chamber as a function of mass.]{The total stopped ions in the gas cell as a function of fragment mass, showing about 9 \% relative difference. The mean ion kinetic energy was simulated. Error bars are plotted assuming Poisson statistics ($\sigma = \sqrt{N}$). The line is plotted to guide the eye.}
\label{Fig11}
\end{figure}

\begin{figure}[b!]
\centering
    \includegraphics[width=0.95\textwidth]{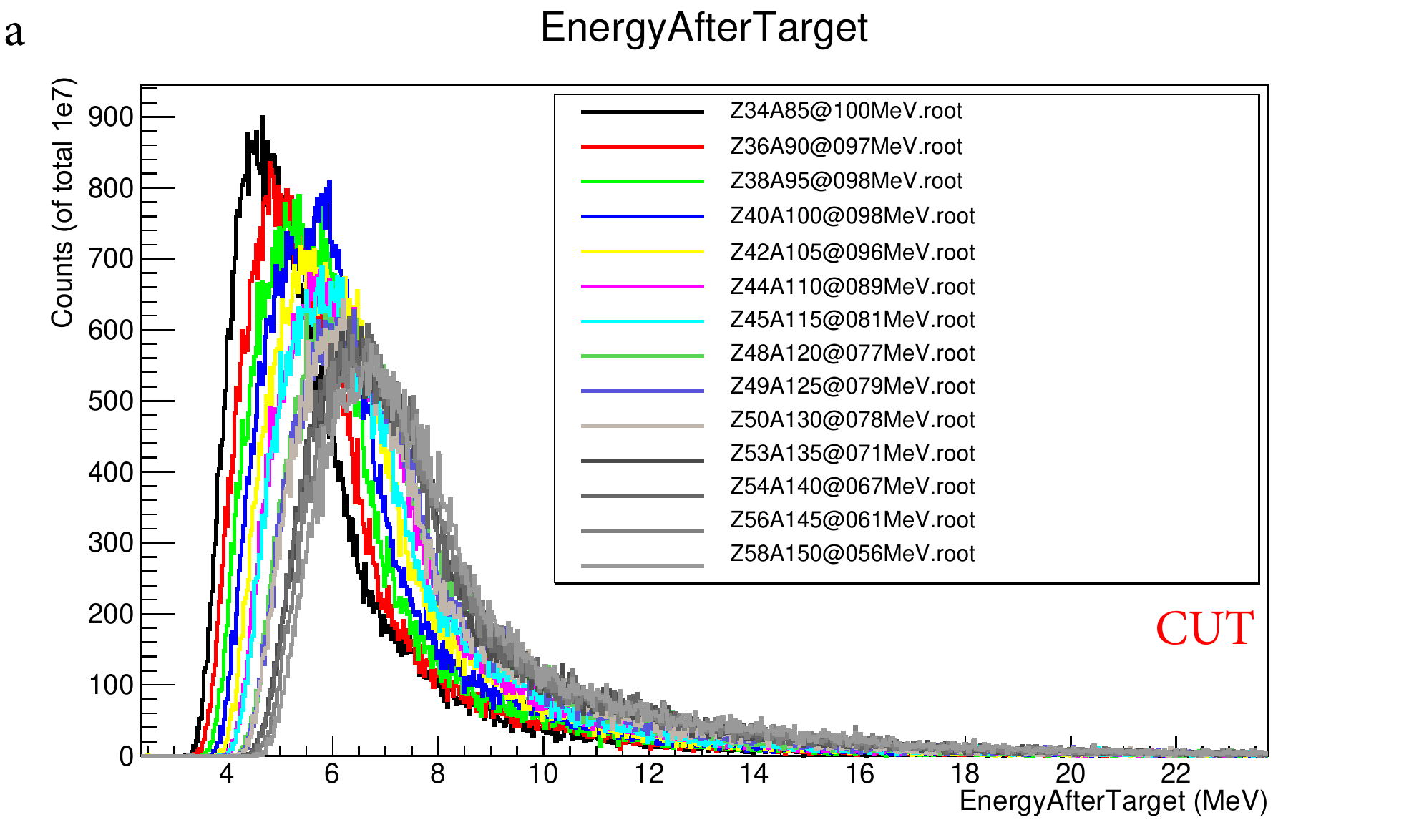}
    \includegraphics[width=0.95\textwidth]{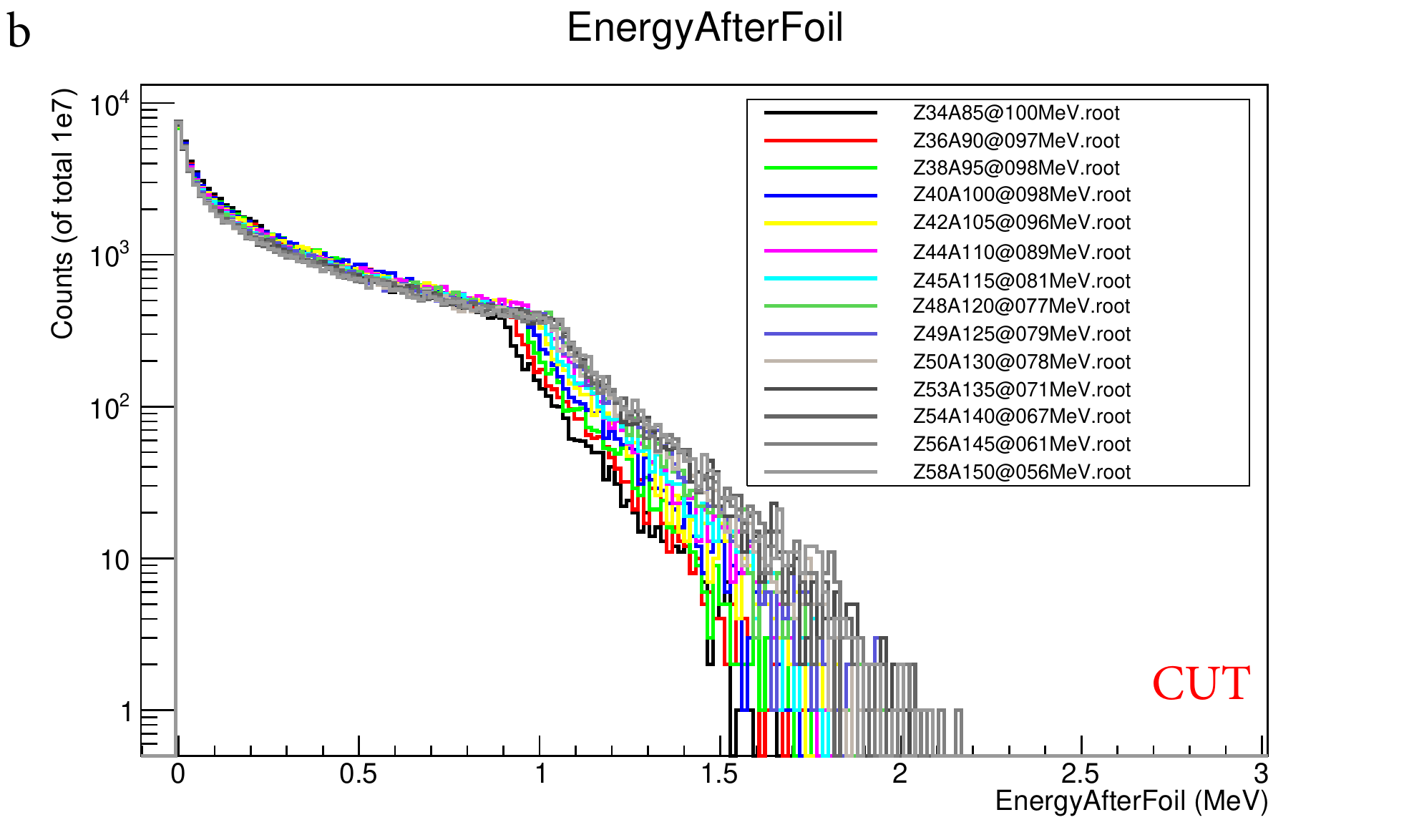}
\caption[The energy after leaving the target/foil.]{The ions stopped in the gas cell. a) The ion energy after leaving the Uranium target. (b) The ion energy after leaving the Ni foil. The ions have the mean fragment energies.}
\label{Fig12}
\end{figure}

\begin{landscape}
\begin{figure}[b!]
\centering
    \includegraphics[scale=0.56]{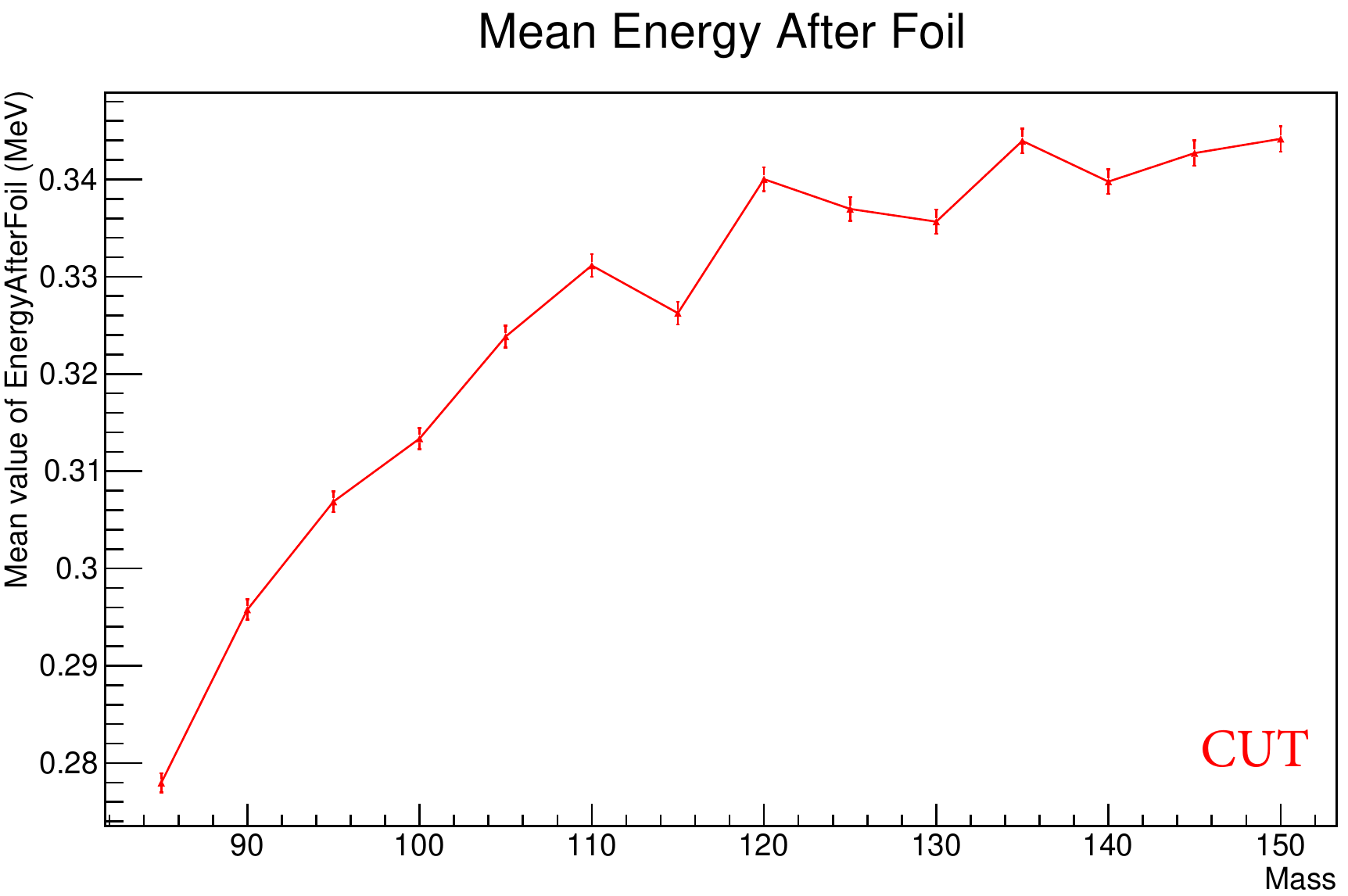}
    \includegraphics[scale=0.56]{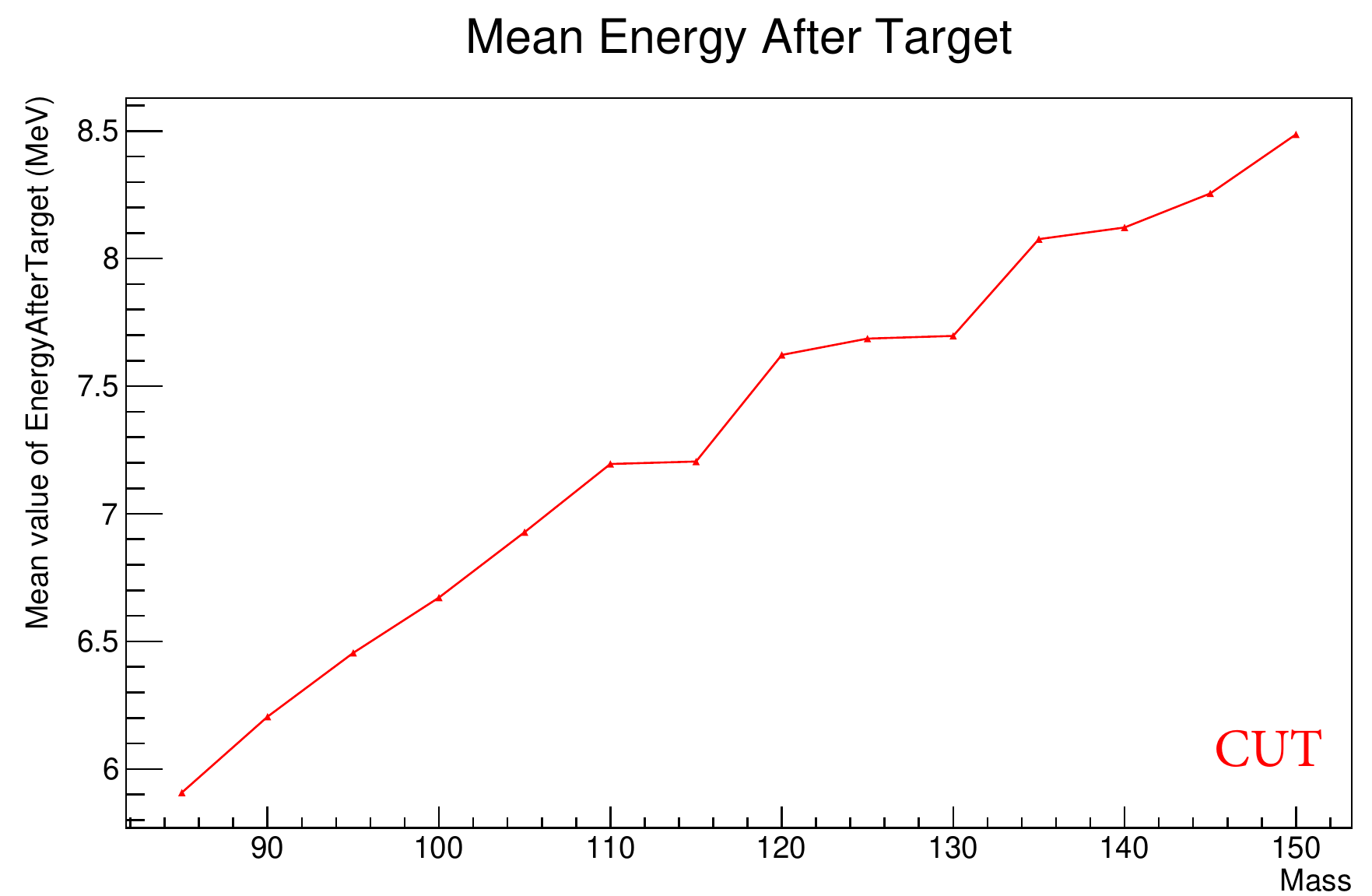}
    \includegraphics[scale=0.56]{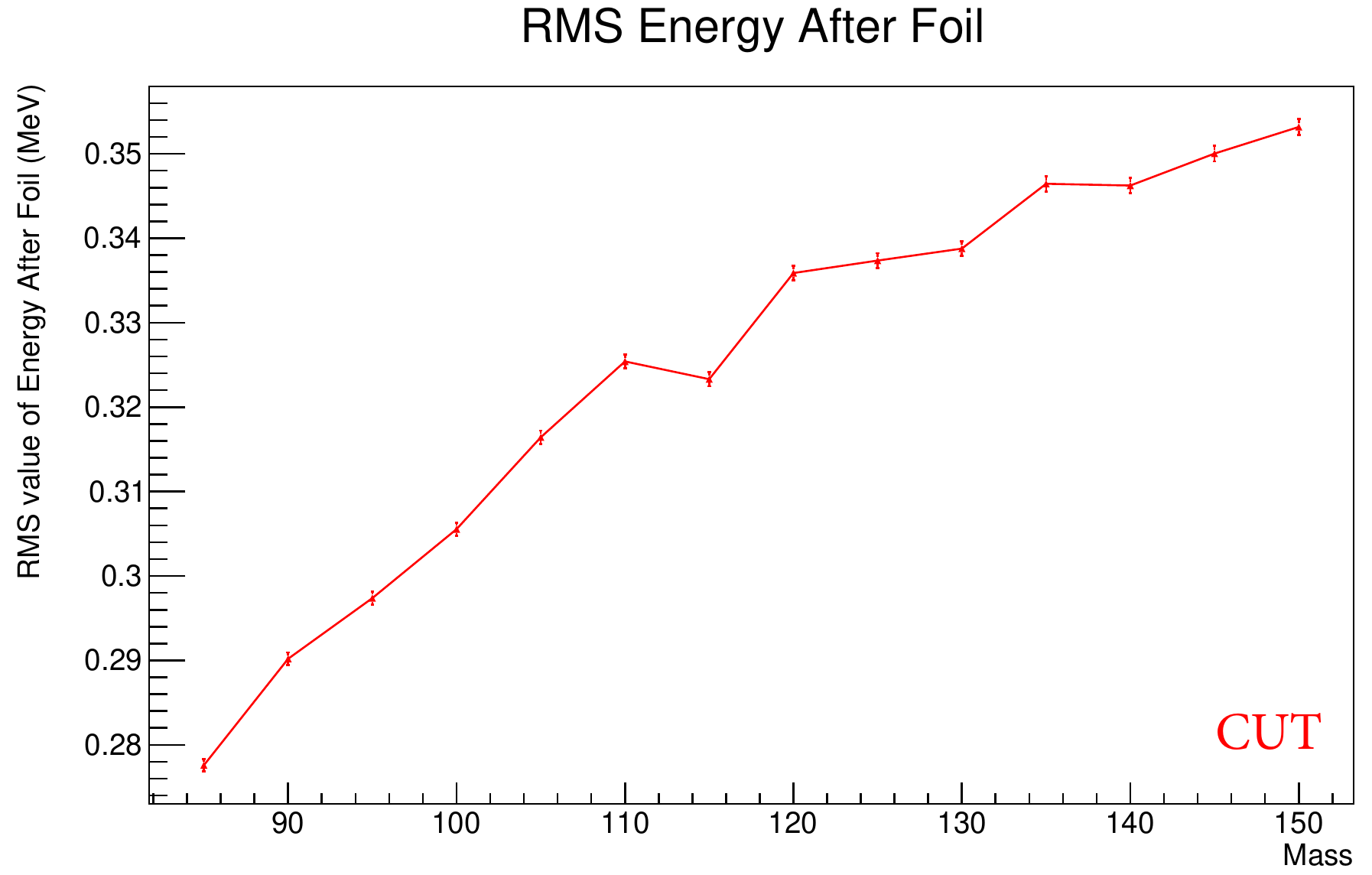}
    \includegraphics[scale=0.56]{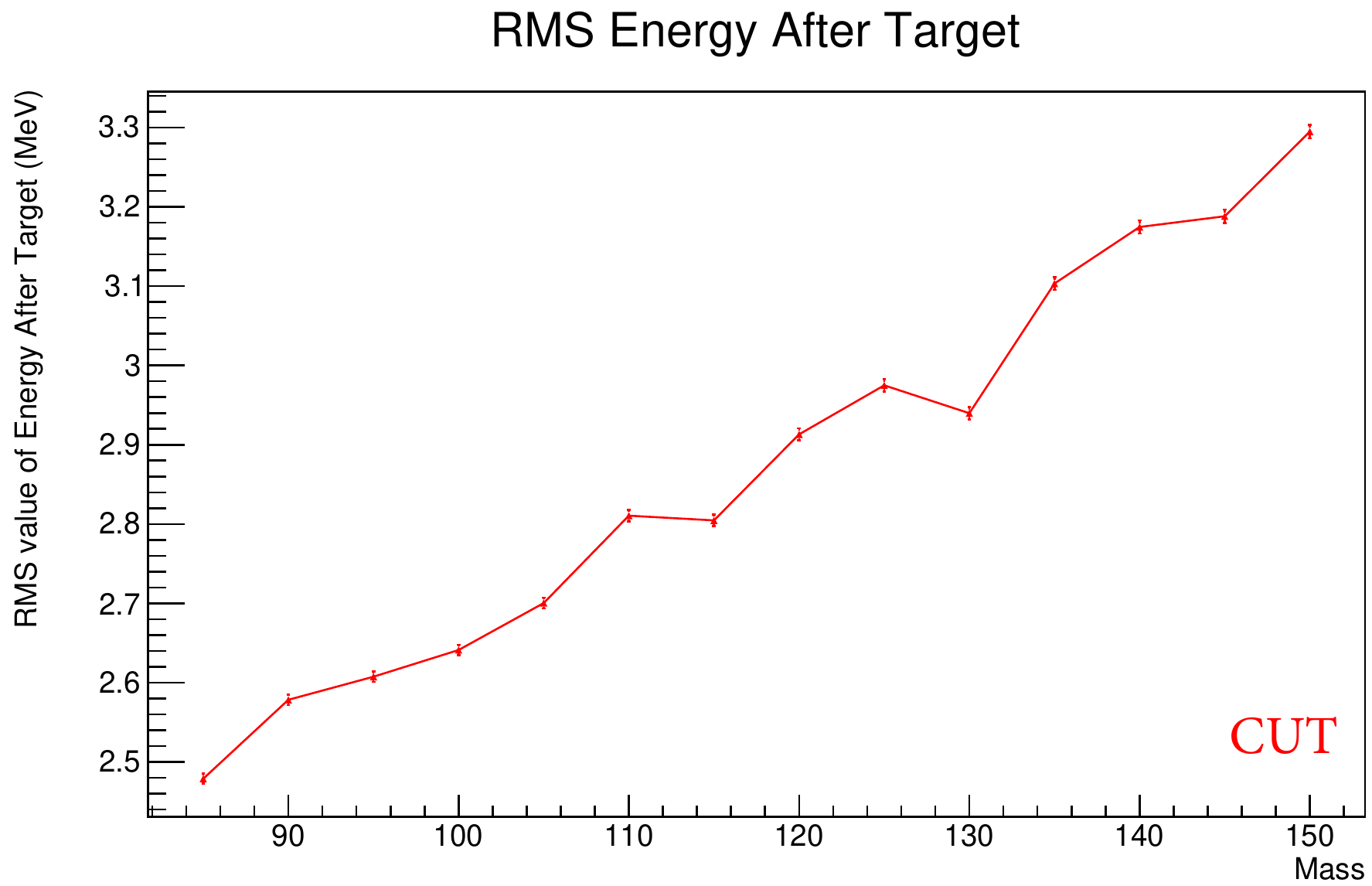}
\caption[The mean and RMS as a function of fragment mass.]{The mean values and RMS (distribution widths) as a function of fragment mass, obtained from the distributions in Fig.~\ref{Fig12}. The fragments were emitted at mean fragment kinetic energy.}
\label{Fig13}
\end{figure}  
\end{landscape}

The simulations were performed for the nuclides listed in Table \ref{table1}. As stated above, in all instances about 0.5\% of all created fission fragments are stopped by the He gas in the stopping volume. For the mean fragment energies (i.e. most probably kinetic energies), the stopped ions are plotted as a function of fragment mass in Fig.~\ref{Fig11}. The fission products have fairly equal chances of being collected independent of their charge or mass. However, a trend is observed, namely that heavier masses have a slightly lower stopping probability compared to the lighter ones. The relative difference is about 9\% between the fission fragments around $A = 100$ and $A = 150$. The energies after target and Ni foil are shown in Fig.~\ref{Fig12}. The mean energies increase as a function of mass as seen in Fig.~\ref{Fig13}. Heavy fragments must have larger energy than light ones after leaving the Uranium target, in order to be stopped in the gas cell. Also the distribution width (RMS) increases as a function of mass (see Fig.~\ref{Fig13}).  

\begin{landscape}
\begin{figure}[b!]
\centering
    \includegraphics[scale=0.56]{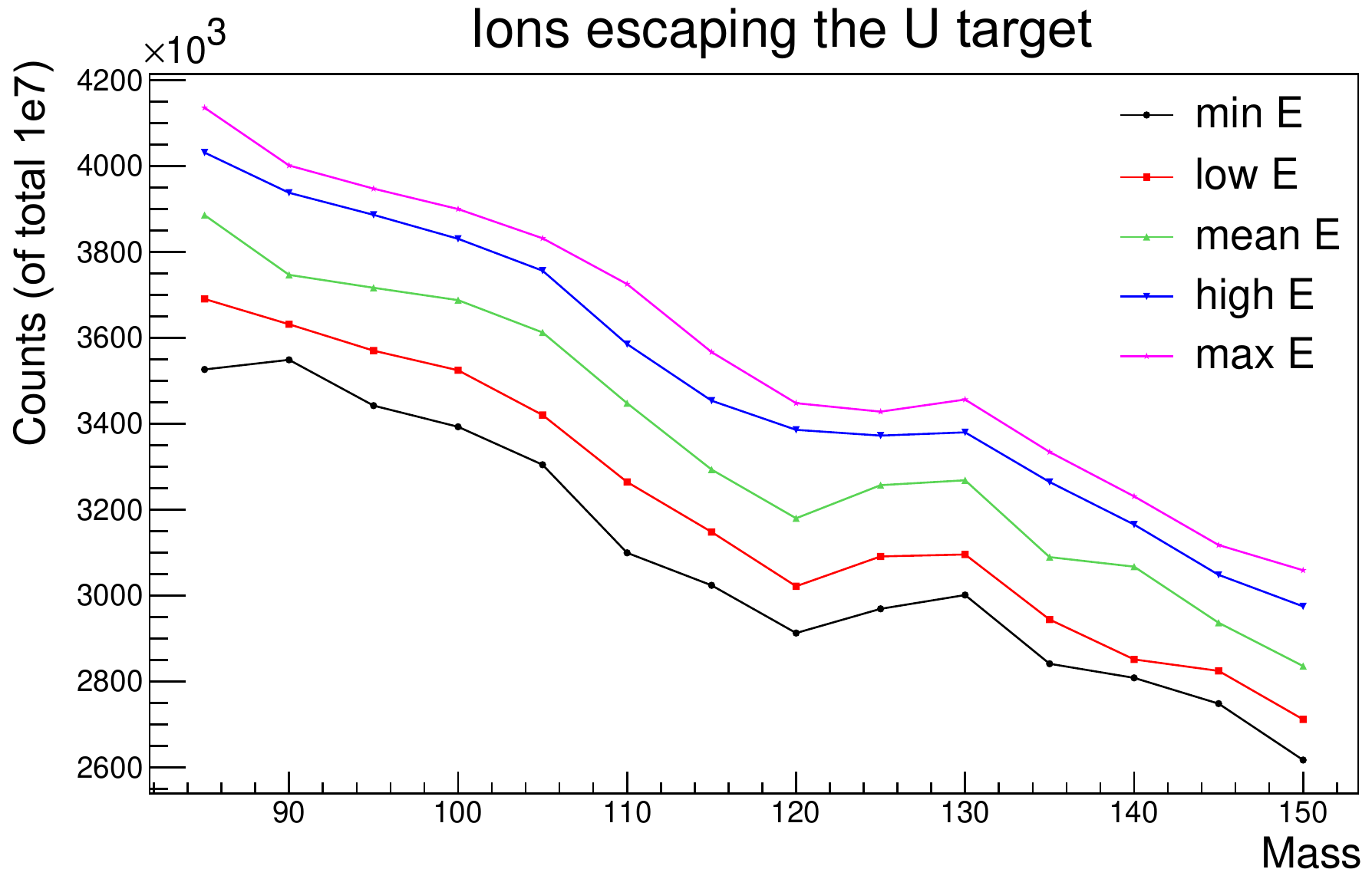}
    \includegraphics[scale=0.56]{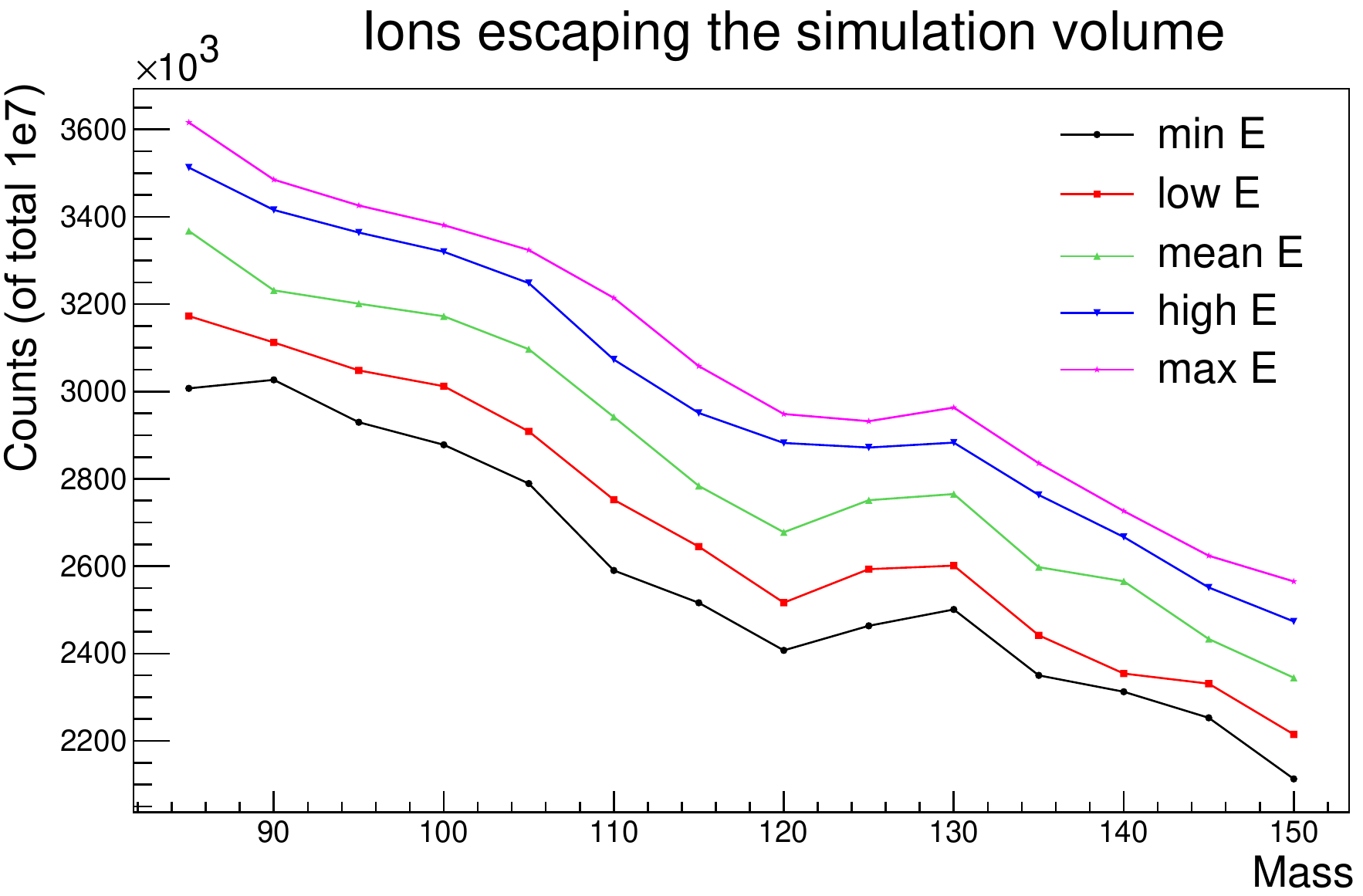}
    \includegraphics[scale=0.56]{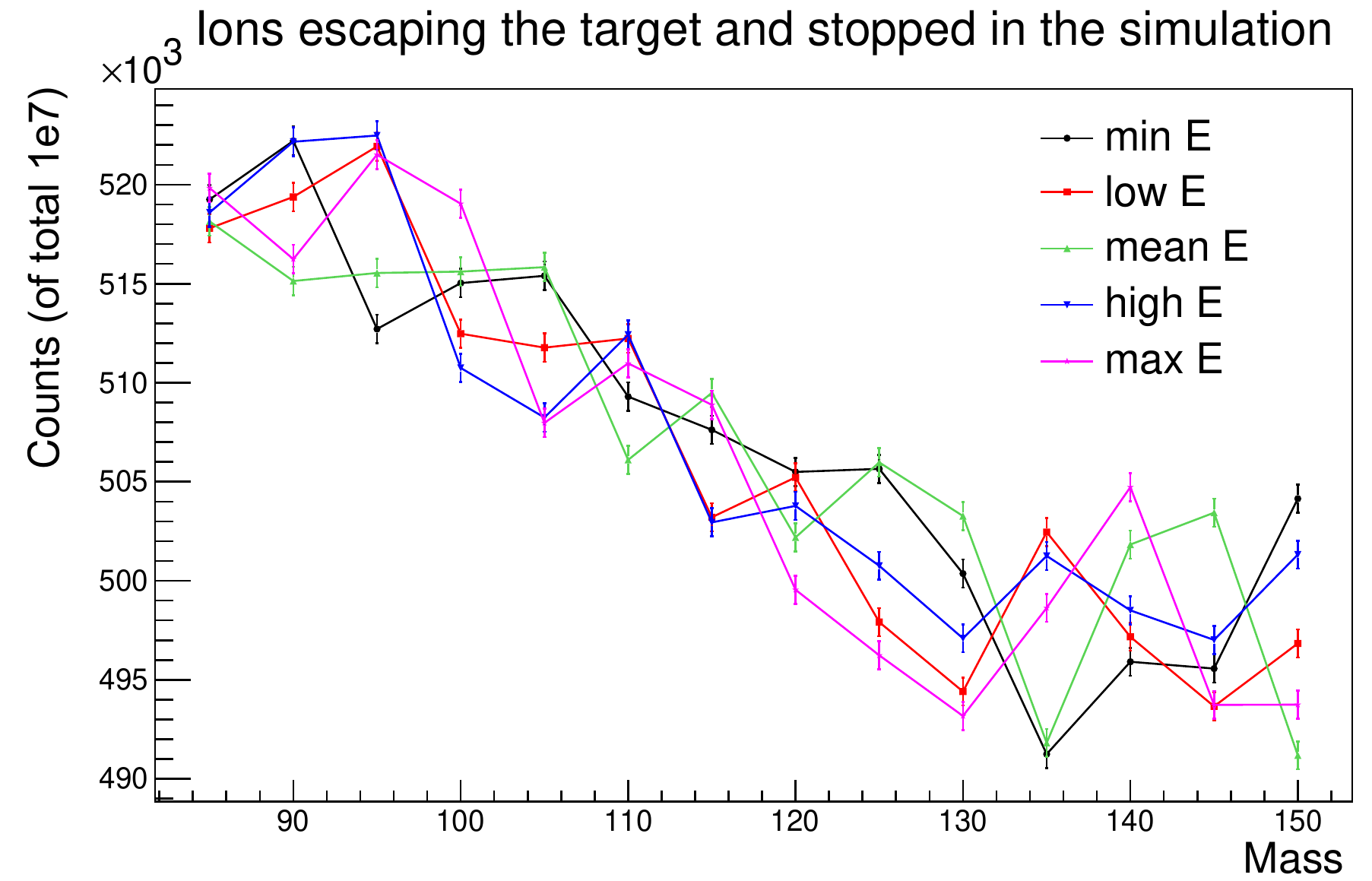}
    \includegraphics[scale=0.56]{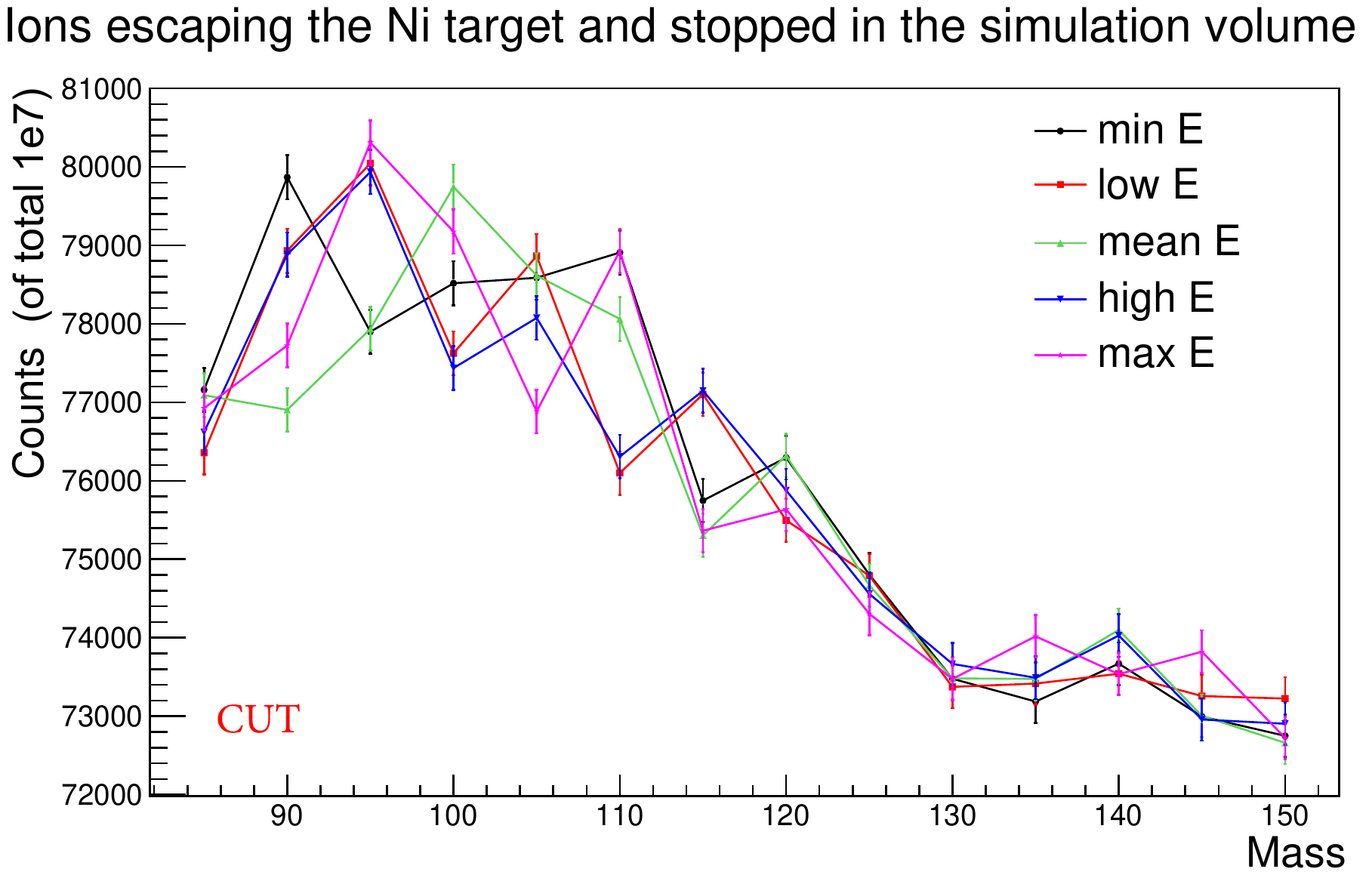}
\caption[Ions that stop in the U target or leave the volume.]{1e7 ions are emitted in 2$\pi$ for each nuclide. (a) The ions escaping in the U target; lighter ions show a greater chance escaping than the heavy ions. (b) The number of ions leaving the simulation boundaries. The light fragments are less prone to be stopped in the He gas than the heavy fragments. (c) The number of ions that escape the target and are stopped within the volume. Note that the ions stopped in the Ni foil are also included. (d) The ions that stop in the gas show a smaller mass-dependence due to the compensating effects in (a), (b) and (c). }
\label{Fig14}
\end{figure}    
\end{landscape}

The reason for the relatively small mass dependence is a compensation effect. Figure~\ref{Fig14}a shows the number of ions that successfully exit the U target. The light ions have a larger chance escaping due to their lower charge and higher average kinetic energy. However, when entering the stopping chamber cell the lighter fragments have a higher tendency of escaping, for the same reason (see  Fig.~\ref{Fig14}b). The sum of these competing effects gives a nearly mass-independent stopping efficiency. Figure~\ref{Fig14}c shows all ions exiting the target and stopping in the chamber (including the ones stopped in the Ni foil). Figure~\ref{Fig14}d shows the collected ions in the gas after passing the Ni foil. As seen these trends are valid for all fragment energies. The weighted ion counting for all kinetic energies is plotted in Fig. \ref{Fig1e61e7}a (i.e. a weighted average of Fig.\ref{Fig11}d). Clearly the overall trend averaged over the fragments energies gives a slightly larger stopping efficiency for lighter fragments.  

\subsection{Statistical uncertainties} 
\begin{figure}[htb]
\centering
    \includegraphics[scale=0.65]{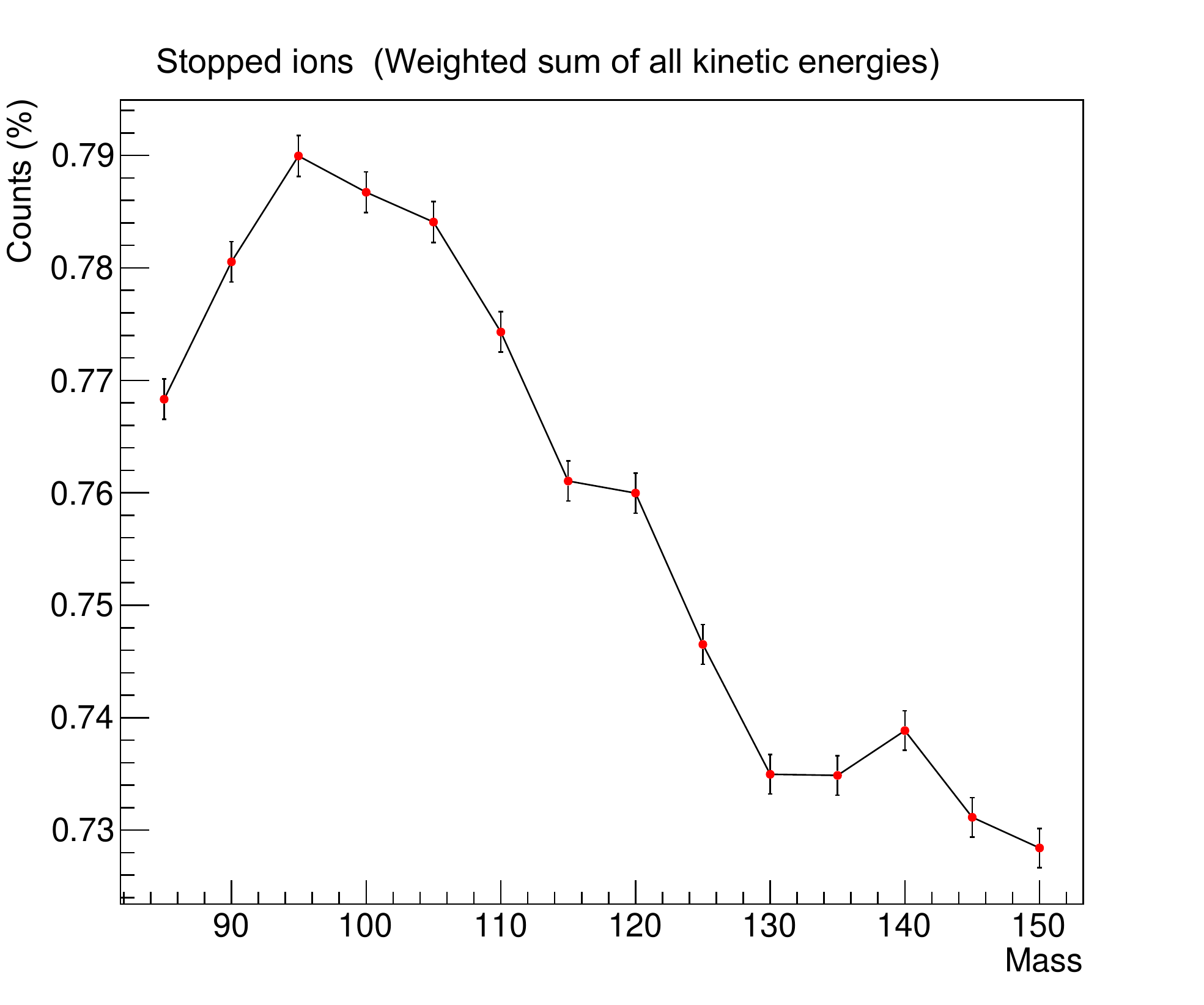}
    \includegraphics[scale=0.65]{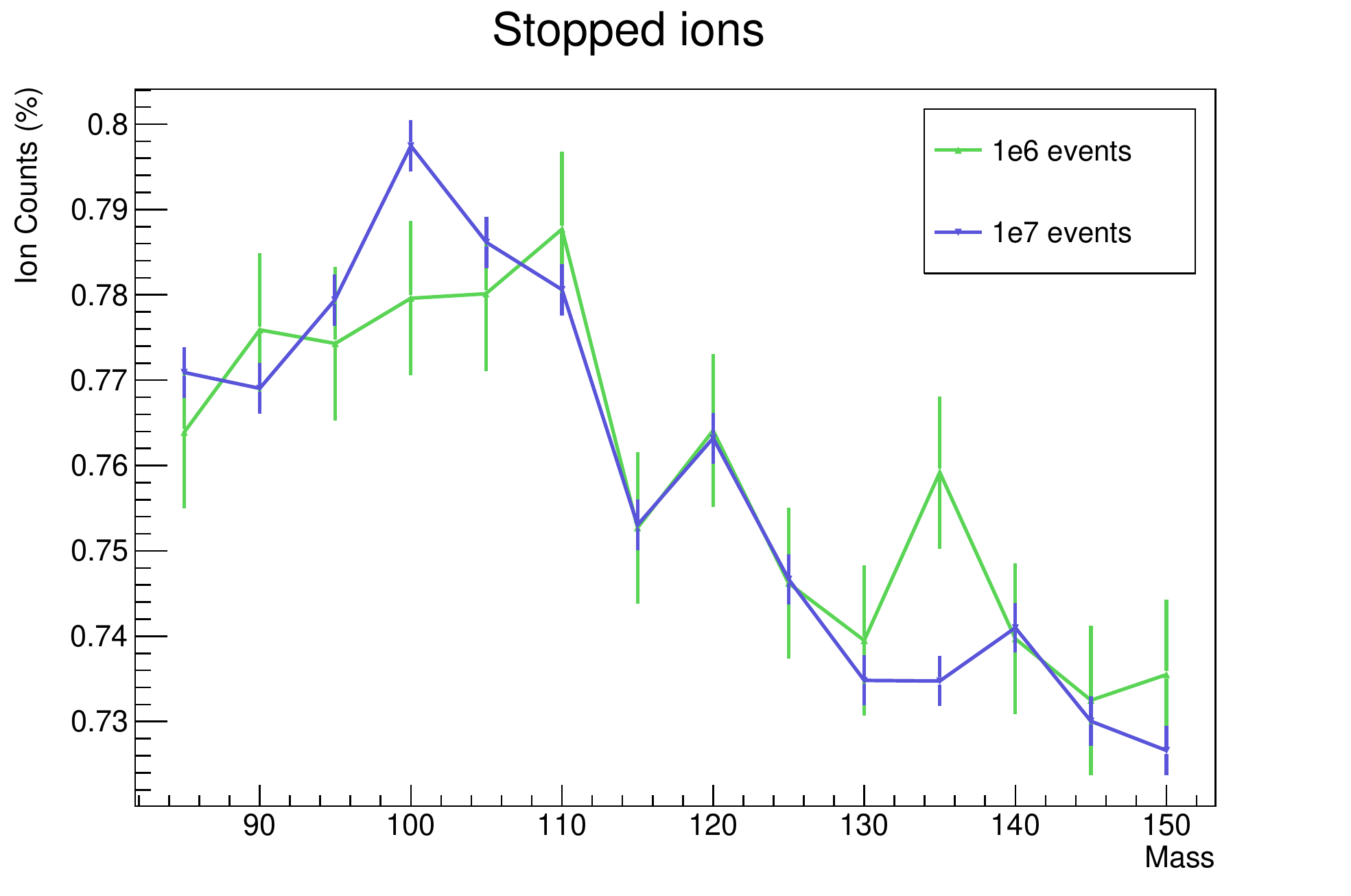}
\caption[Statistical variations in the ion counting.]{a: The ion counting efficiency in 2$\pi$ calculated as the weighted sum over all kinetic energies. b: The counting statistical variation shown in two independent calculations (1e6 versus 1e7 simulated ions in 2$\pi$). Error bars calculated by $\sigma = \sqrt{N}$.}
\label{Fig1e61e7}
\end{figure}

\begin{figure}[b!]
\centering
    \includegraphics[width=0.8\textwidth]{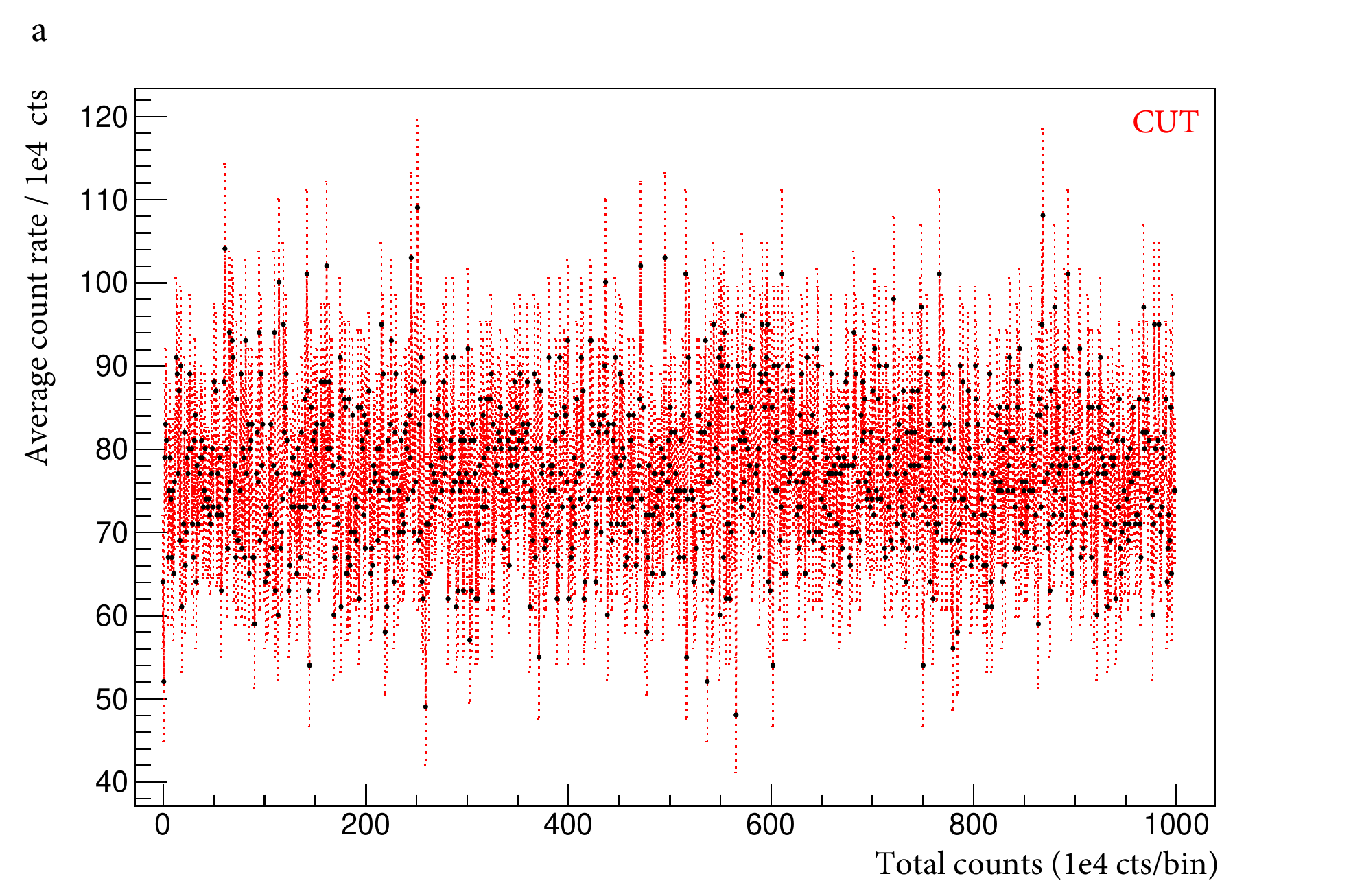}
    \includegraphics[width=0.78\textwidth]{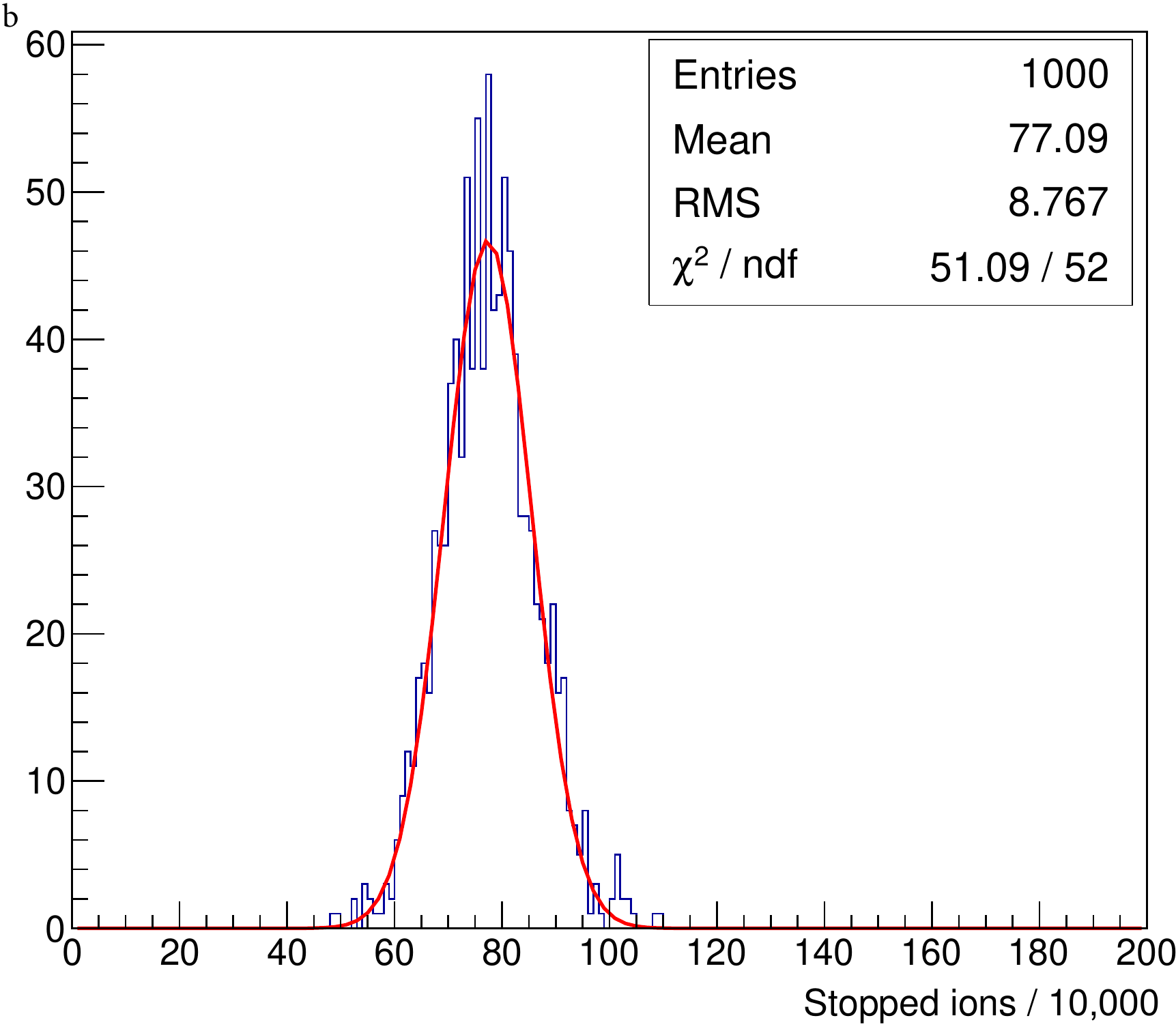}
\caption[Counting errors based on Poisson statistics.]{The statistical uncertainty is estimated by the variance of the ion count rate for (Z = 34, A = 85, E = 100 MeV) ion. a) The number of stopped ions is binned as a function of the total number of simulated ions in 2$\pi$ (bin size = 10000 ions/bin). b) The count rates in (a) plotted in a histogram, which gives a mean rate of 77 ions per 10000. The counting errors are fairly well described by the Poisson statistics since $\sigma \approx 8.77$ which follows the relation $\sigma=\sqrt{77}\approx8.77$.}
\label{Figstat}
\end{figure}

In figure \ref{Fig1e61e7}b, the ion counting is shown for two independent runs having 1e6 and 1e7 statistics, respectively. Most data points fall within the margins of the error bars. The statistical uncertainties were estimated assuming Poisson statistics, i.e calculated by $\sqrt{N}$ where $N$ is the ions counts. Figure \ref{Figstat}a shows the entire data set for (Z = 34, A = 85, E = 100 MeV) as a function of time. The number of stopped ions is plotted as a function of the total number of simulated ions (data binned with 10000 ions/bin in the x axis). By dividing into subsets the variance of the data can be studied. The number of counts for each bin is projected in Fig. \ref{Figstat}b with a mean around 77.09 counts/bin. The variance is in the order of $\sigma=8.77$ which agrees with the expected $\sqrt{77}\approx8.77$.

\clearpage

\section{Energy dependence of ion counting}
\begin{figure}[b!]
\centering
    \includegraphics[scale=0.81]{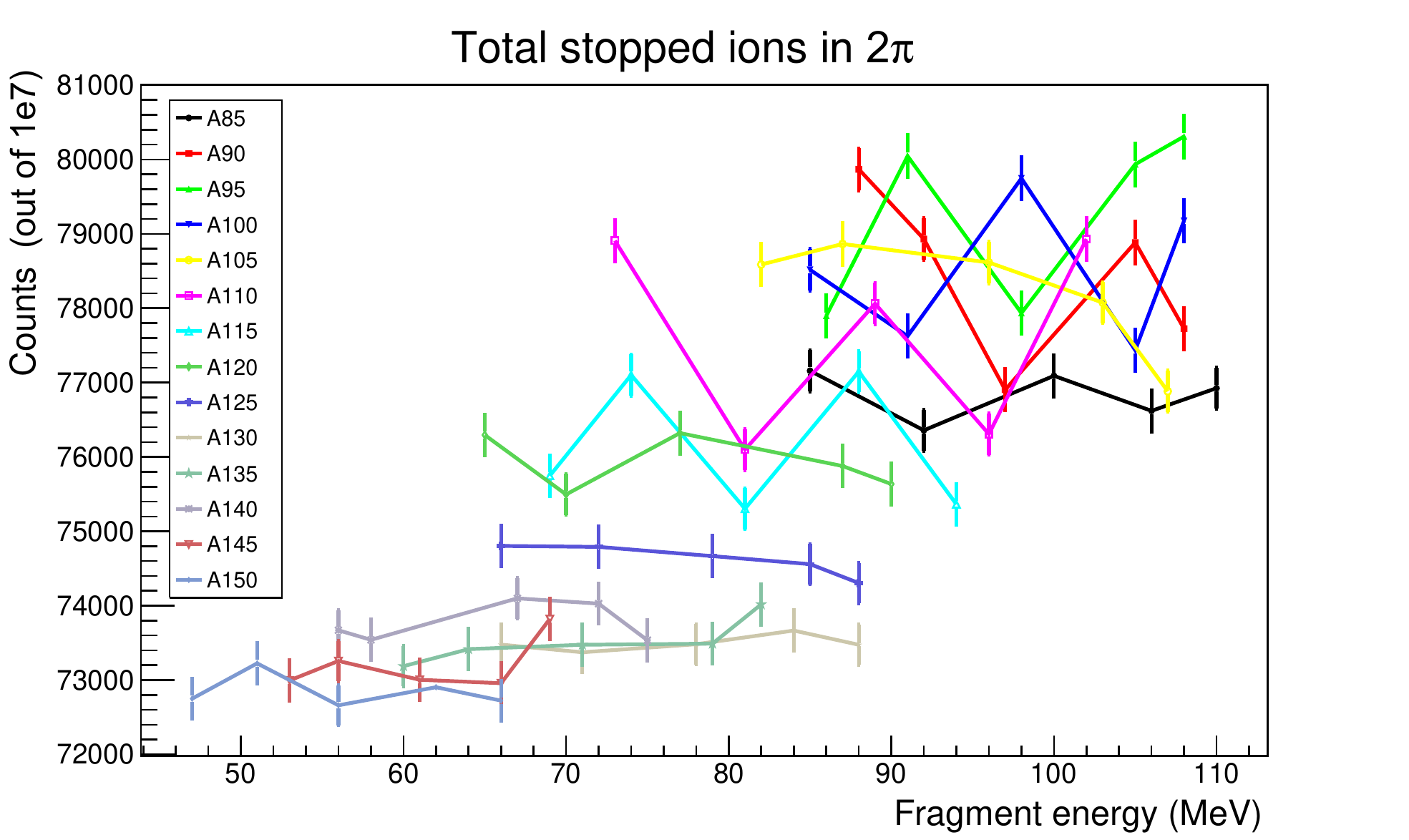}
\caption[The number of stopped ions as a function of fission fragment energy.]{The number of stopped ions as a function of fission fragment energy (out of 1e7 emitted ions in 2$\pi$ for each nuclide+energy). The variations as a function of kinetic energies do not show clear rends as compared to those between different masses. The error bars presented are calculated based on Poisson counting statistics.}
\label{Fig15}
\end{figure}

\begin{figure}[htb]
\centering
    \includegraphics[width=0.9\textwidth]{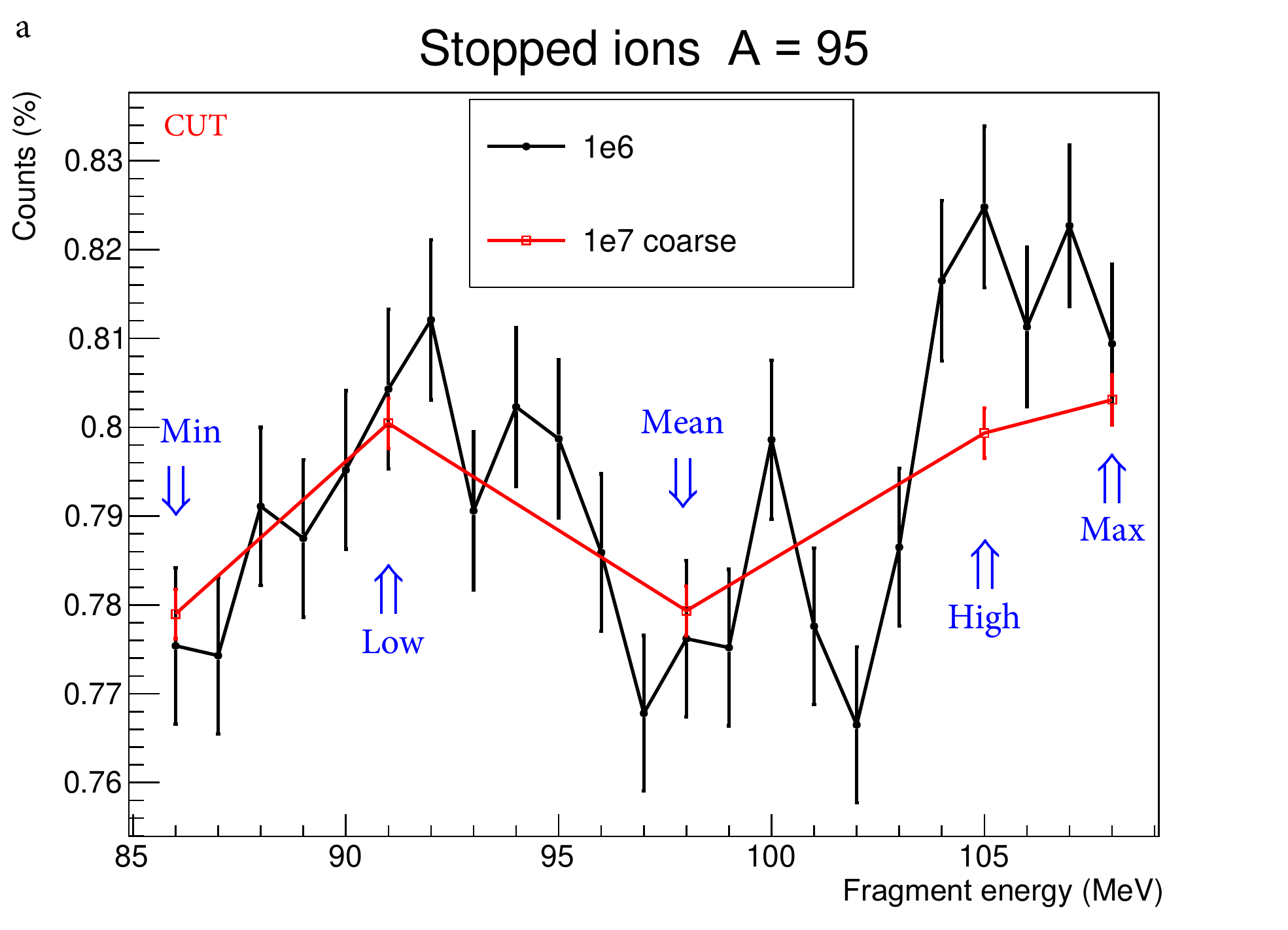}
    \includegraphics[width=0.9\textwidth]{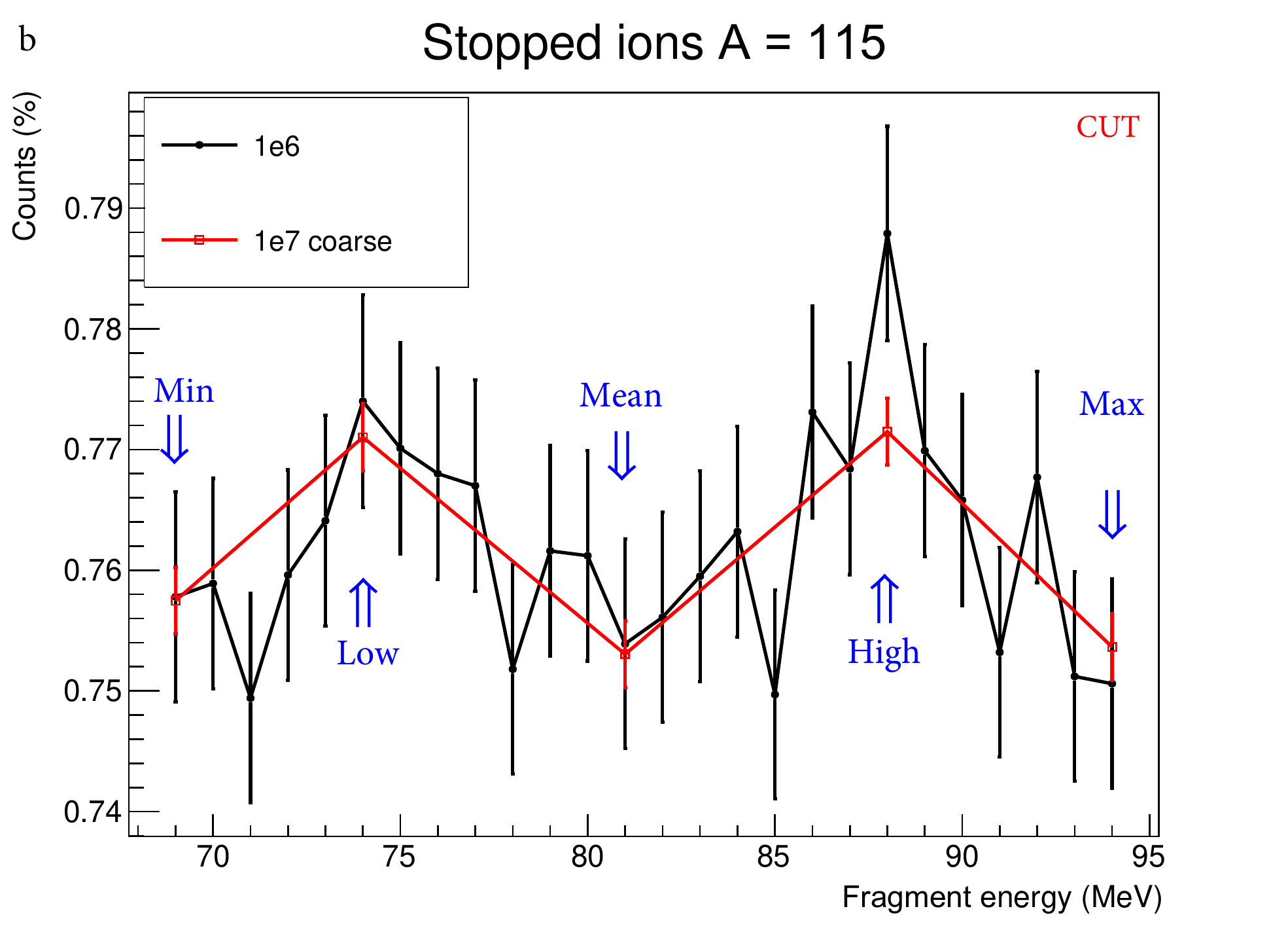}    
\caption[Detailed energy dependences.]{The energy dependence of the ion counting for A = 95 and 115 amu. The staggering effect observed in the coarse runs (1e7 ions) can also be seen in the fine runs (1e6 ions), thus they are attributed to the model.}
\label{Figdetail}
\end{figure}

The five kinetic energies for each isotope, listed in Tab. \ref{table1} were used as initial energies for the selected 14 masses. The energy dependences on the stopping efficiency were smaller than those of the mass/charge. Figure~\ref{Fig15} shows the counting efficiency as a function of the kinetic energy and no clear trends were observed, except for a staggering effect. The observed staggering is larger than the error bars and seem to be stronger in lighter fragments. To investigate whether the staggering was due to physical model or due to statistical fluctuations, two masses (A = 95 and 115 amu) were simulated with 1 MeV energy steps. The results shown in Fig. \ref{Figdetail} reveal that the staggering behaviour is still present at higher energy resolution and could probably be attributed to the model. Despite the staggering effect, the ion energy is of less importance to the ion stopping compared to the mass-dependence.

 \clearpage
\section{Origin of ions in the Uranium target} 
The randomly chosen start positions $(X_0,Y_0,Z_0)$ of each ion are extracted to determine the ion origin in the target. Since the target is tilted, $Z_0$ has to be related to the center of the target at a given X$_0$ and Y$_0$ in order to be suitable for plotting. The start position relative to the target center plane, $\Delta Z_0$, can be approximately given by: 

\begin{equation}
\label{Eq1}
\Delta Z_0 = Z_0 - (\textbf{central target plane}) \approx Z_0 - \left(X_0 \tan 7^\circ - 21\,\text{mm} \right).
\end{equation}

\begin{figure}[b!]
\centering
    \includegraphics[width=0.95\textwidth]{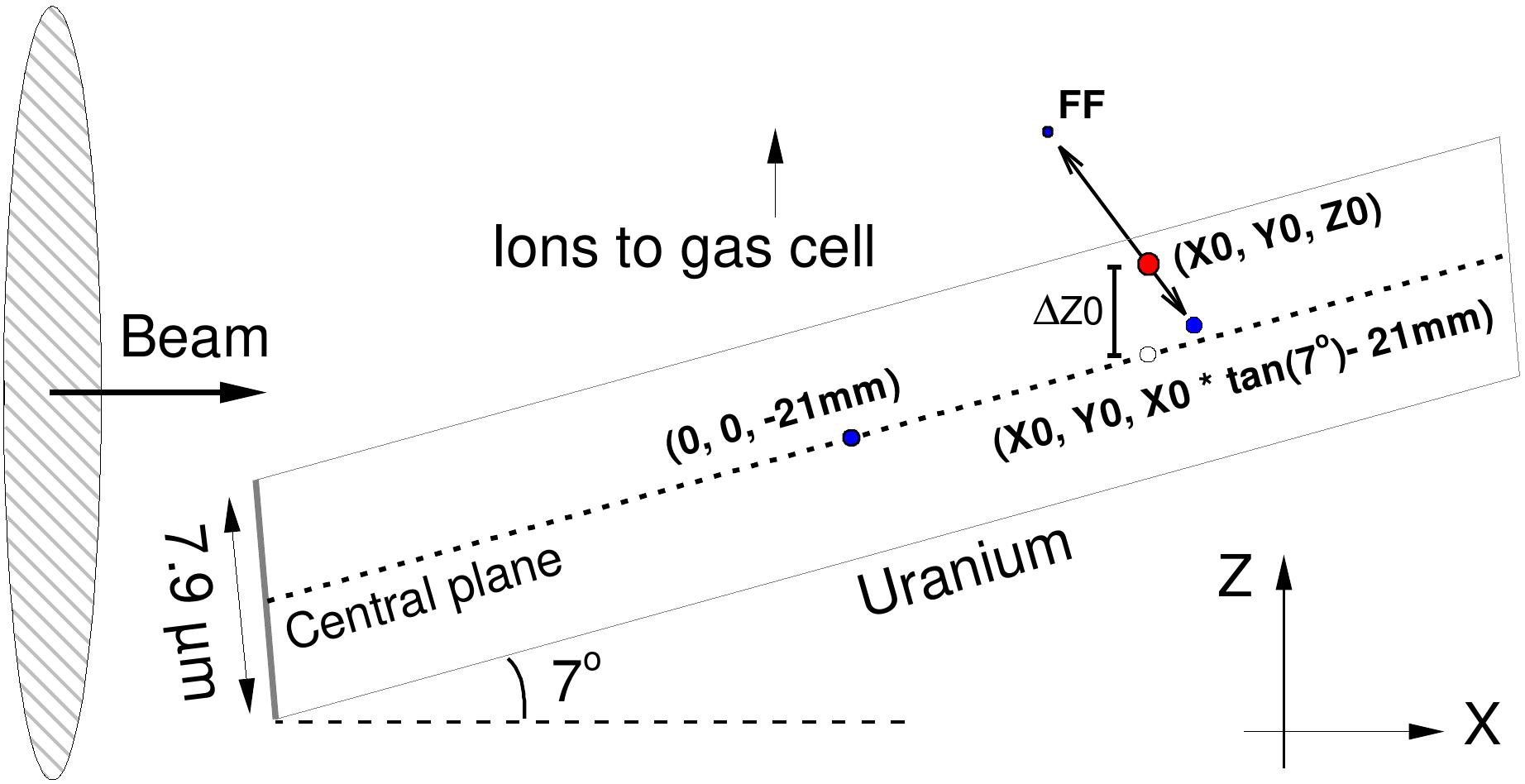}
\caption[Schematic view of the Uranium target.]{Schematic view of the Uranium target with the target thickness of 7.9\,$\mu$m. Two back-to-back fission fragments are emitted following a fission event (red dot). $\Delta Z0$ gives approximately the distance to the central target plane. }
\label{Fig16}
\end{figure}

Note that 21\,mm is the offset of the center of the target relative to the center of the entire volume. Figure~\ref{Fig16} shows an illustration of the target and one example event (in red). To determine the starting position relative to the central plane, $\Delta Z_0$, $X_0$ is used from the ion position vector.   

\begin{figure}[b!]
\centering
    \includegraphics[width=0.95\textwidth]{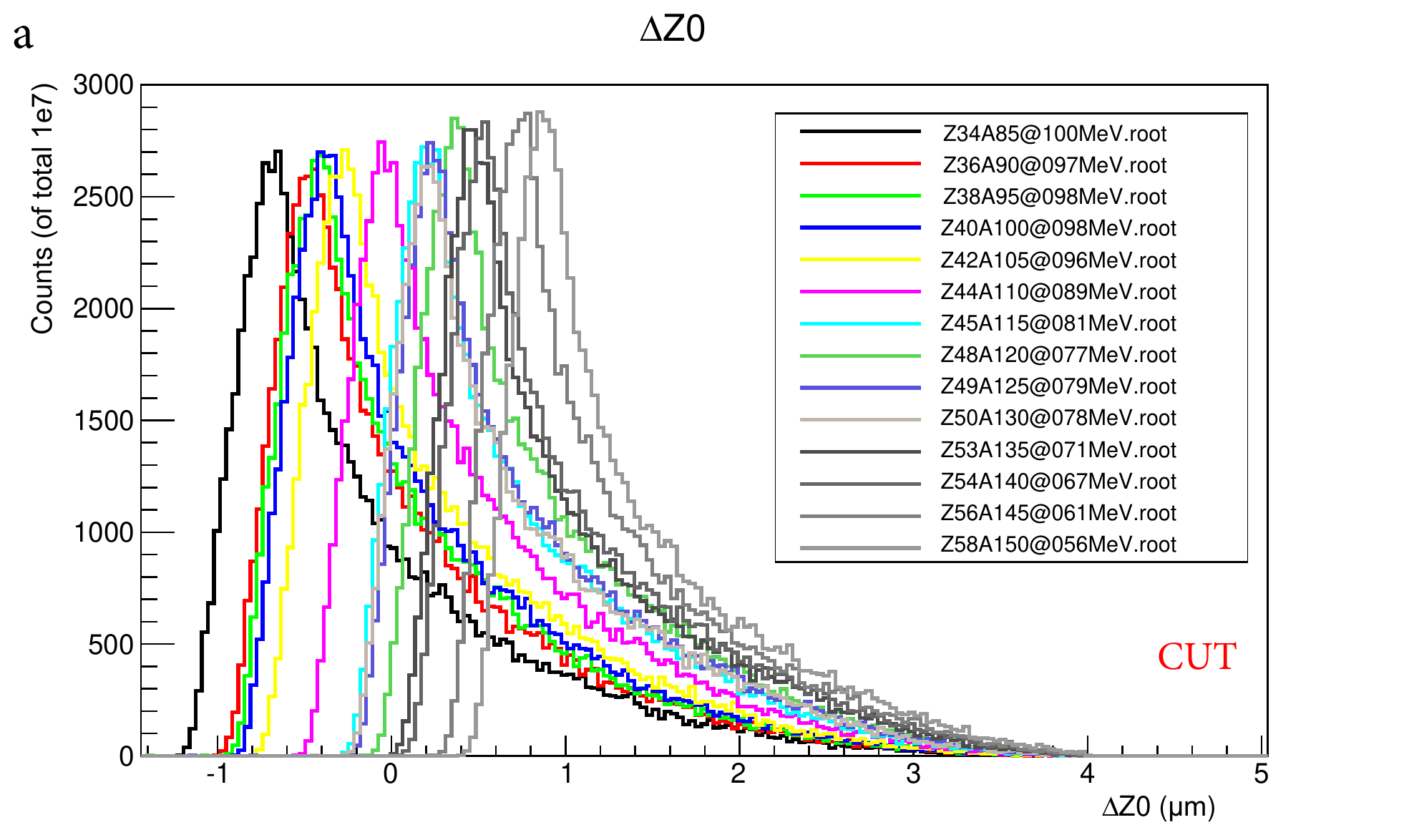}
    \includegraphics[width=0.95\textwidth]{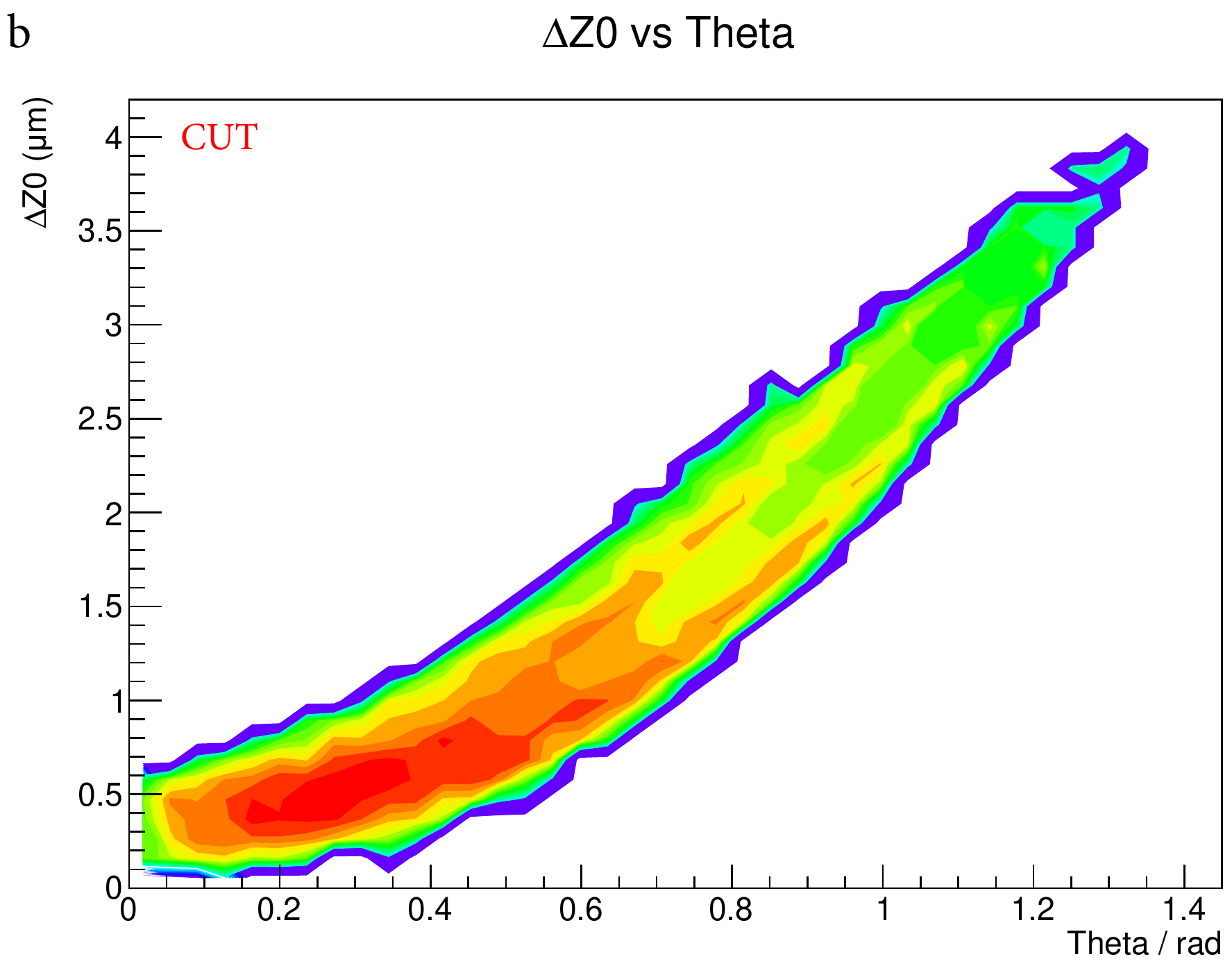}		
\caption[The starting positions ($\Delta Z_0$) in the U target, of the ions stopped in the gas.]{a) The starting positions ($\Delta Z_0$) in the U target, of the ions stopped in the gas. Zero is the center plane of the target seen in Fig.~\ref{Fig16}. b) The place of origin as a function of emission angle $\theta$ for $A = 140$, $Z = 54$ and $E = 67$\,MeV. Ions that are emitted at higher angles are produced closer to the stopping chamber.}
\label{Fig17}
\end{figure}

\begin{figure}[b!]
\centering
    \includegraphics[width=0.95\textwidth]{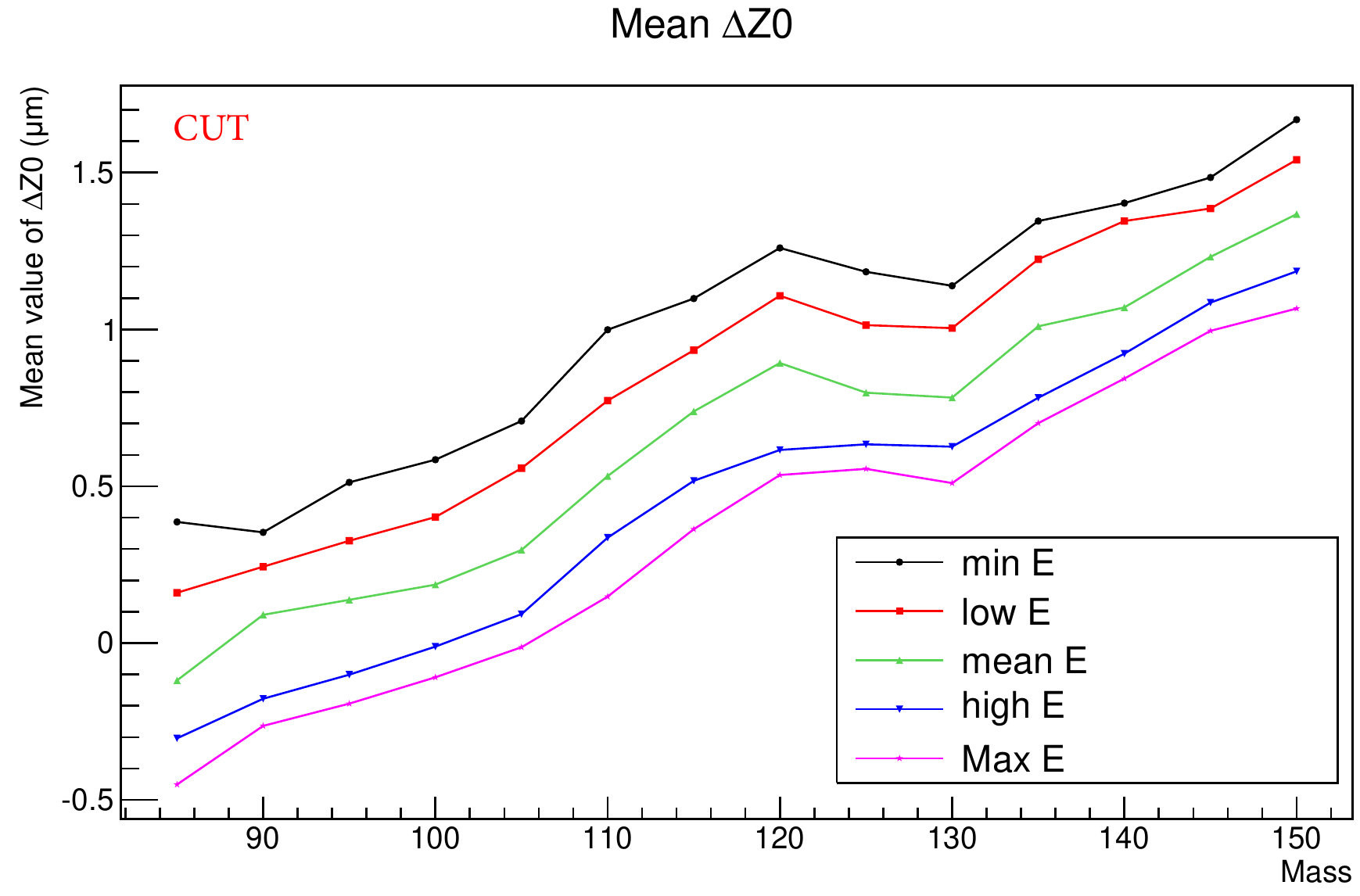}	
    \includegraphics[width=0.95\textwidth]{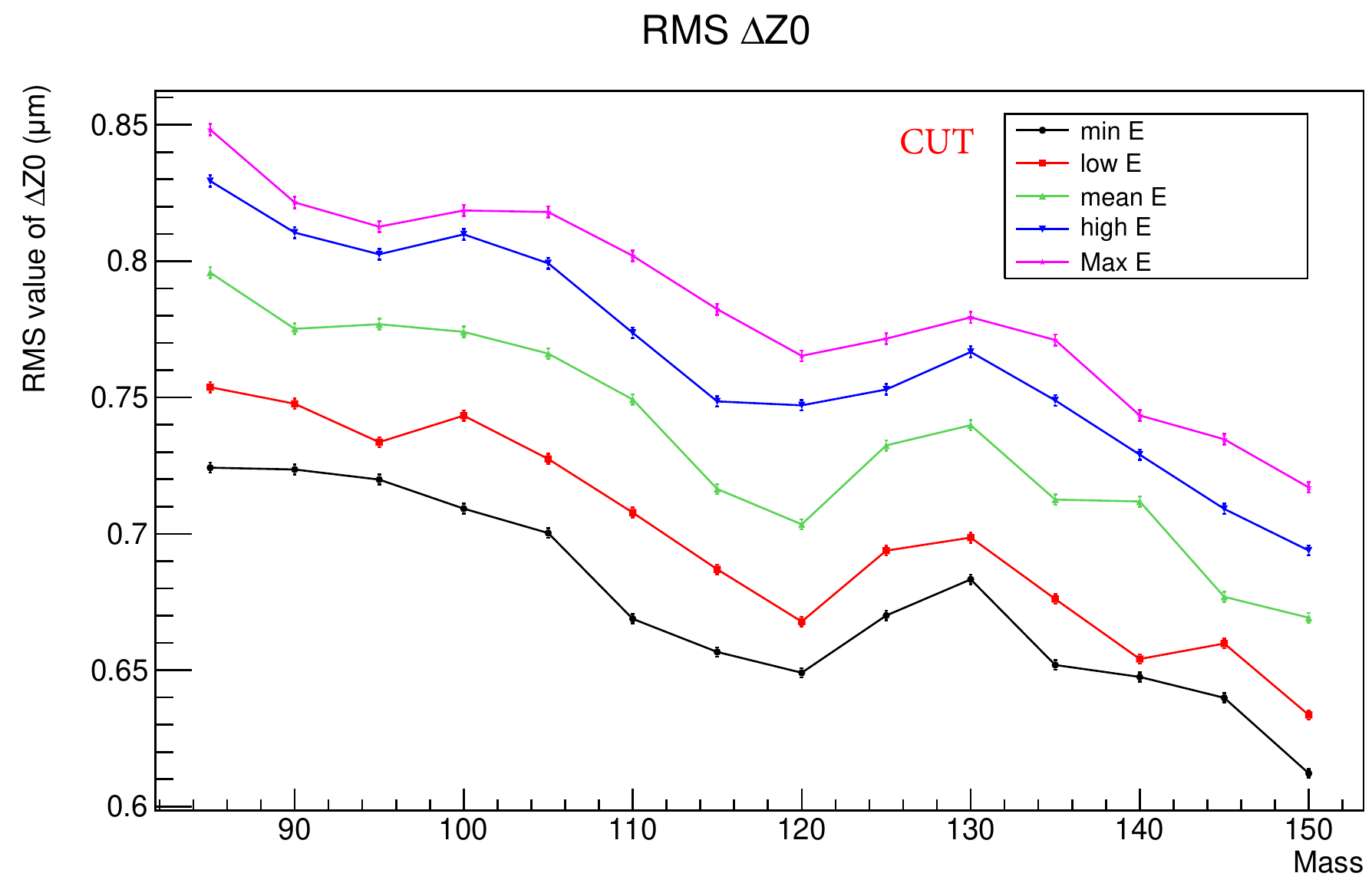}
\caption[The mean values and RMS of the $\Delta Z_0$.]{The mean values and RMS of the $\Delta Z_0$ distributions as a function of mass and energy.}
\label{Fig18}
\end{figure}

The starting positions of the mean energy ions (stopped in the chamber) are shown in the histograms of Fig.~\ref{Fig17}. The ions stopped in the gas are originating from the last 5\,$\mu$m of the U target. No ions originates from the 3\,$\mu$m furthest away from the gas. However, when simulating higher energetic fission fragments, the ions start to appear from deeper in the target, up to a maximum change of about 1\,$\mu$m in the mean starting position. The shape of the distributions can be understood considering higher emission angles as seen in Fig.~\ref{Fig17}b. Figure~\ref{Fig18} shows the mean values of $\Delta Z_0$, from all energies and masses. The same trends were seen in overall, namely that the mean increases as a function of mass. Higher ion energies give a lower mean $Z_0$ since the ions need to lose more energy to be stopped in the gas. 

\clearpage
\section{Ni Foil Thickness} 
\begin{figure}[b!]
\centering
    \includegraphics[width=0.95\textwidth]{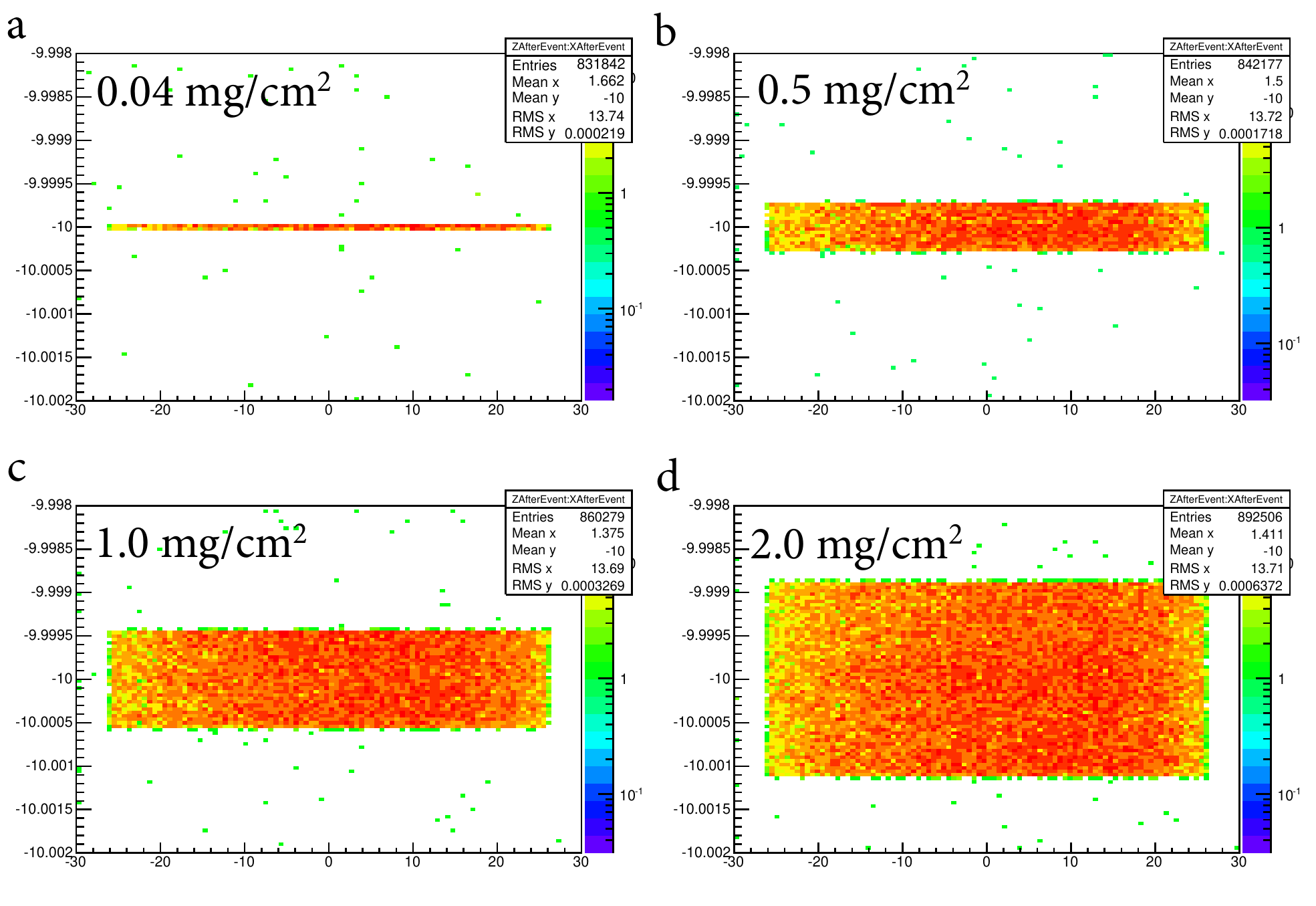}	
\caption[The simulated Ni foil thicknesses.]{The Ni foil thicknesses simulated, $Z$ plotted versus $Y$. The figures show the ions stopped in the Ni foil for the thicknesses 0.04\,mg/cm$^2$ (a), 0.5\,mg/cm$^2$ (b), 1.0\,mg/cm$^2$ (c) and 2.0\,mg/cm$^2$ (d).}
\label{Fig19}
\end{figure}

\begin{figure}[b!]
\centering
    \includegraphics[width=0.95\textwidth]{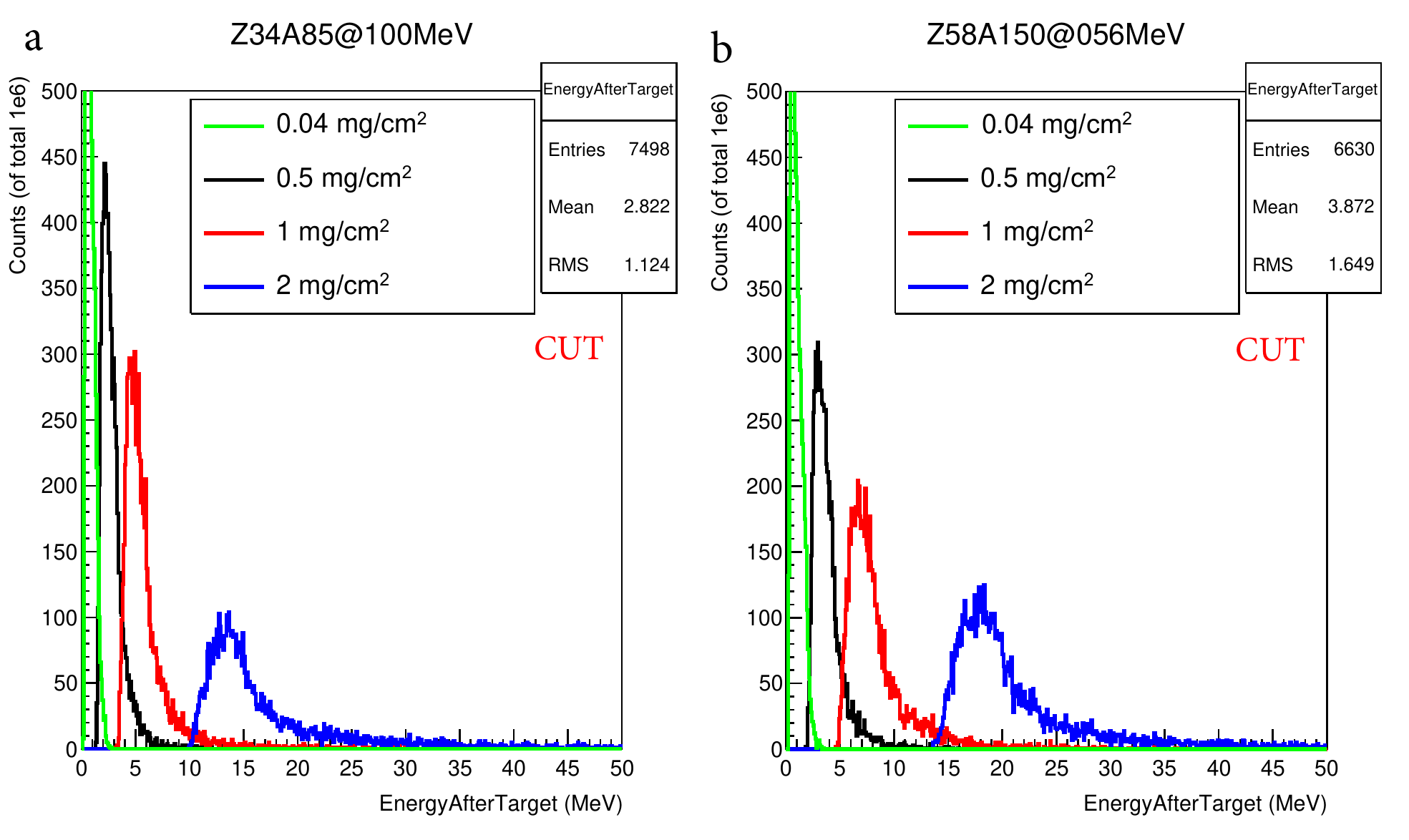}
    \includegraphics[width=0.95\textwidth]{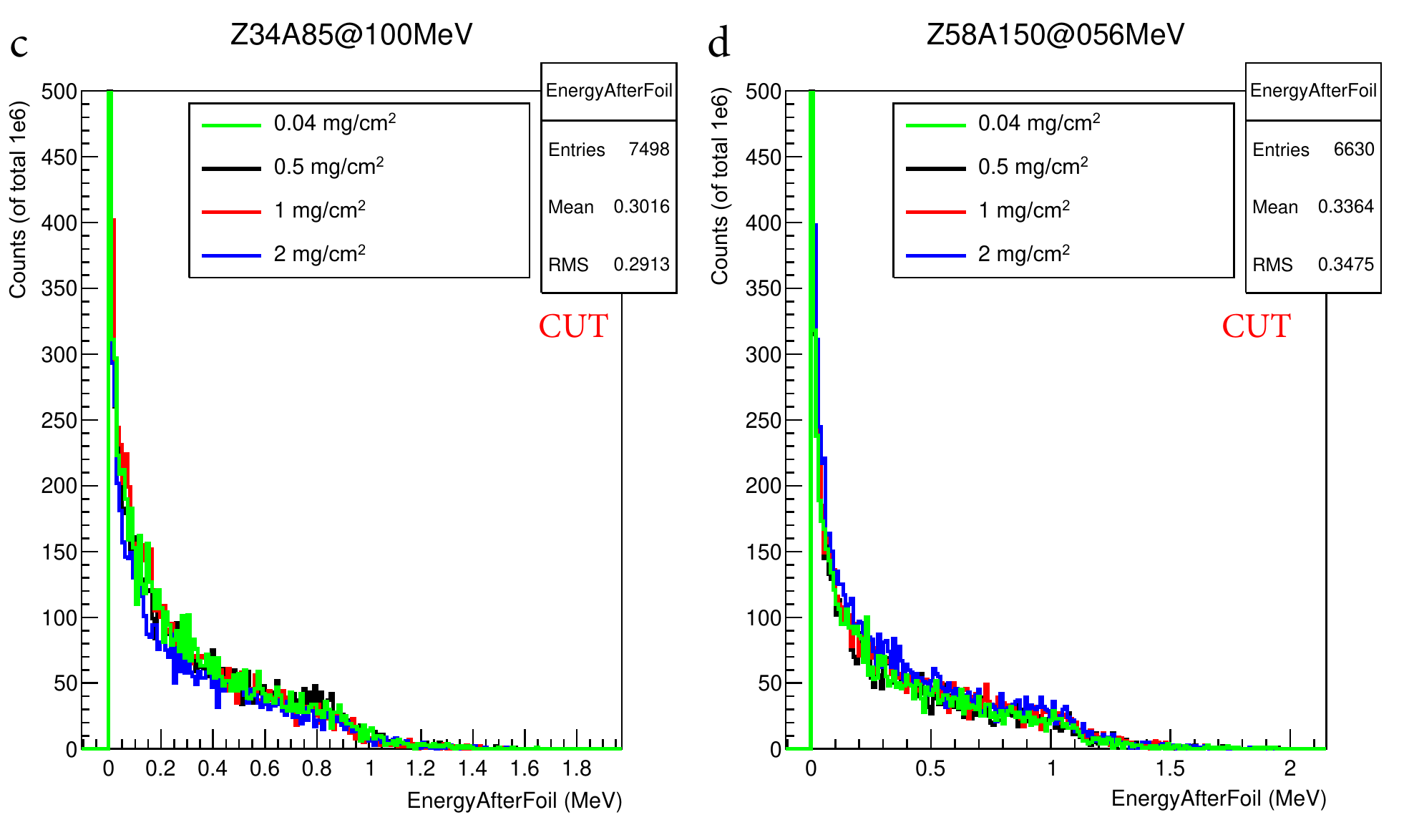}	
\caption[The energy after foil after target for different Ni foils.]{The energy after target (a and b) and energy after foil (c and d) shown for the lightest and heaviest masses, A = 85 (left) and 150 (right). The simulated Ni foil thicknesses are 0.04\,mg/cm$^2$, 0.5\,mg/cm$^2$, 1.0\,mg/cm$^2$ and 2.0\,mg/cm$^2$.}
\label{Fig20}
\end{figure}

\begin{figure}[b!]
\centering
    \includegraphics[width=0.95\textwidth]{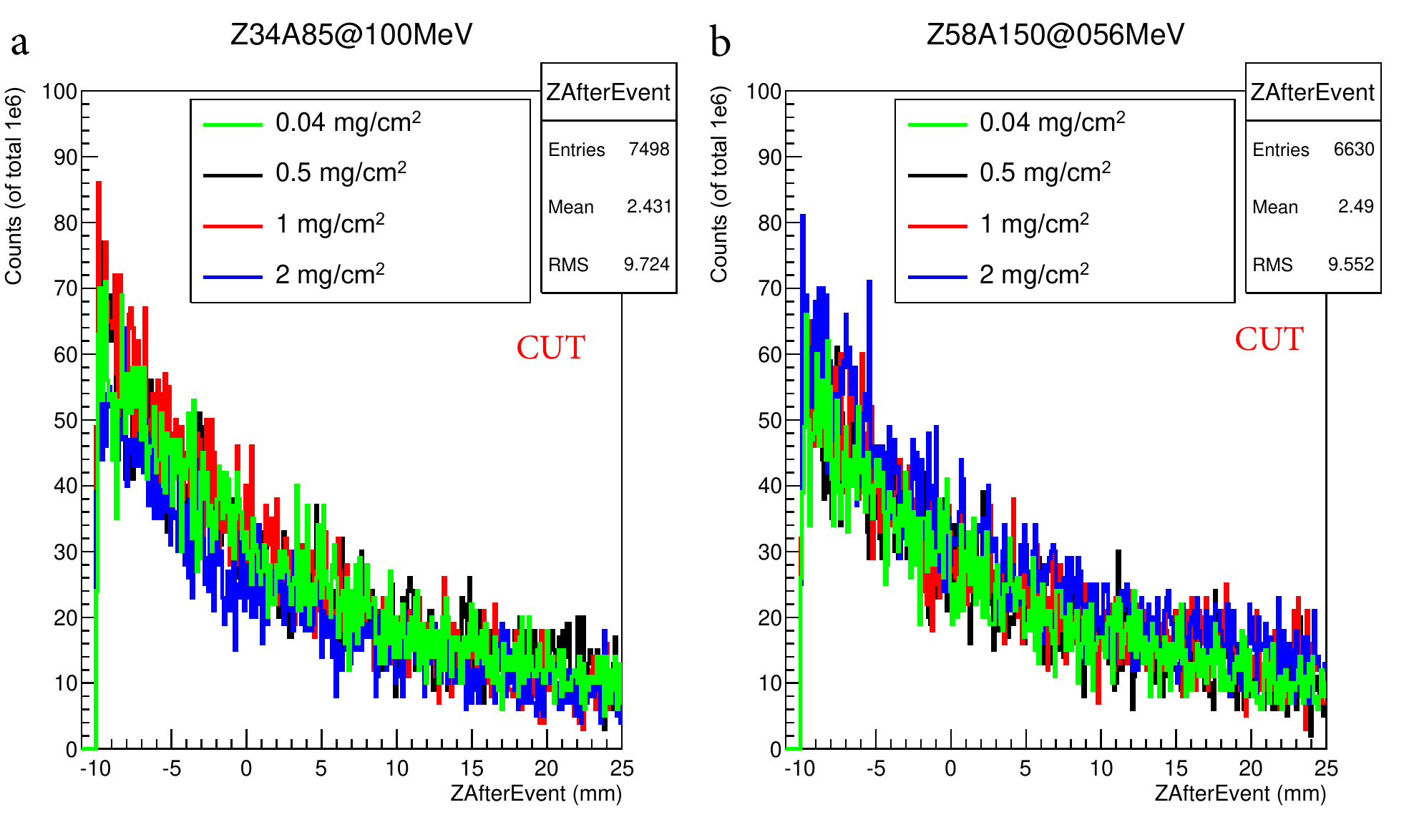}
    \includegraphics[width=0.95\textwidth]{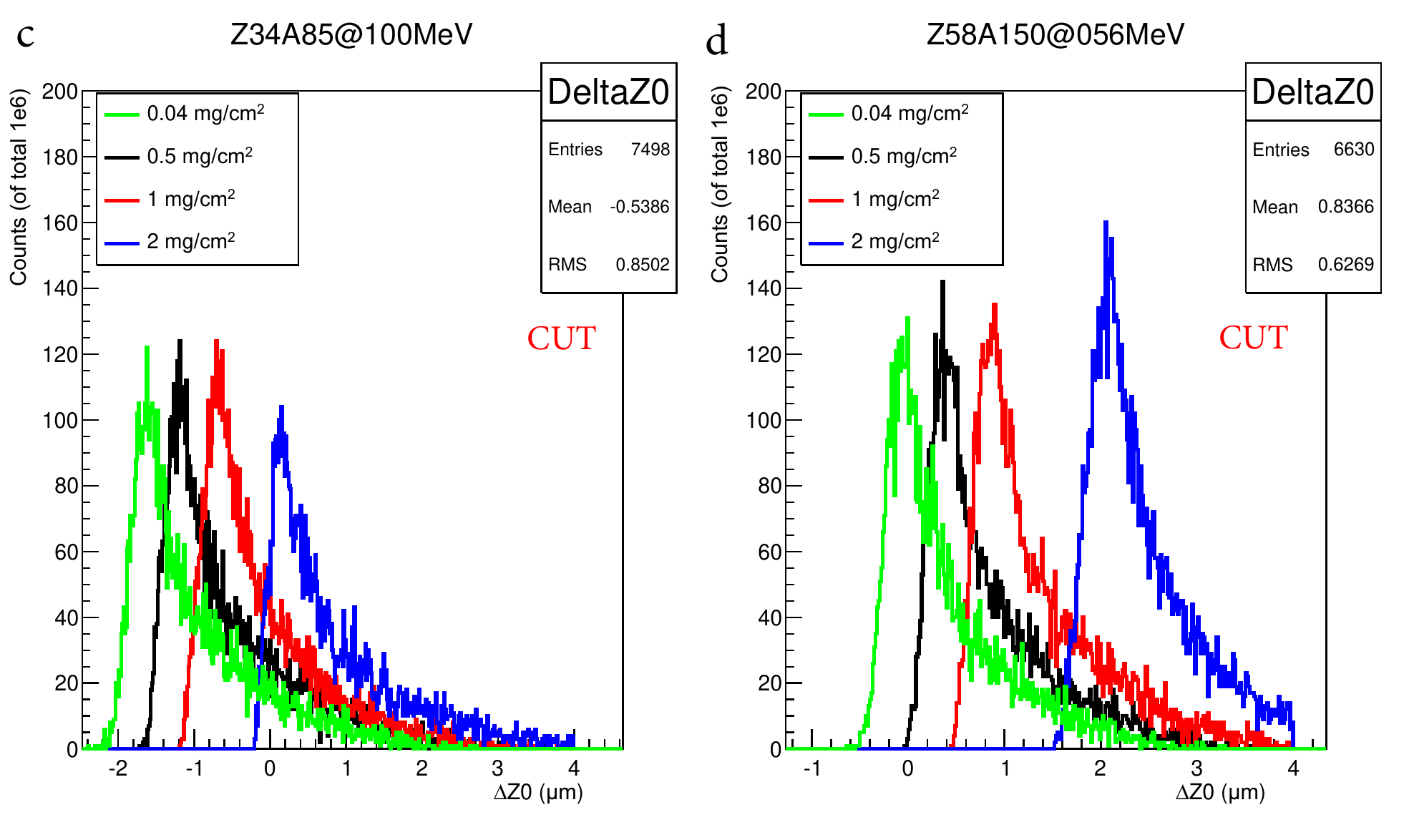}   	
\caption[The place of origin for ions passing through different Ni foils.]{The place of origin in the target (upper) and stopping range (lower) shown for the lightest and heaviest masses, $A = 85$ and 150. The simulated Ni foil thicknesses are 0.04\,mg/cm$^2$, 0.5\,mg/cm$^2$, 1.0\,mg/cm$^2$ and 2.0\,mg/cm$^2$.}
\label{Fig21}
\end{figure}

The Ni foil has a thickness of 1\,mg/cm$^2$ which translates into roughly 1.1\,$\mu$m. Four thicknesses were simulated to study their impact on the ion stopping efficiency. The mean energy of the fragments were used as seen in Fig.~\ref{Fig19} for the thicknesses (0.04, 0.5, 1.0 and 2.0\,mg/cm$^2$). Figures~\ref{Fig20}a and b show the energy of the lightest and heaviest fragments after leaving the target and foil for the four thicknesses. Thicker Ni layer means the energy after target has to be higher and more widely spread in order for the fragments to pass through the foil. Figure~\ref{Fig21}a shows the range in the He gas, displayed for the lightest and heaviest masses. The place of origin is shown in Fig.~\ref{Fig21}b.

\begin{figure}[b!]
\centering
    \includegraphics[width=0.49\textwidth]{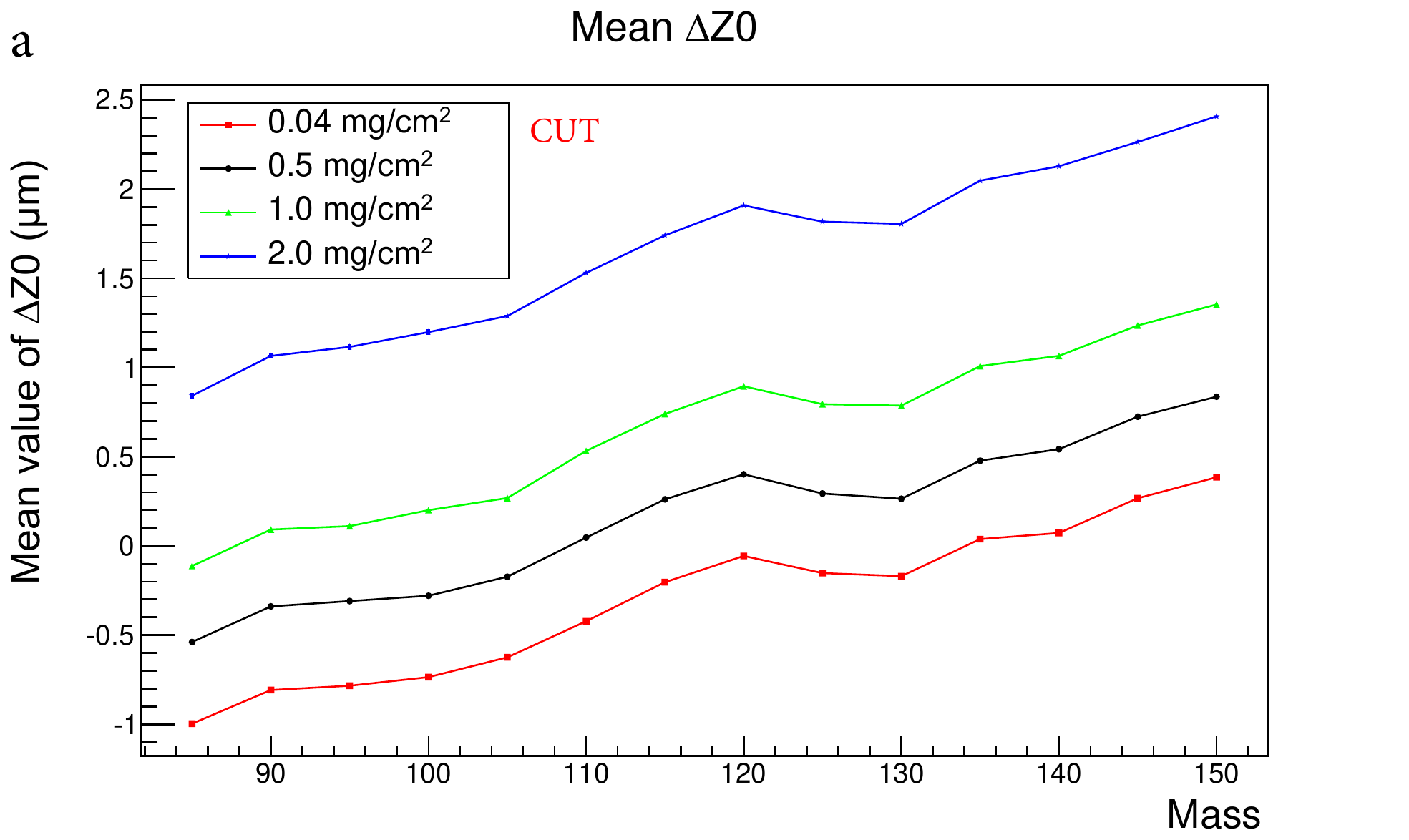}	
    \includegraphics[width=0.49\textwidth]{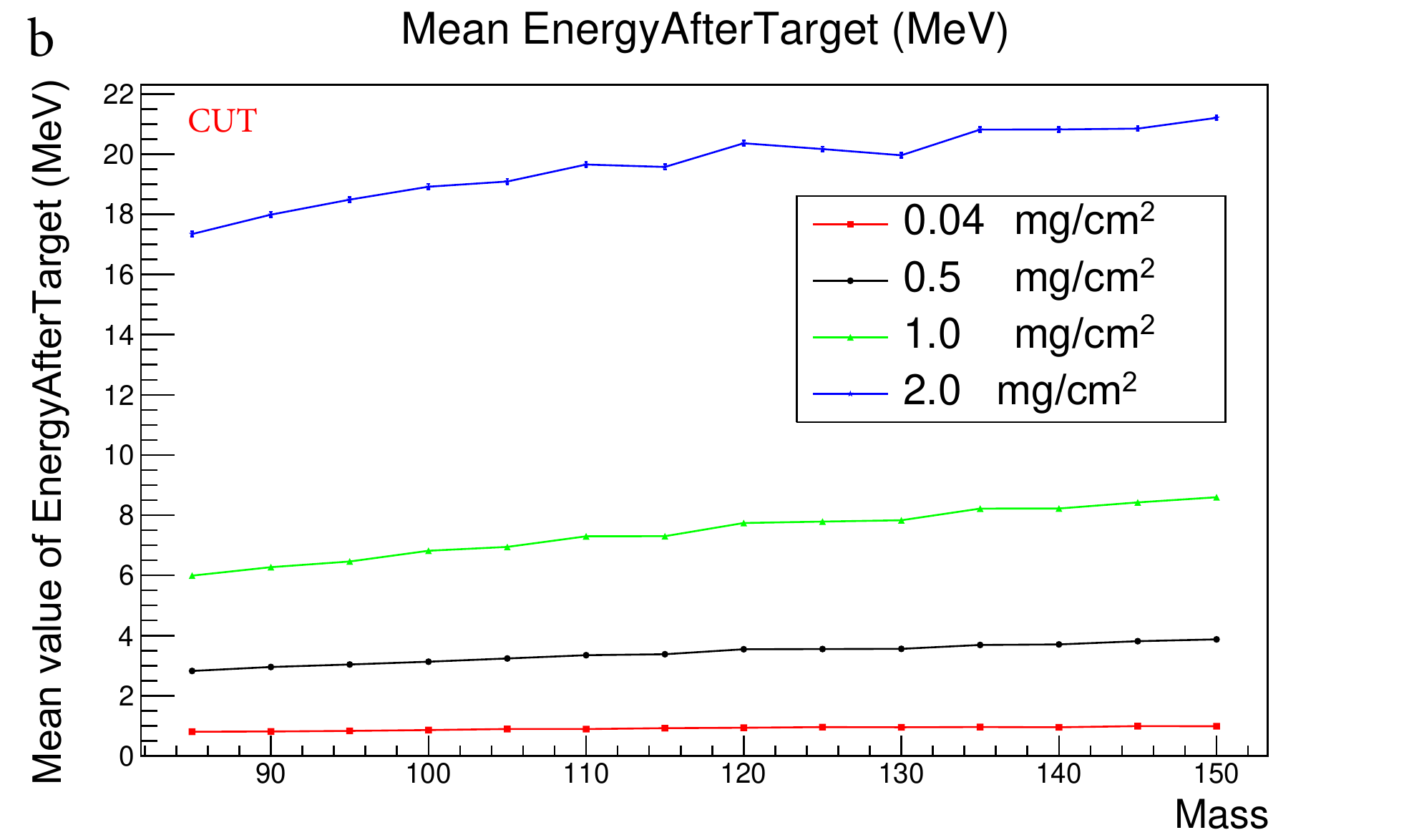}
    \includegraphics[width=0.49\textwidth]{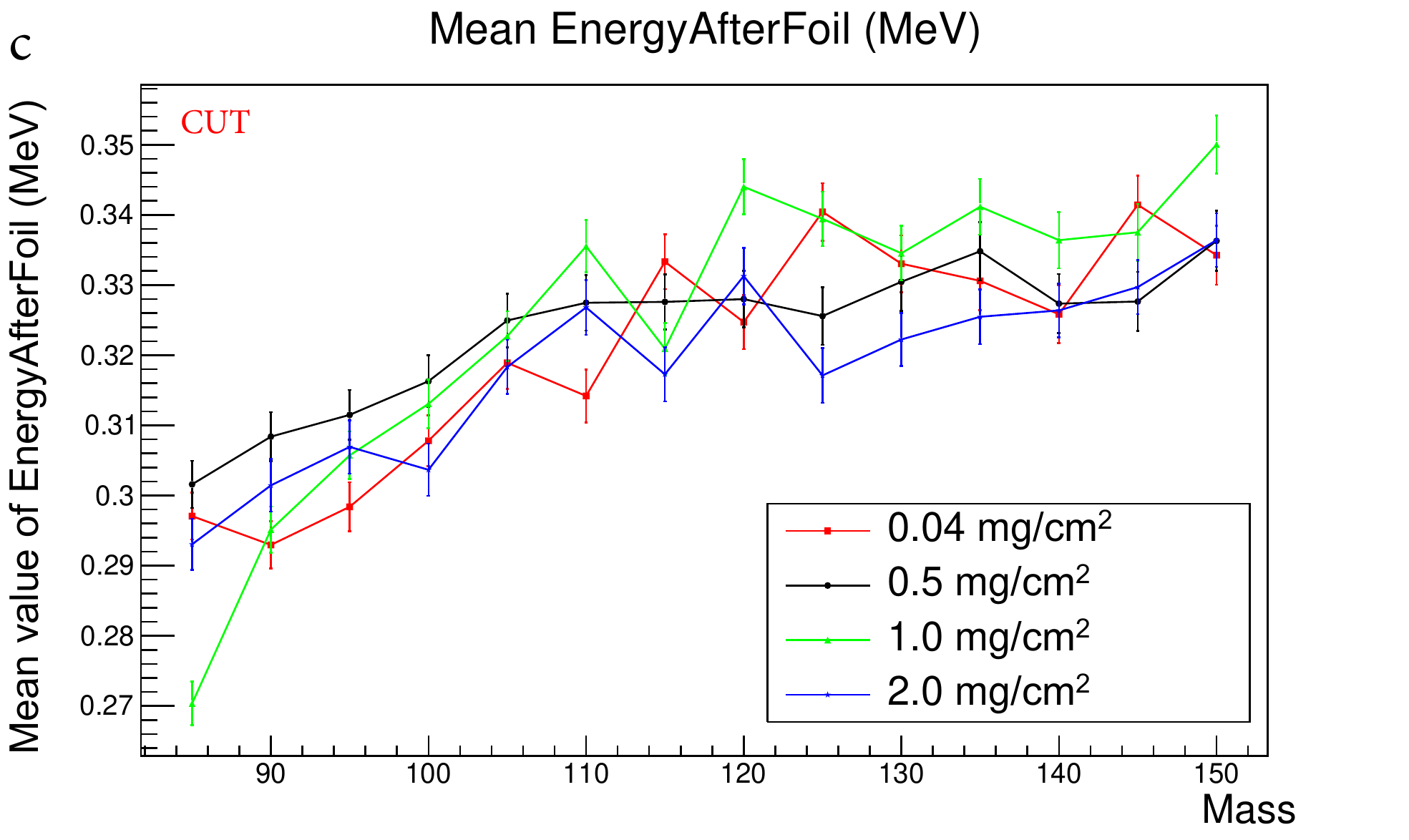}
\caption[The mean values as a function of mass and Ni thickness.]{The mean values as a function of mass and Ni thickness for (a) place of ion origin n the U target, (b) Energy after target and (c) Energy after foil.}
\label{Fig22}
\end{figure}

\begin{figure}[b!]
\centering
    \includegraphics[width=0.9\textwidth]{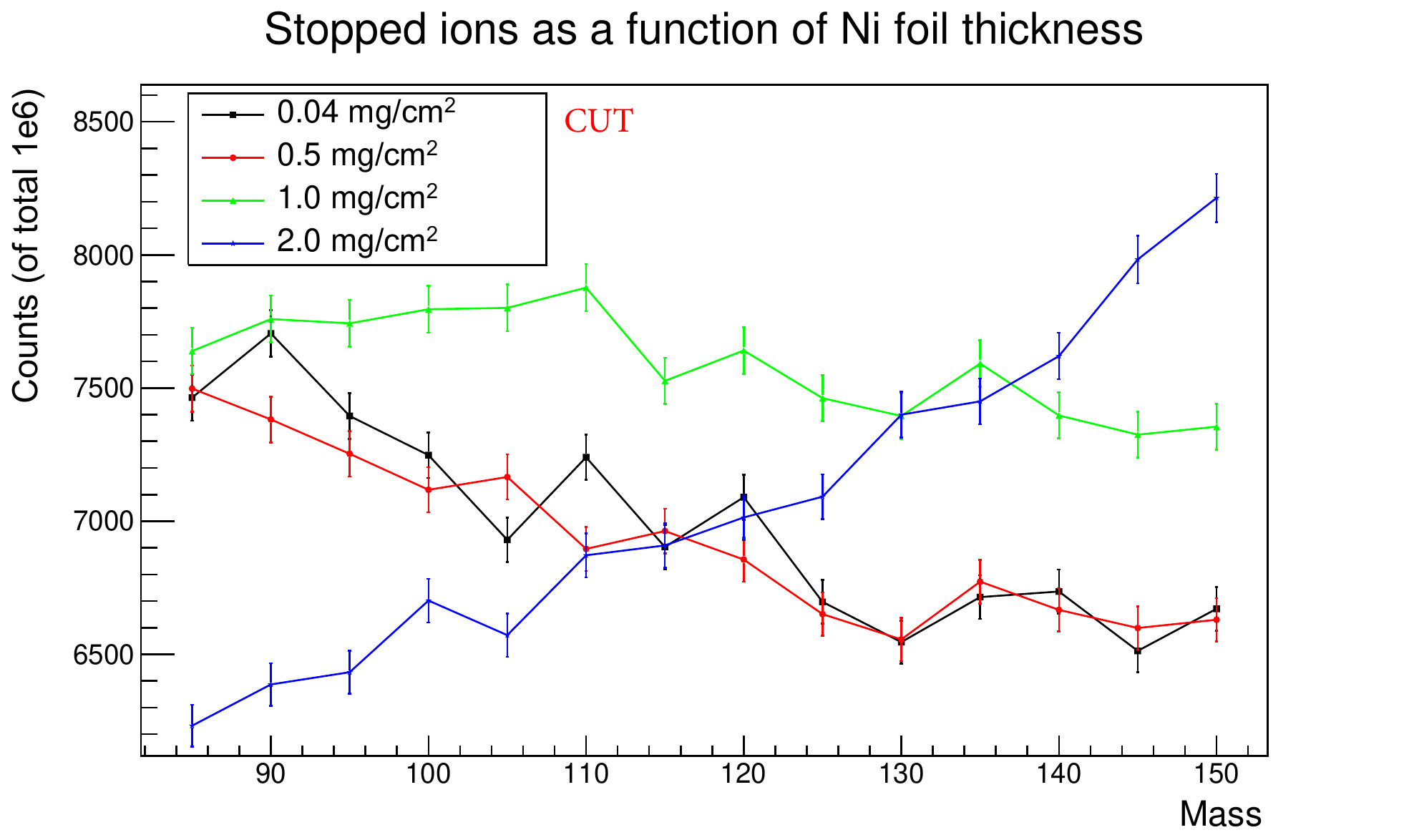}	
\caption[The ions stopped as a function of mass and Ni foil thickness.]{The ions stopped in the gas volume as a function of mass and Ni foil thickness. The mass-dependence is smallest for the 1.0 mg/cm$^2$ Ni foil.}
\label{Fig23}
\end{figure}

In Fig.~\ref{Fig22}, the mean trends are shown. The mean start position in the U target moves in the direction of the stopping chamber as a function of mass. When having thicker Ni foil the average start position also moves in the direction of the stopping chamber. Since the ions lose more energy in the foil they need to lose less energy in the U target. Therefore, the average energy after target is larger with higher Ni thickness as seen in Fig.~\ref{Fig20}b. The total ion counting as a function of mass and Ni foil thickness is plotted in Fig.~\ref{Fig23}. The largest mass dependences were found for the 2.0\,mg/cm$^2$ foil, while the 1.0 mg/cm$^2$ foil gave the smallest dependency. However, it seems the Ni foil thickness may be fine-tuned in order to shift the mass dependence. Thicker foils seem to favour the heavy fragments over the light fragments. 

\clearpage
\section{Gas pressure} 
\begin{figure}[b!]
\centering
    \includegraphics[width=0.91\textwidth]{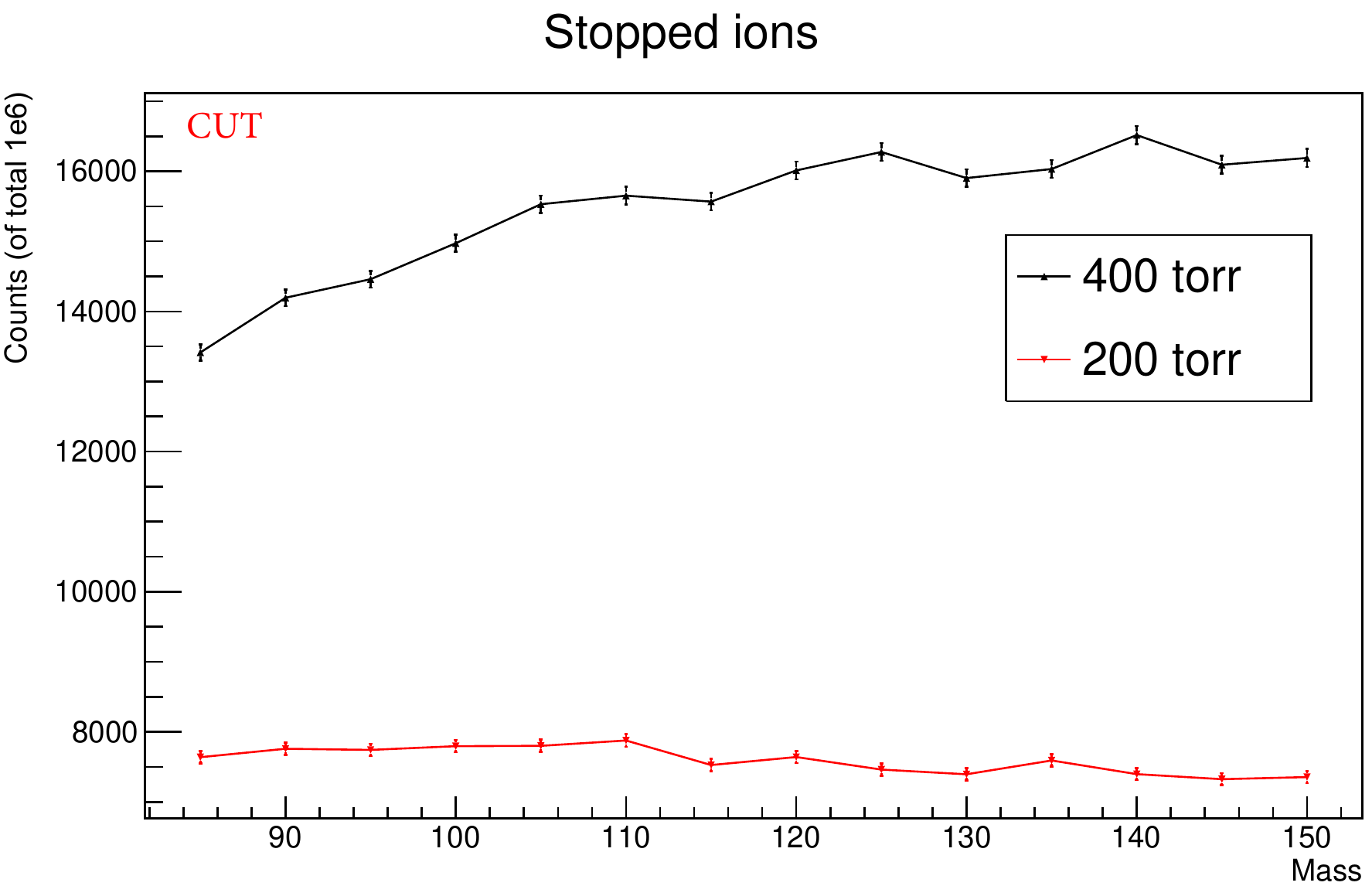}	
\caption[Number of stopped ions as a function of gas pressure.]{Number of stopped ions as a function of gas pressure. At 400\,torr twice the count rate was found, however also giving larger relative mass dependence.}
\label{Fig24}
\end{figure}

\begin{figure}[b!]
\centering
    \includegraphics[width=0.95\textwidth]{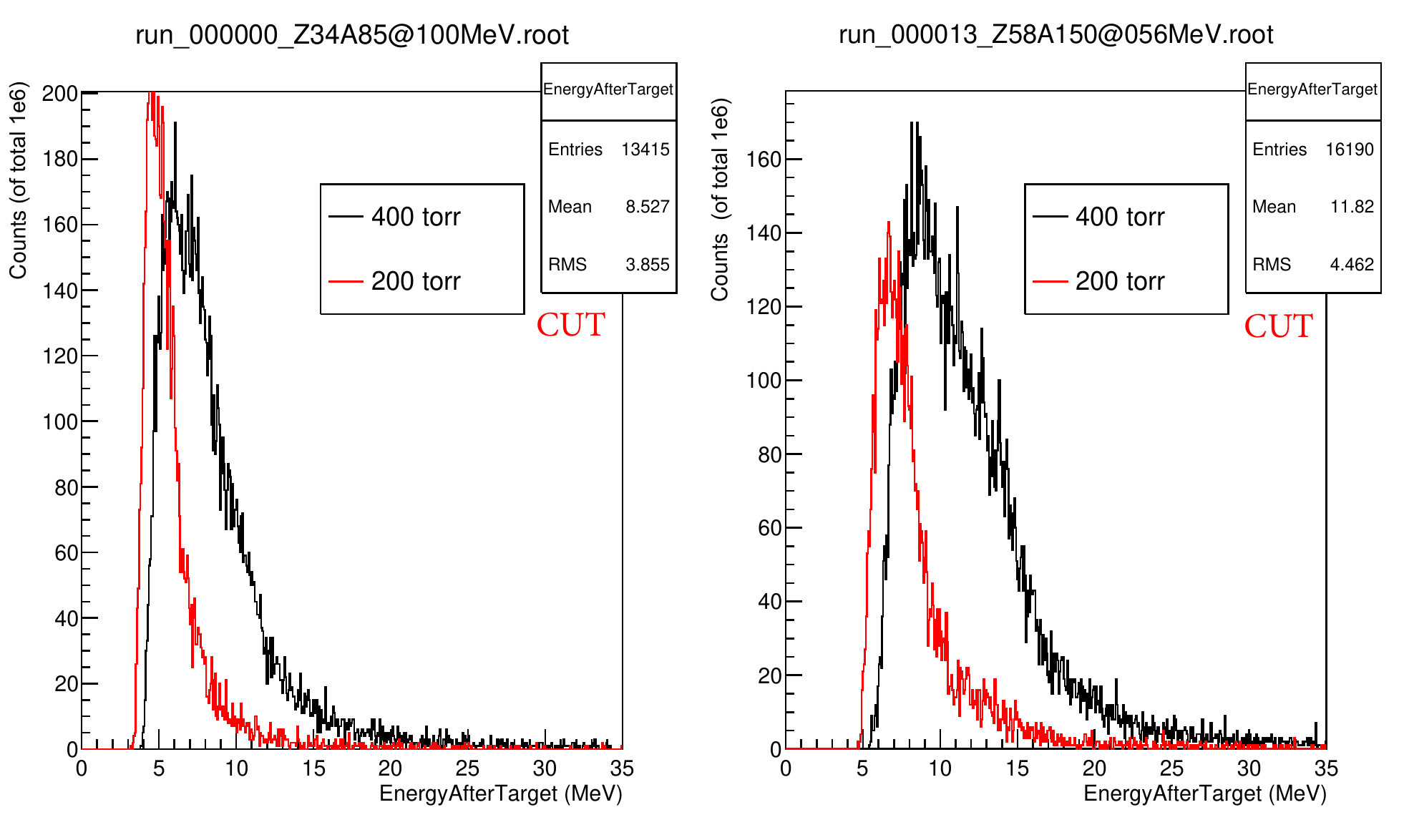}	
    \includegraphics[width=0.95\textwidth]{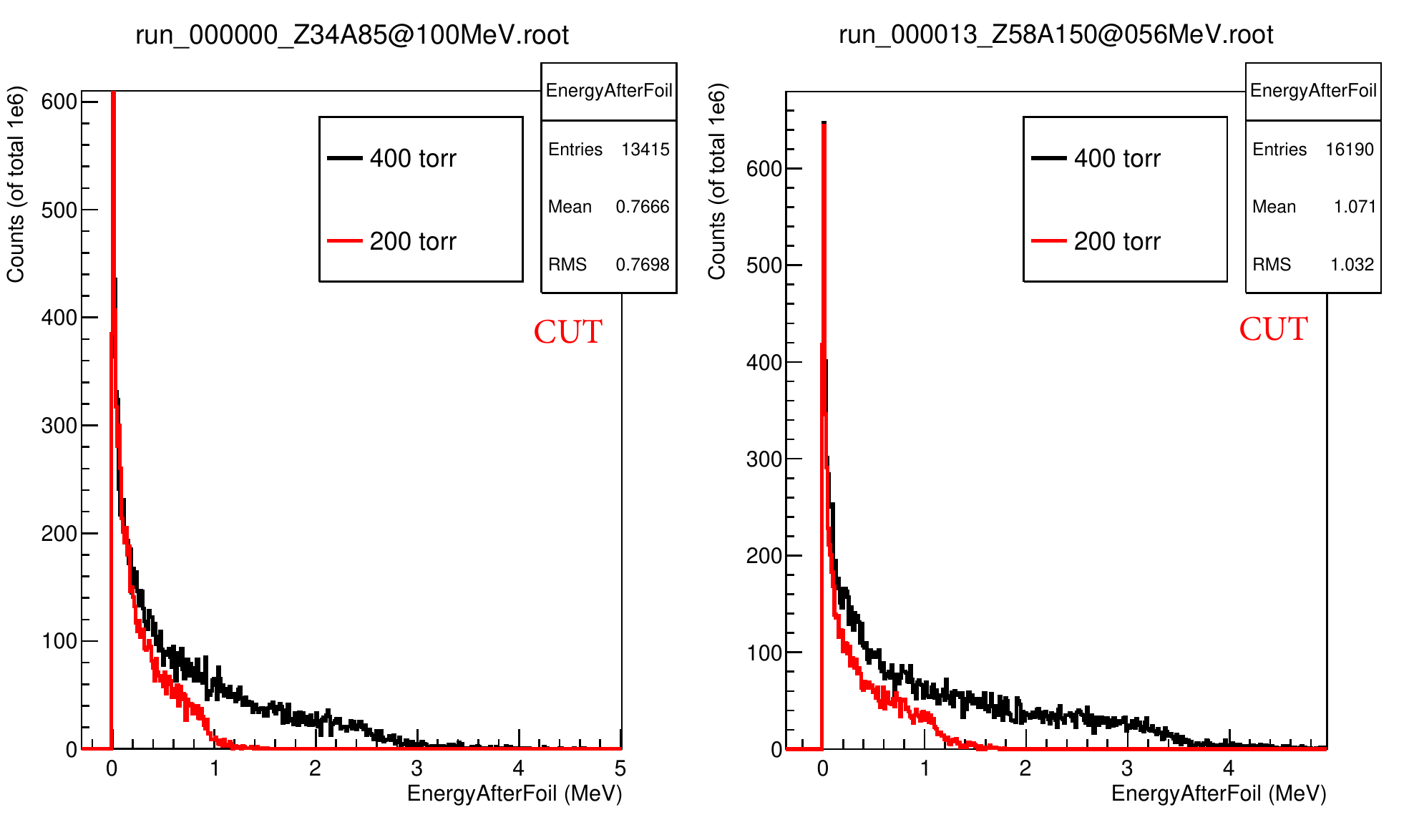}	
\caption[Ions energies for 400\,torr gas pressure.]{The energy after target (upper) and after Ni foil (lower) shown for the lightest and heaviest masses, $A = 85$ and 150 for 200 and 400\,torr gas pressure.}
\label{Fig25}
\end{figure}

\begin{figure}[b!]
\centering
    \includegraphics[width=0.95\textwidth]{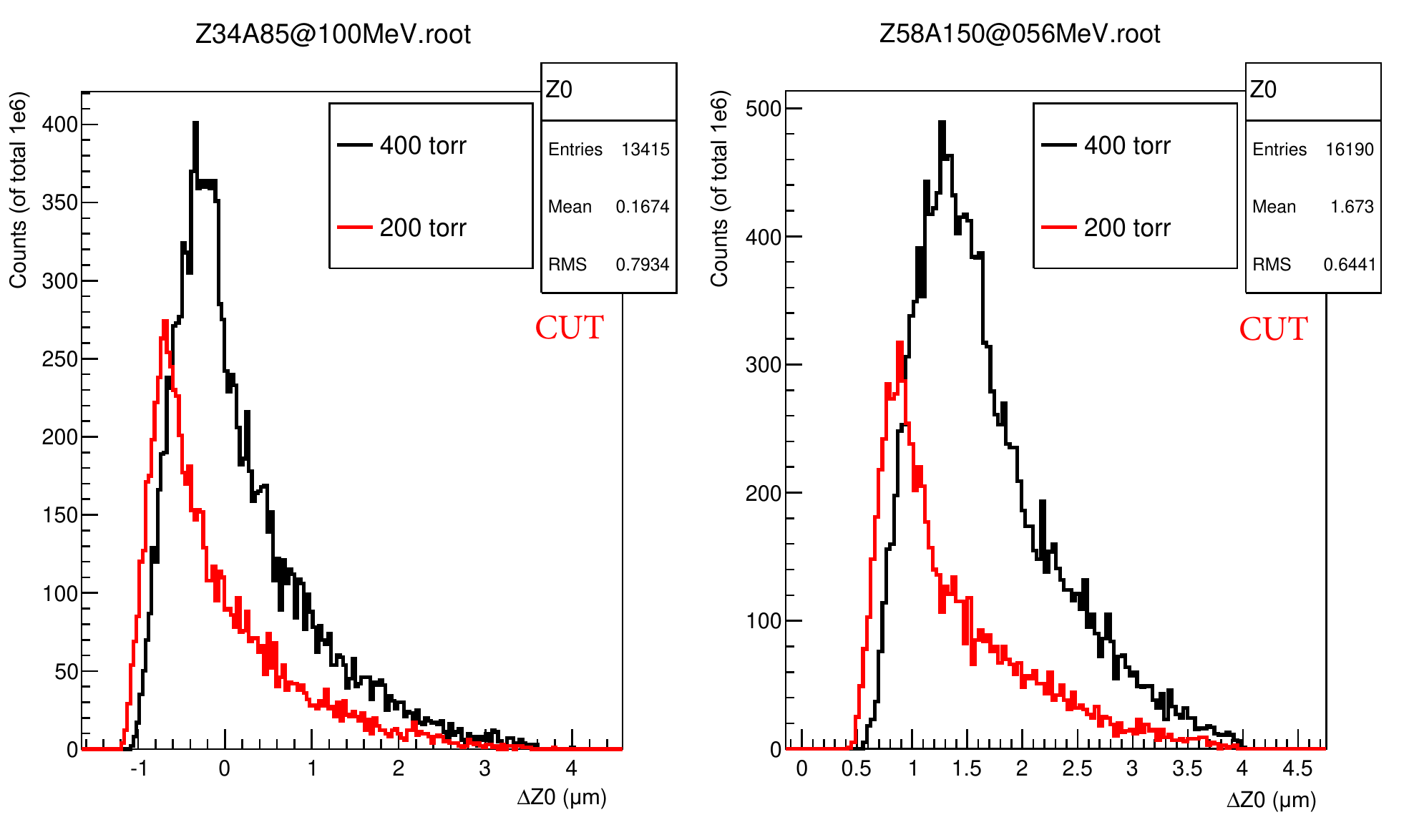}	
    \includegraphics[width=0.95\textwidth]{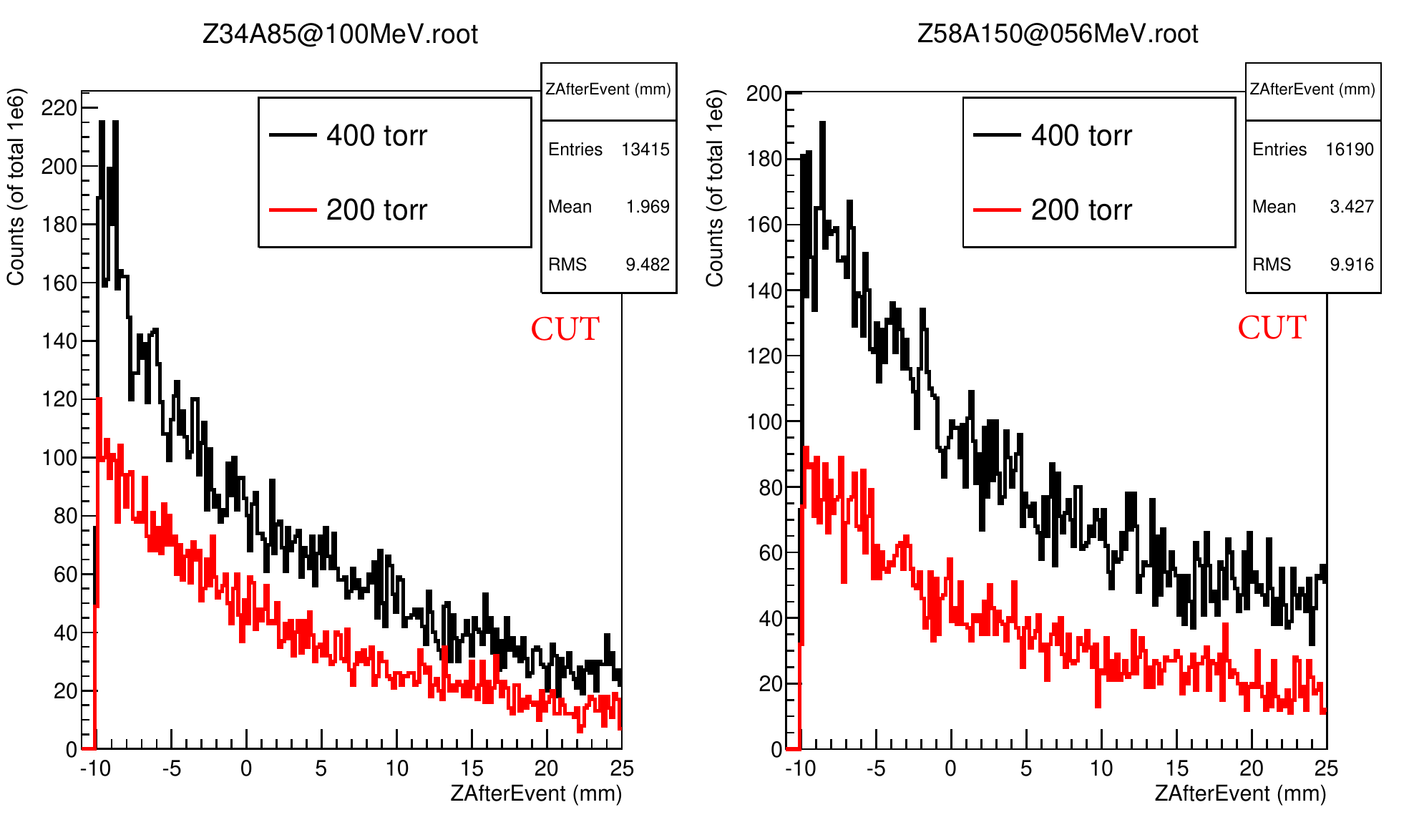}	
\caption[The place of origin for 400\,torr gas pressure]{The place of origin for the stopped ions (a) and the stopping range (b) for the lightest and heaviest masses, $A = 85$ and 150 amu for 200 and 400\,torr gas pressure.}
\label{Fig26}
\end{figure}

Higher gas pressure (400\,torr) was simulated in comparison to the nominal 200\,torr. By increasing the pressure, more fragments were stopped in the gas. Figure~\ref{Fig24} shows the mass-dependence of the stopping efficiency rate. About twice as many fragments were stopped but at a price of stronger mass-dependence (from 9\% negative trend to about 20\% positive trend). Higher gas pressure increases the number of heavier fragments. The following plots show the energy after the Ni foil, energy after target, $\Delta Z_0$ and $Z$ for the lightest and heaviest masses. At a higher gas pressure more ions are accepted with higher exit energies from the U target as well as from the Ni foil (Fig.~\ref{Fig25}). Collected ions can originate from other regions in the target, closer to the stopping chamber (Fig.~\ref{Fig26}).  

\begin{figure}[b]
\centering
    \includegraphics[width=0.95\textwidth]{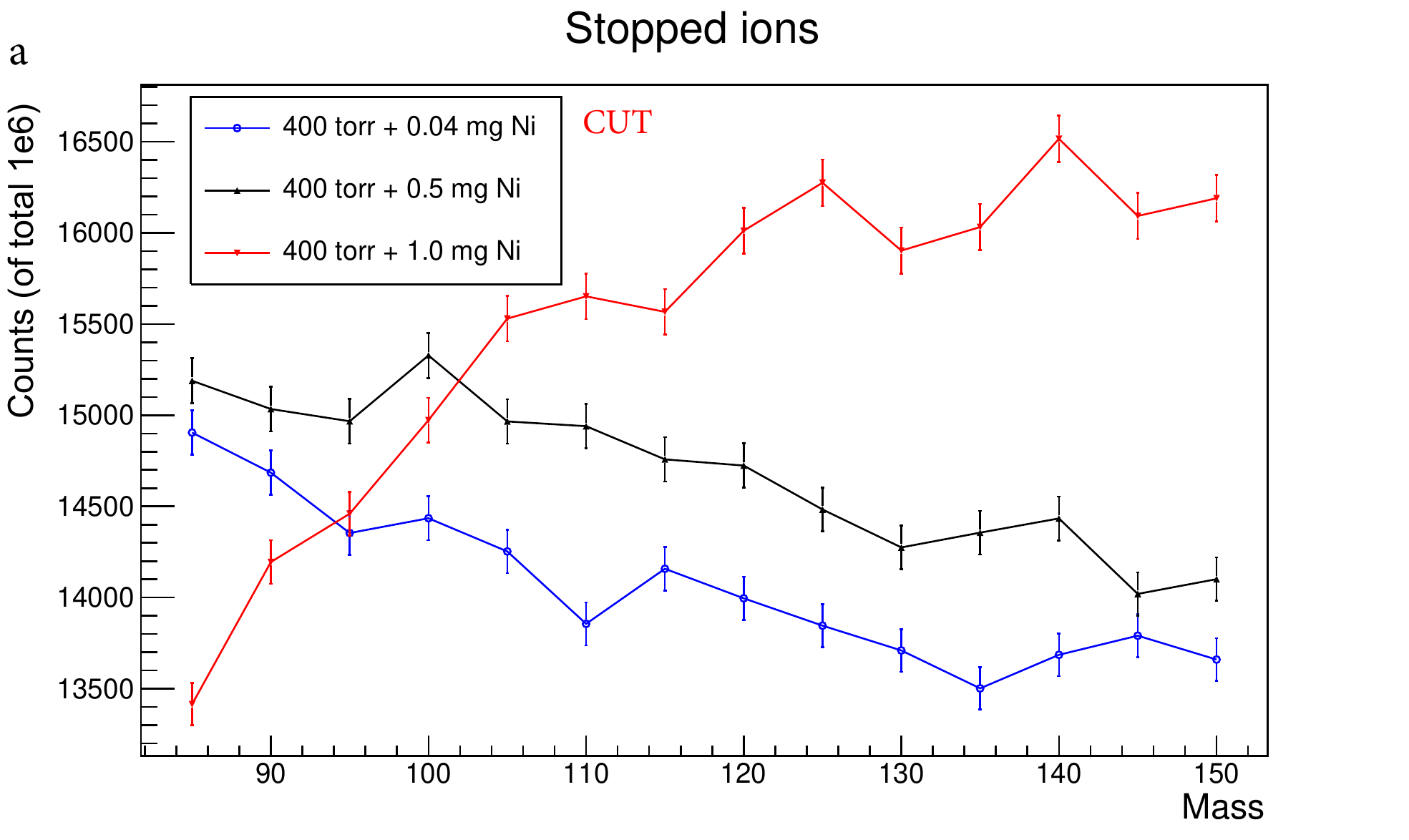}	
    \includegraphics[width=0.95\textwidth]{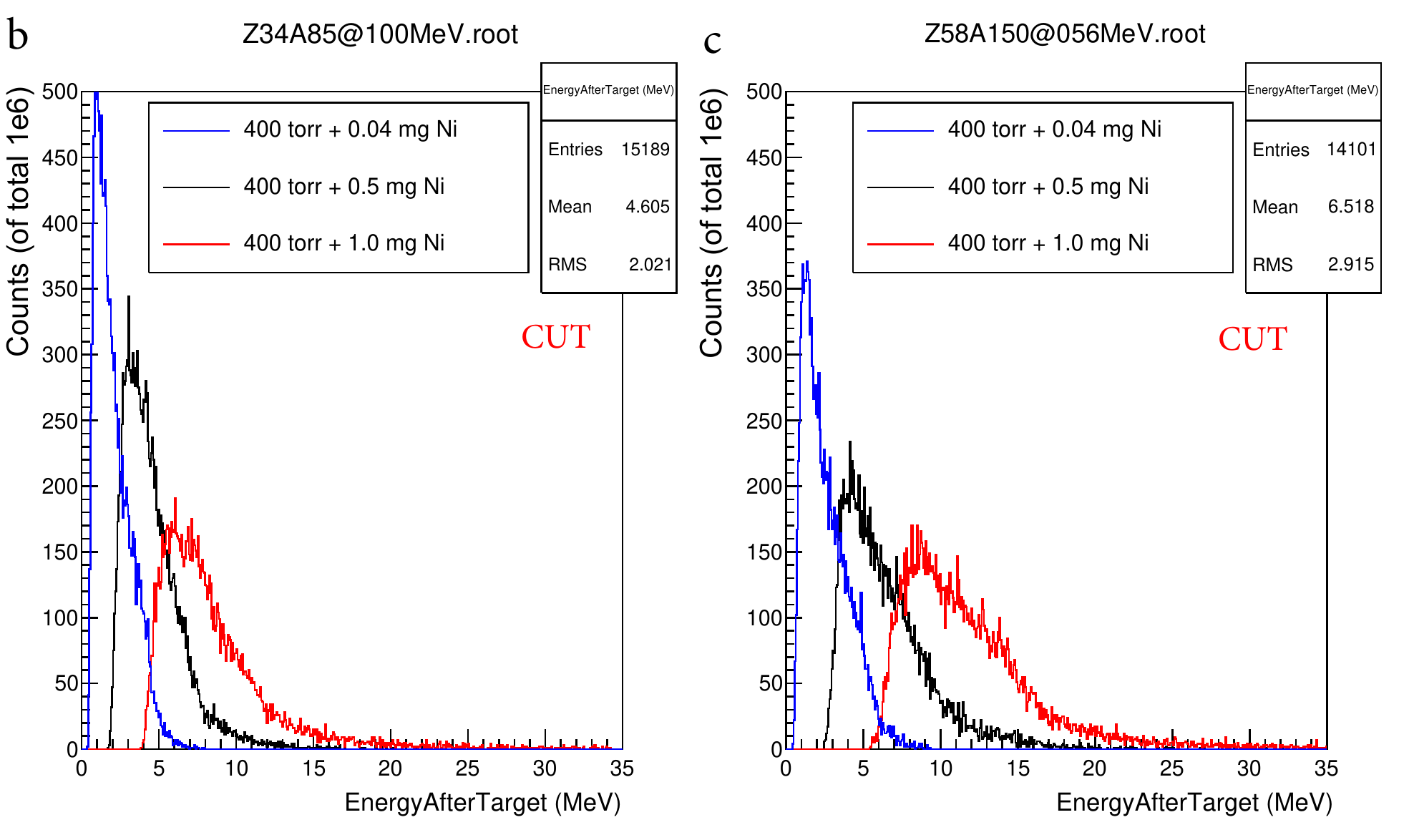}
\caption[Ion counting for 400 torr + 0.04\,mg/cm$^2$, 0.5\,mg/cm$^2$ and 1\,mg/cm$^2$.]{(a) Stopped ion for the 400 torr runs, with Ni thicknesses of 0.04, 0.5 and 1\,mg/cm$^2$. The thinner foils gives a smaller mass-dependence (about 8 \%) in contrast to the relative difference of about 20 \% for the 1\,mg/cm$^2$ foil. (b) energy after target for the stopped ions, for $A = 85$ and 150 amu. }
\label{Fig27}
\end{figure}

 The Ni foil thickness was found to affect the mass-dependency in the ion counting (Fig.~\ref{Fig23}). A thinner layer of Ni decreased the stopping of heavy fragments to a larger extent than lighter fragments. Therefore, in view of the trend observed for the higher gas pressure, one could search for a balance between the gas pressure and Ni foil thickness, to obtain twice the count rate but with a reduced mass-dependence. Using thinner Ni foils together with a He gas pressure of 400\,torr, a smaller mass dependence was achieved as seen in Fig.~\ref{Fig27}a. The spread of the energy distribution is reduced by having a thinner Ni foil as can be seen in Fig.~\ref{Fig27}b for the energy after target. One could obtain a nearly mass-independent trend with a Ni foil thickness between 0.5 and 1.0\,mg/cm$^2$.
 
 \clearpage

\section{\label{sectUTarg}Uranium target thickness}
\begin{figure}[H]
\centering
    \includegraphics[width=0.95\textwidth]{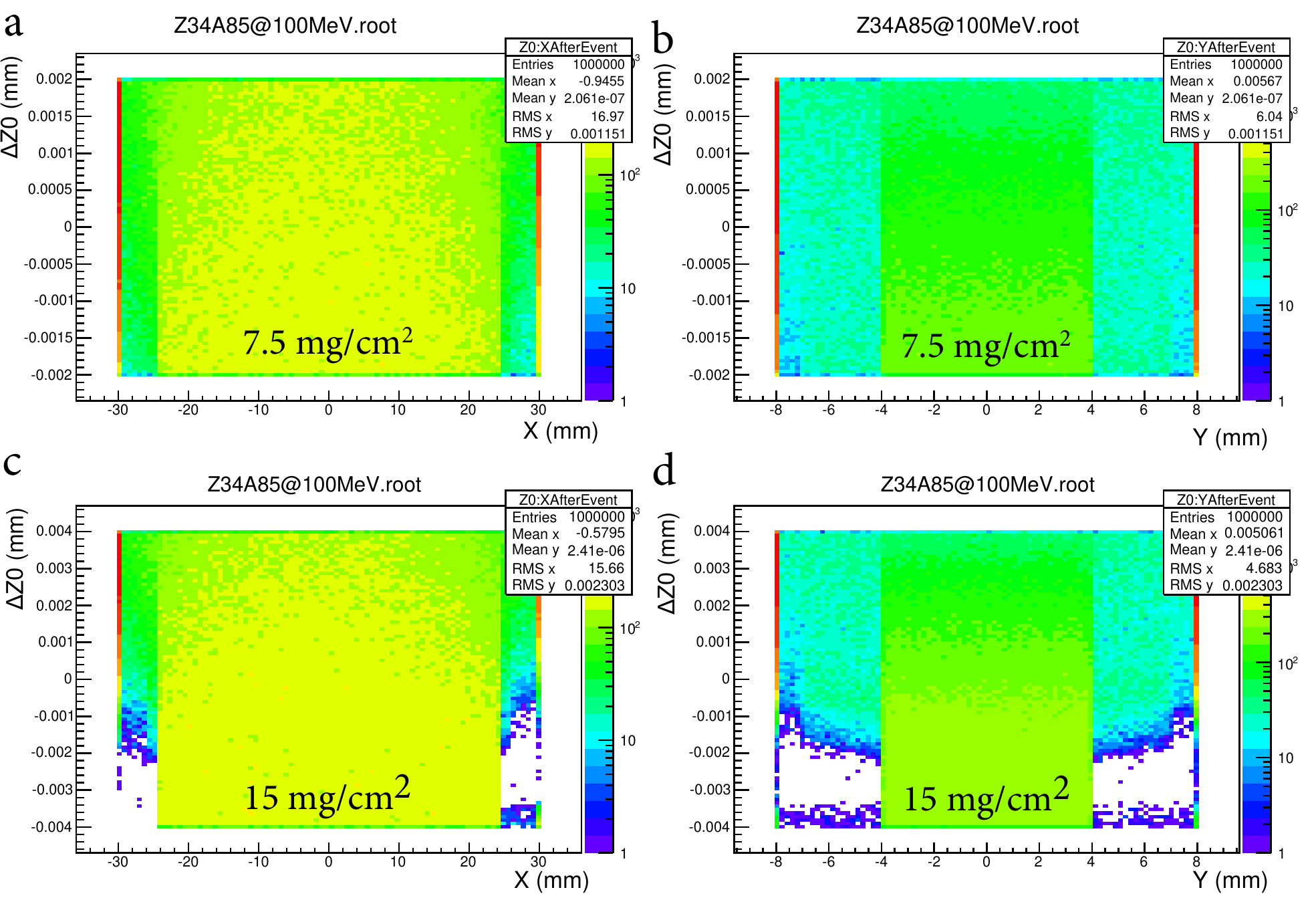}	
\caption[Simulated U target thicknesses.]{The U target thicknesses projected in xz-plane (left) and yz-plane (right), for 7.5\,mg/cm$^2$ (a,b) corresponding to $\sim$4 $\mu$m and 15\,mg/cm$^2$ (c,d) corresponding to $\sim$8 $\mu$m.}
\label{Fig28}
\end{figure}

\begin{figure}[b!]
\centering
    \includegraphics[width=0.9\textwidth]{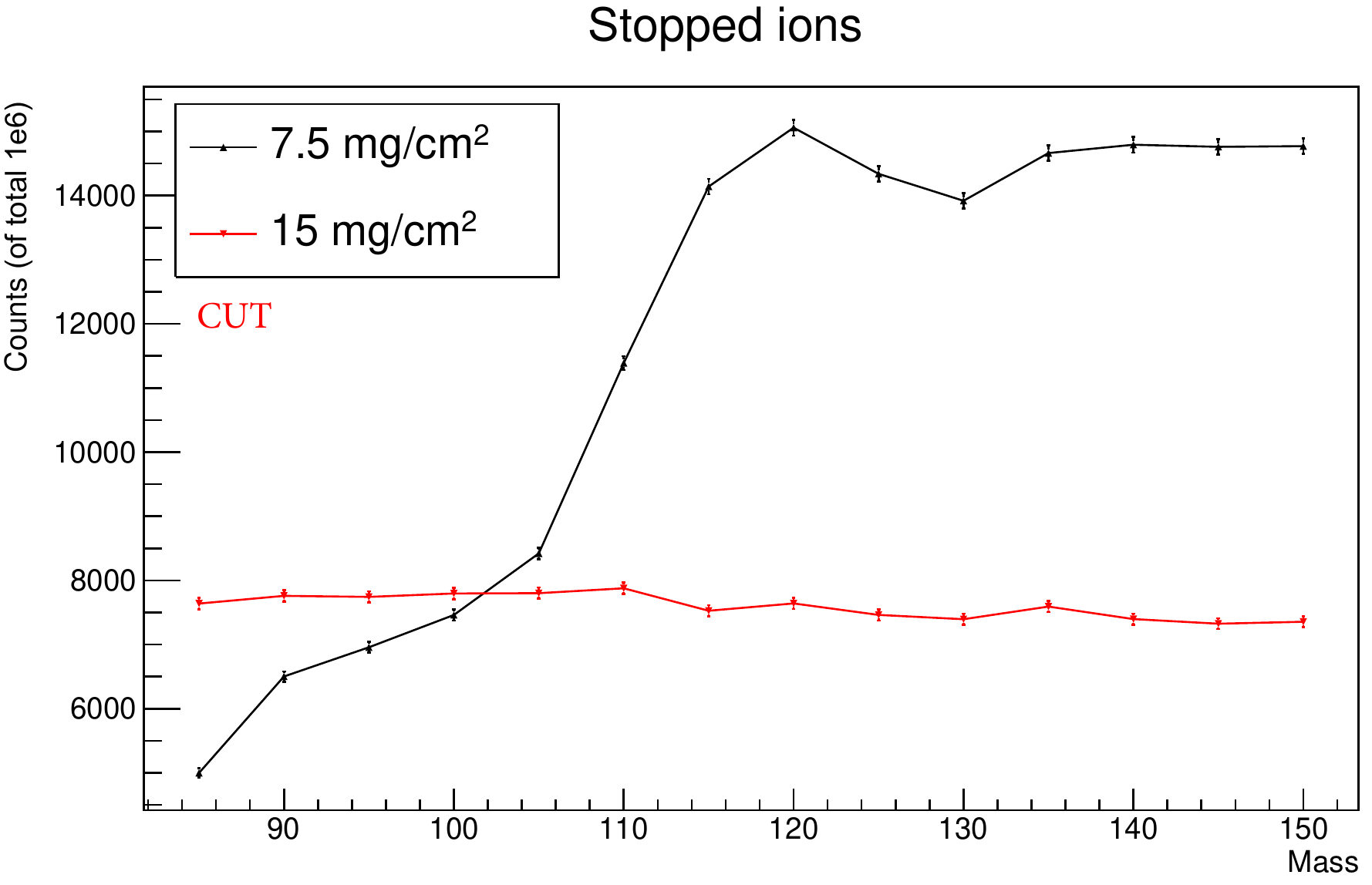}	
    \includegraphics[width=0.98\textwidth]{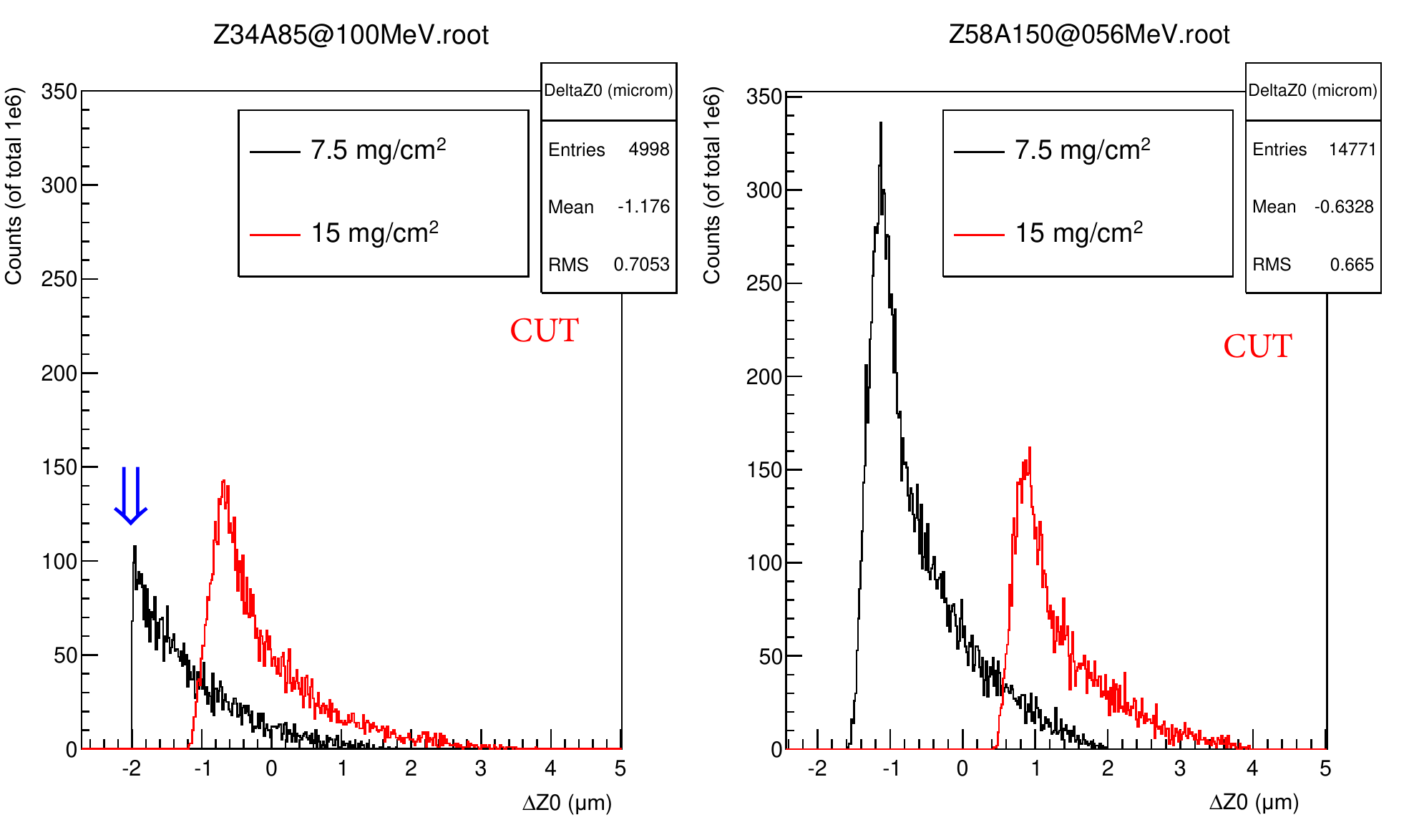}
\caption[Ion counting for different U thicknesses.]{(a) The ion counting as a function of fragment mass. The ion counting becomes strongly mass dependent for a thinner Uranium target. (b) The place of origin for the ions. For the lighter fragments, there is not enough material to stop sufficient amount of ions (at -2 $\mu$m).}
\label{Fig29}
\end{figure}

It is also of interest to study the effect of a thinner Uranium target. The target thickness was changed from 8\,$\mu$m to 4\,$\mu$m (7.5\,mg/cm$^2$). The target thicknesses can be seen in Fig.~\ref{Fig28}. By simulating half the U thickness the ion counting showed a large mass dependence. For the light fragments the counting was half of that of the heavier fragments (See Fig.~\ref{Fig29}a).

This is due to the fact that the lighter fragments need larger energy losses to be fully stopped in the stopping chamber. Looking at the place of origin, the lighter fragments originate from the end of the target and need in fact more target material to be at nominal energy after leaving target. We conclude that the there is a minimum target thickness required and that 15\,mg/cm$^2$ is enough to provide the entire spectrum of fragment energies that ensures a fair chance of stopping all fragments. 

\begin{figure}[b!]
\centering
    \includegraphics[width=0.95\textwidth]{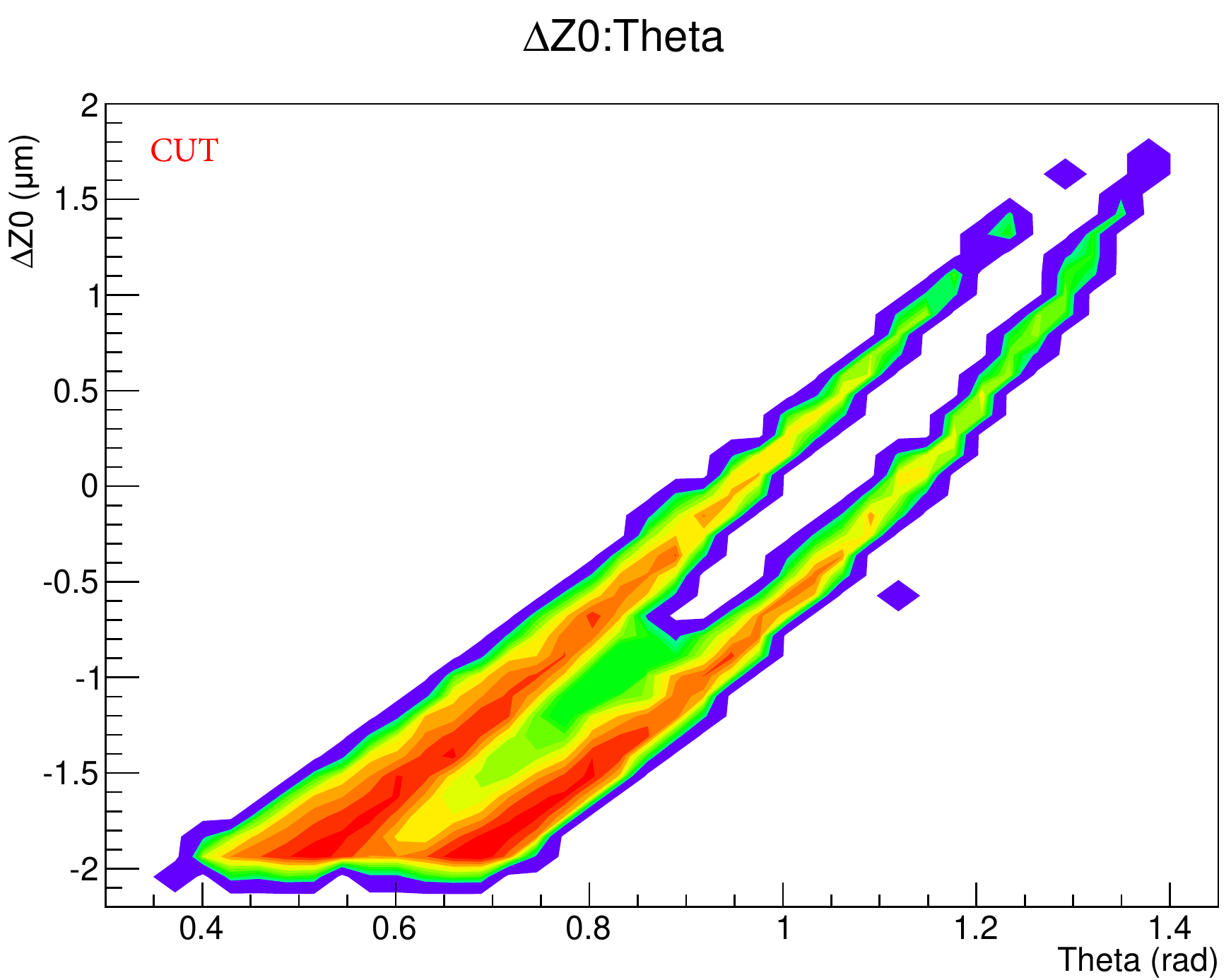}	
\caption[The place of origin as a function of emission angle.]{The place of origin as a function of emission angles of ions stopped in the gas cell, for the 7.5\,mg/cm$^2$ thick target. Forward emission angles are excluded since the energy losses are too low to stop the fragments, in contrast to Fig.~\ref{Fig17}b.}
\label{Fig30}
\end{figure}

Thinner samples will give large mass dependencies of the ion counting. Figure~\ref{Fig30} shows that ions with forward emission angles are not stopped in the chamber as compared with Fig.~\ref{Fig17}b. Only fragments with higher emission angles are stopped since they travel through enough material. 

\clearpage

\section{Ion charge} 
\begin{figure}[b!]
\centering
    \includegraphics[width=0.45\textwidth]{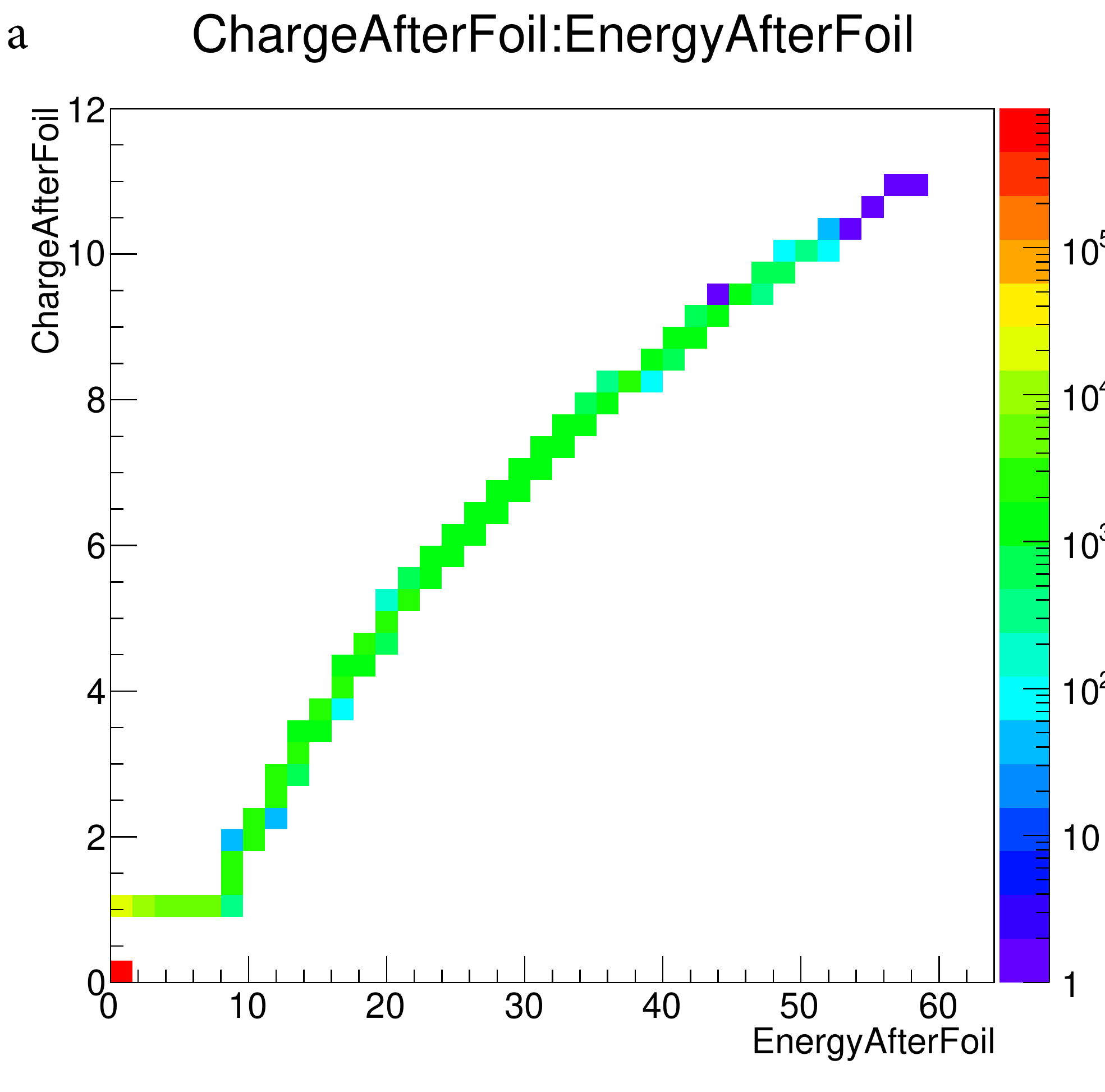}
    \includegraphics[width=0.45\textwidth]{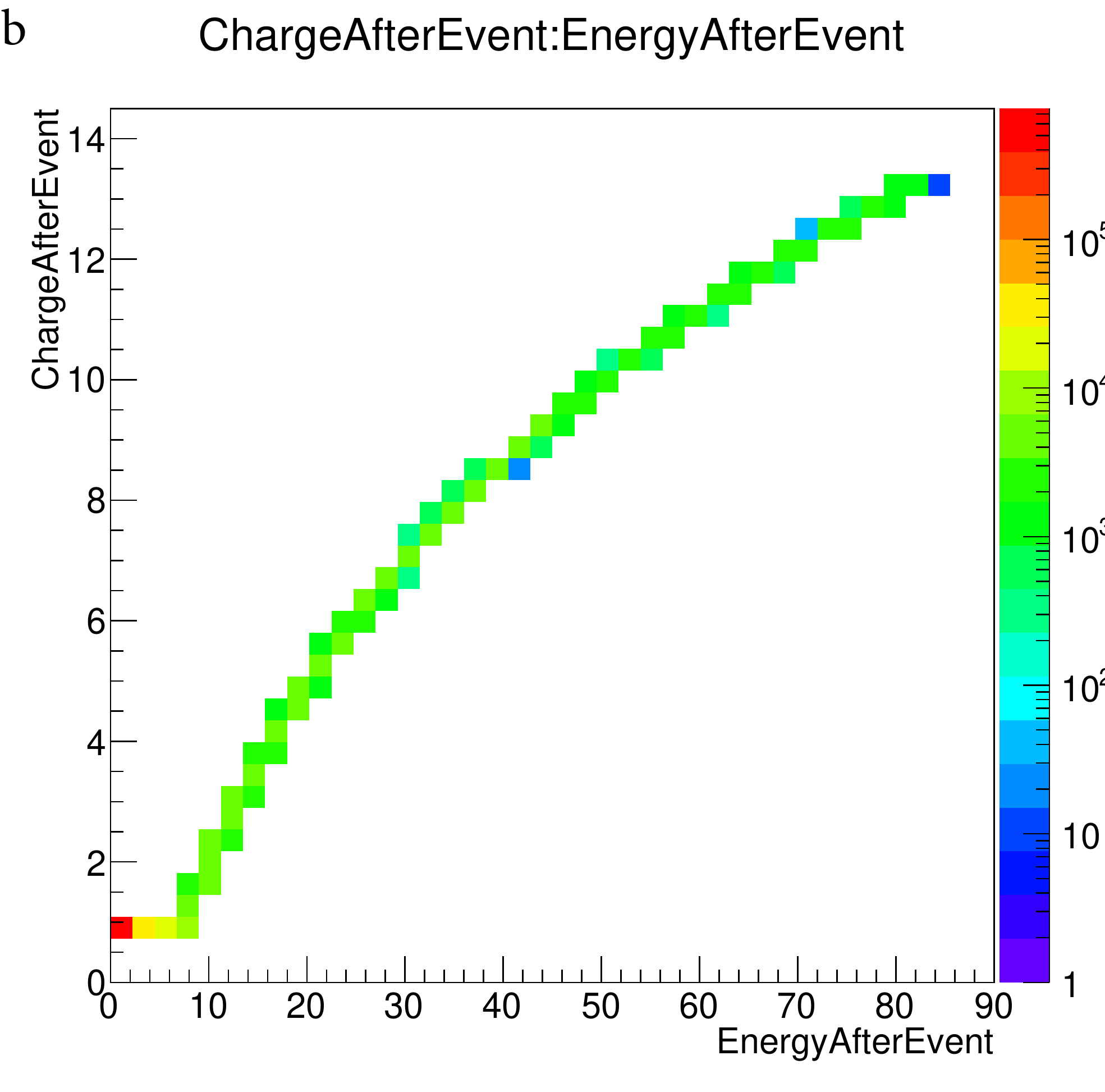}   
    \includegraphics[width=0.45\textwidth]{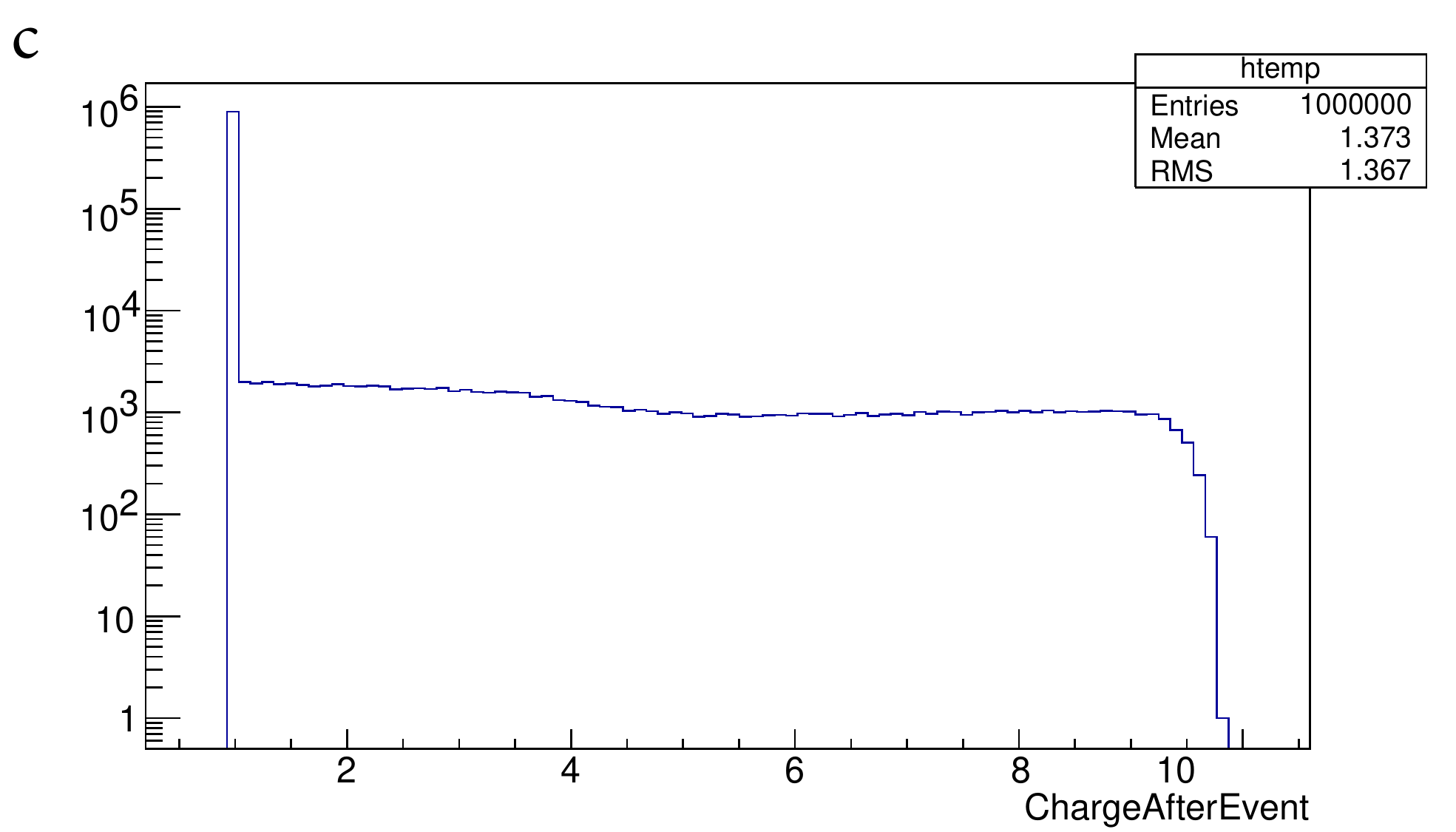}
    \includegraphics[width=0.45\textwidth]{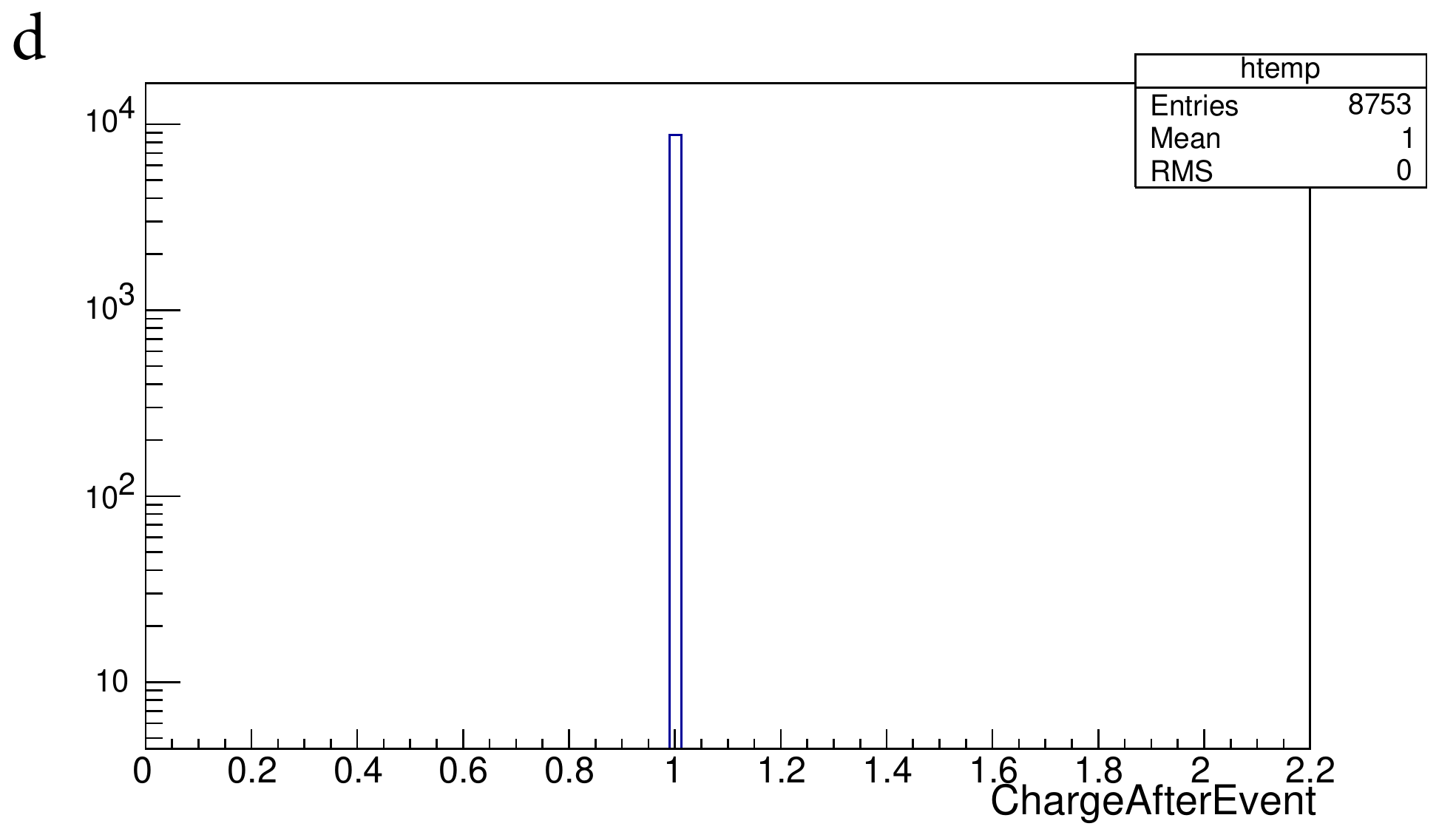}      	
\caption[Ion charge state]{(a) The ion effective charge as a function of kinetic energy after the foil. The zero charges correspond to ions that have not entered the foil (b). The charge of all ions can be seen in (c) where the higher charges belong to ions escaping the volume. (d) the ions stopped in the volume, all obtaining charge 1+.}
\label{Fig31}
\end{figure}

Geant4 uses the effective charge model (implemented by the class \class{G4EmCorrections} and used by the stopping power model) \cite{Ziegler85}. The ion charge is calculated as a function of the ion kinetic energy as described in Ref.~\cite[p. 216]{Manual}. Figure~\ref{Fig31}a shows the ion charge when leaving the Ni foil as a function of kinetic energy. Ions with charge 0 did not enter the Ni foil whereas charge 1+ is given to ions stopped in the foil. Higher charge is given to ions leaving the foil. Figure~\ref{Fig31}b shows the charge after the ion has stopped inside the volume (q=1+) or left the volume (q>1+). Thus, the charge is calculated one-to-one based on the energy and hence all stopped ions will get the same charge. For all energies below $\sim$10\,MeV the charge becomes 1+. The final ion charge is seen for all simulated ions in Fig.~\ref{Fig31}c. The higher charges are for those ions that escapes the volume. Figure~\ref{Fig31}d shows all ions stopped in the stopping chamber, which all end up having charge 1+. Therefore, it is not possible to investigate the charge state distribution of the stopped ions, at least not with the simulation packages currently available in Geant4. 

\clearpage

\section{Lessons learned during the Geant4 simulations}

In the following section we describe some main obstacles faced in the Geant4 simulations and the solutions enforced to bypass them.

\subsection{\label{secGeant4details}Details on the Geant4 simulations}
The Geant4 version used for these simulations was 4-9.6-p02 \cite{GeantNIM03}. The stopping power calculations in Geant4 were implemented through the class \class{G4Ion\-Parametrised\-LossModel} \cite[p. 222]{Manual}. This class uses tables of stopping power from the ICRU~73 report \cite{ICRU73} originally calculated using the PASS code \cite{PASS}. All three materials used in the simulation (Uranium, Nickel and Helium) are part of the ICRU~73 report so the ion stopping powers will be derived from the ICRU~73 data. Any ions not part of the ICRU~73 report also derived their stopping powers from the ICRU~73 data by appropriate scaling.

The elastic scattering processes used in the code were implemented by two different classes: \class{G4hMultipleScattering} and \class{G4Nuclear\-Stopping}. \class{G4h\-Multiple\-Scattering} simulated the accumulated change of angle over multiple elastic scatterings while \class{G4Nuclear\-Stopping} simulated the total energy transfer due to multiple elastic scattering (important to consider for low energetic ions). Single collisions could be implemented using the \class{G4\-Coloumb\-Scattering}, however, with much slower performance. 

In order to improve the simulation speed, all secondary particles were cut, i.e. they were not explicitly simulated but forced to deposit all their energy on the spot where they were created. 

\begin{figure}[b!]
\centering
    \includegraphics[width=0.32\textwidth]{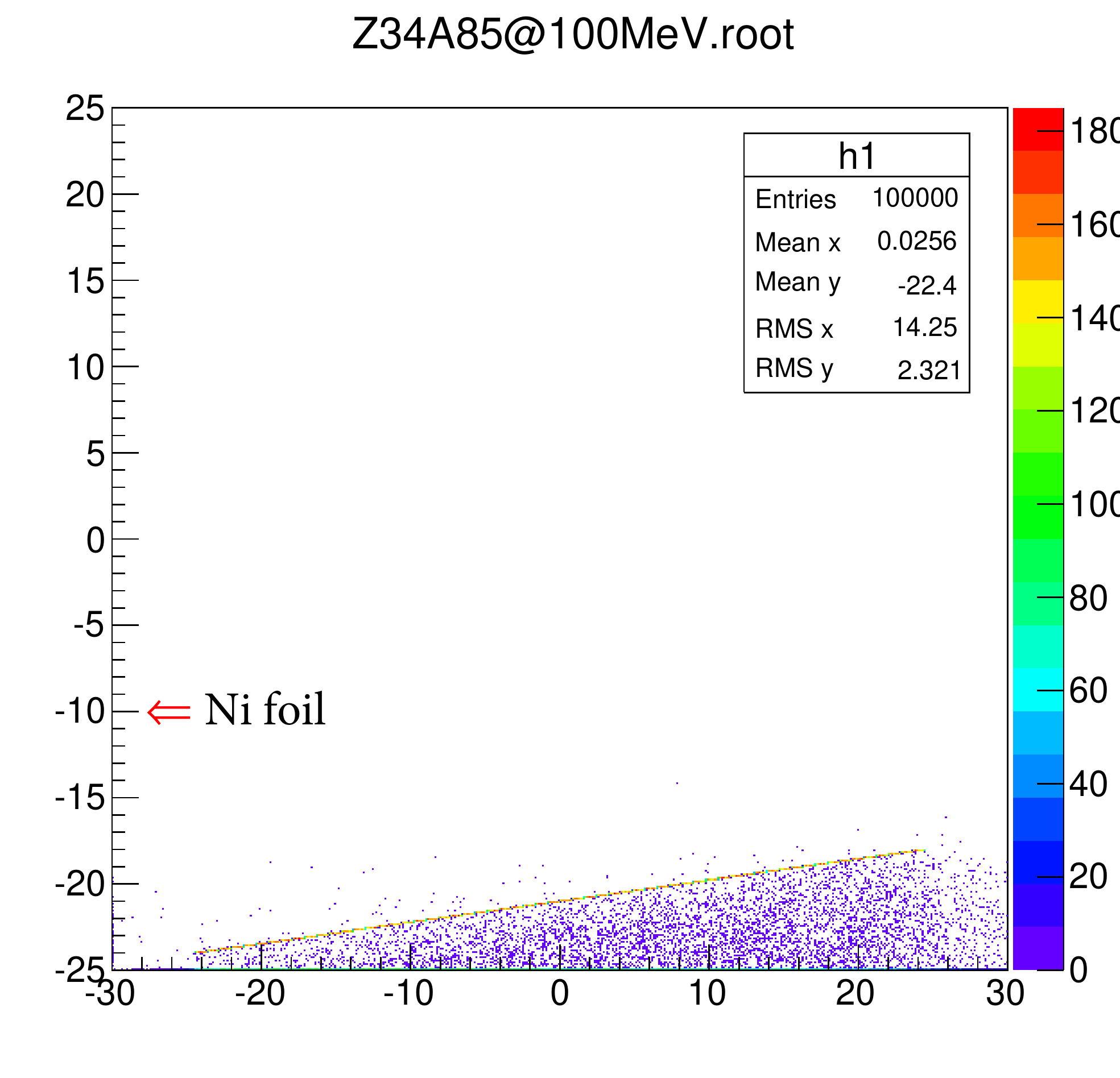}    	
    \includegraphics[width=0.32\textwidth]{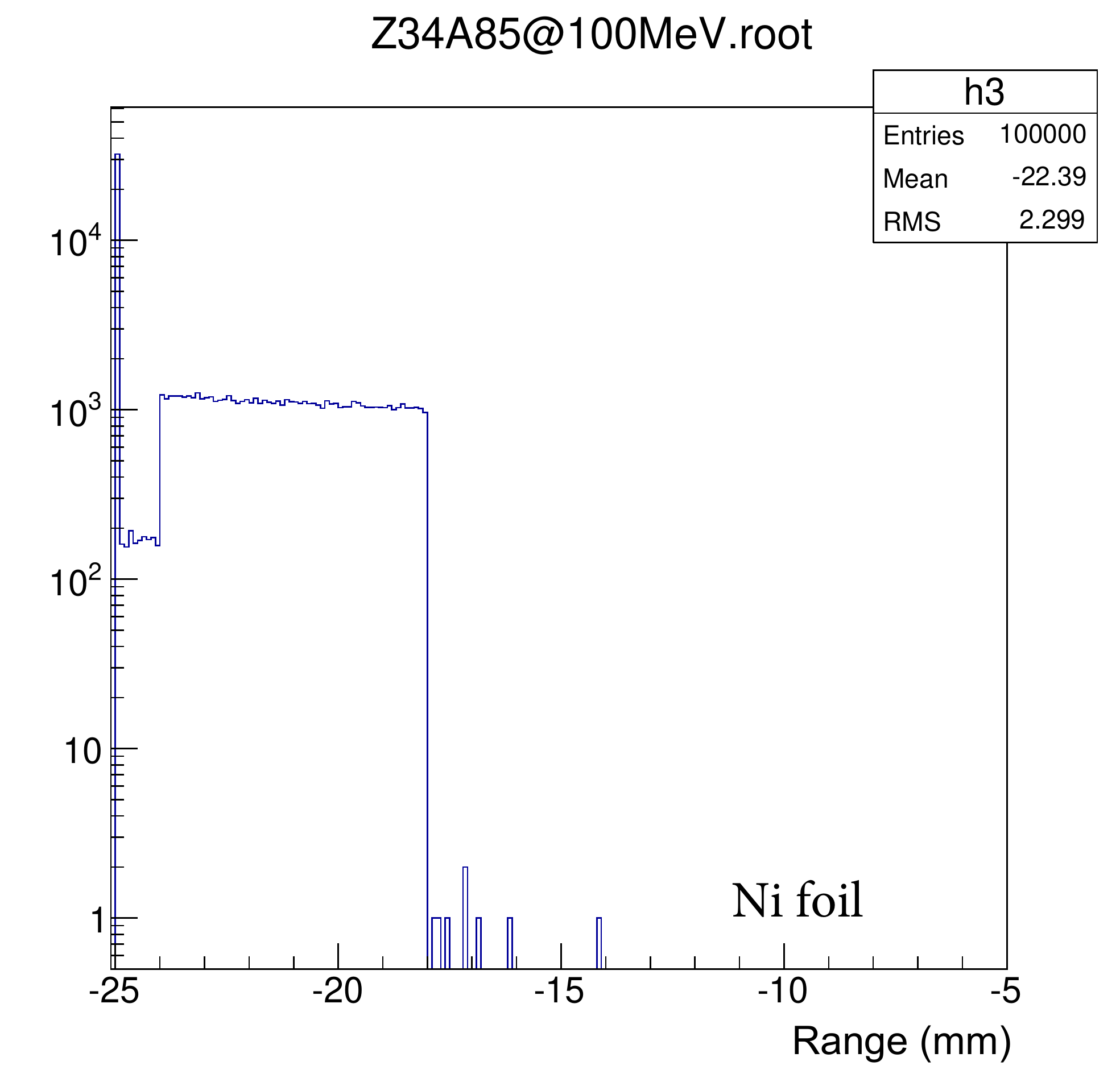}
    \includegraphics[width=0.32\textwidth]{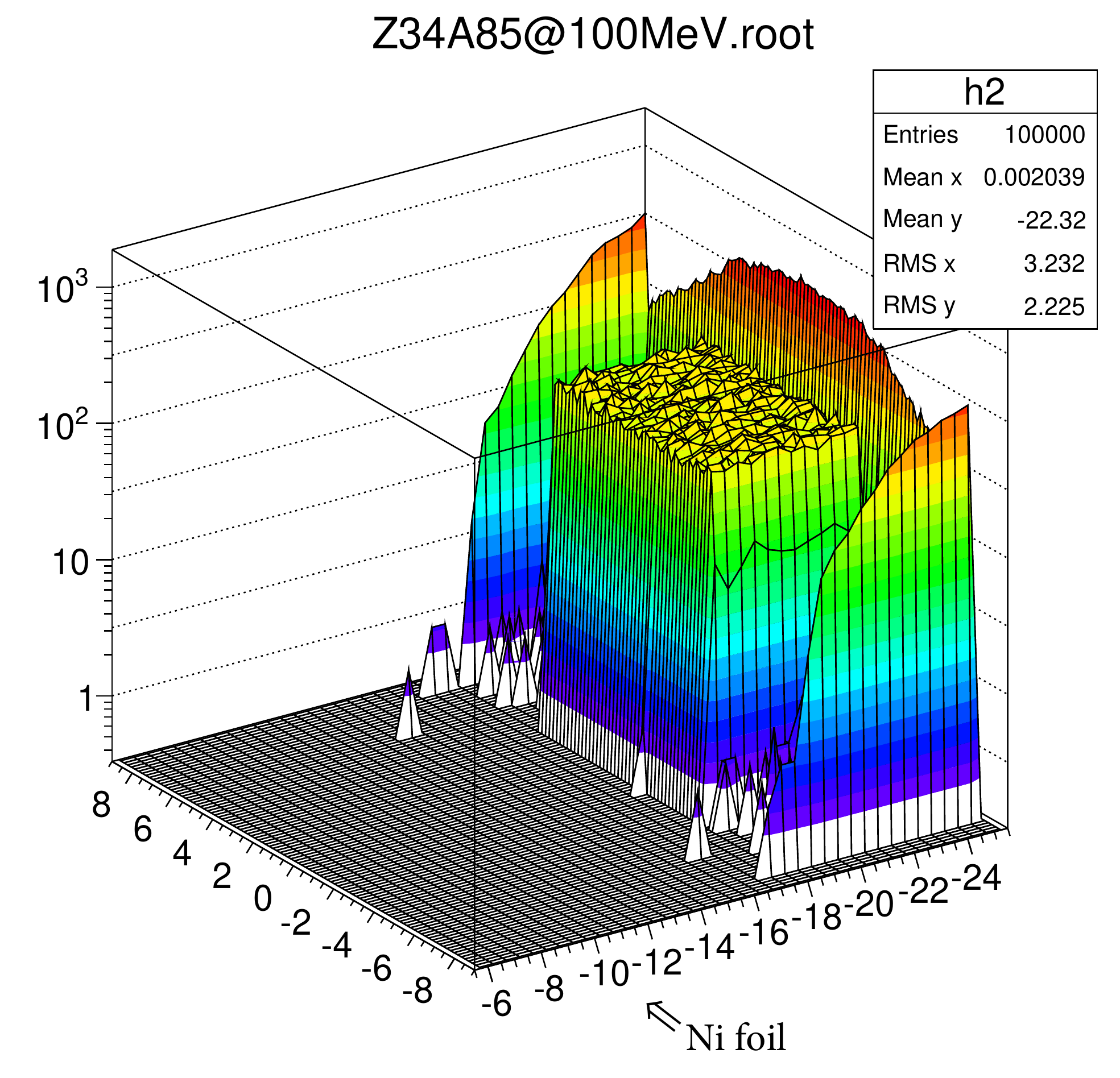}
\caption[Simulation in 2$\pi$.]{The simulations performed for ions with $A = 85$, $E = 100$\,MeV and emission angles $90^\circ \leq \theta \leq 180^\circ$. As can be seen, no ions were backscattered into the stopping chamber.}
\label{Fig32}
\end{figure}

The ion angular emission was selected within 2$\pi$, i.e. only fragments going in the positive $Z$-direction (towards the gas chamber) were simulated. 4$\pi$ selection would require twice the time and it was anticipated that no ions will backscatter to the stopping chamber.  To test this hypothesis and look for possible backscattered events a simulation was performed ($10^5$ ions) with emission angles $90^\circ \leq \theta \leq 180^\circ$. As seen from Fig.~\ref{Fig32}, no backscattered events were able to reach the Ni foil. 

\clearpage


\subsection{\label{sectMultiplescattering}Multiple elastic scattering}

\begin{figure}[b!]
\centering
    \includegraphics[width=0.495\textwidth]{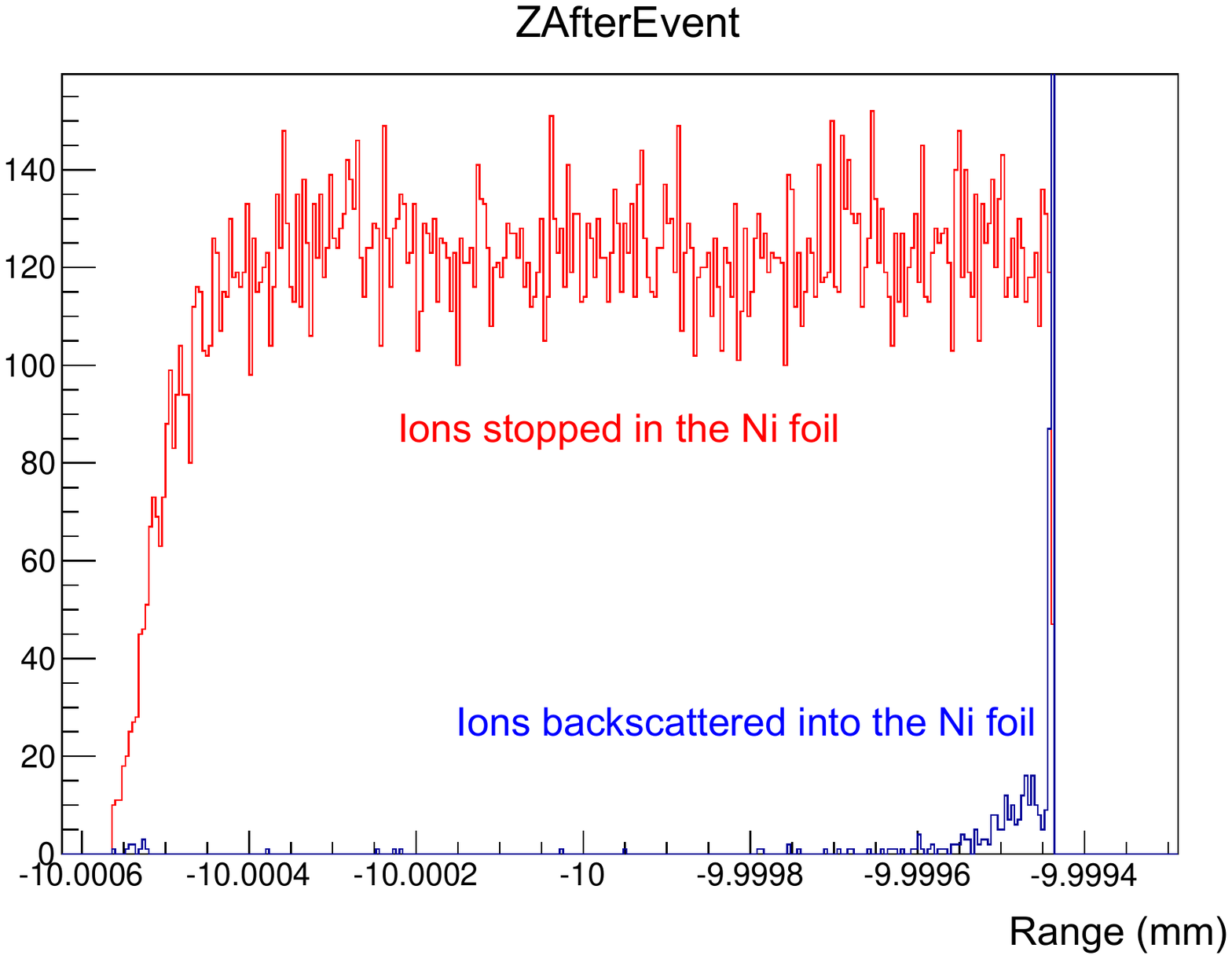}
    \includegraphics[width=0.495\textwidth]{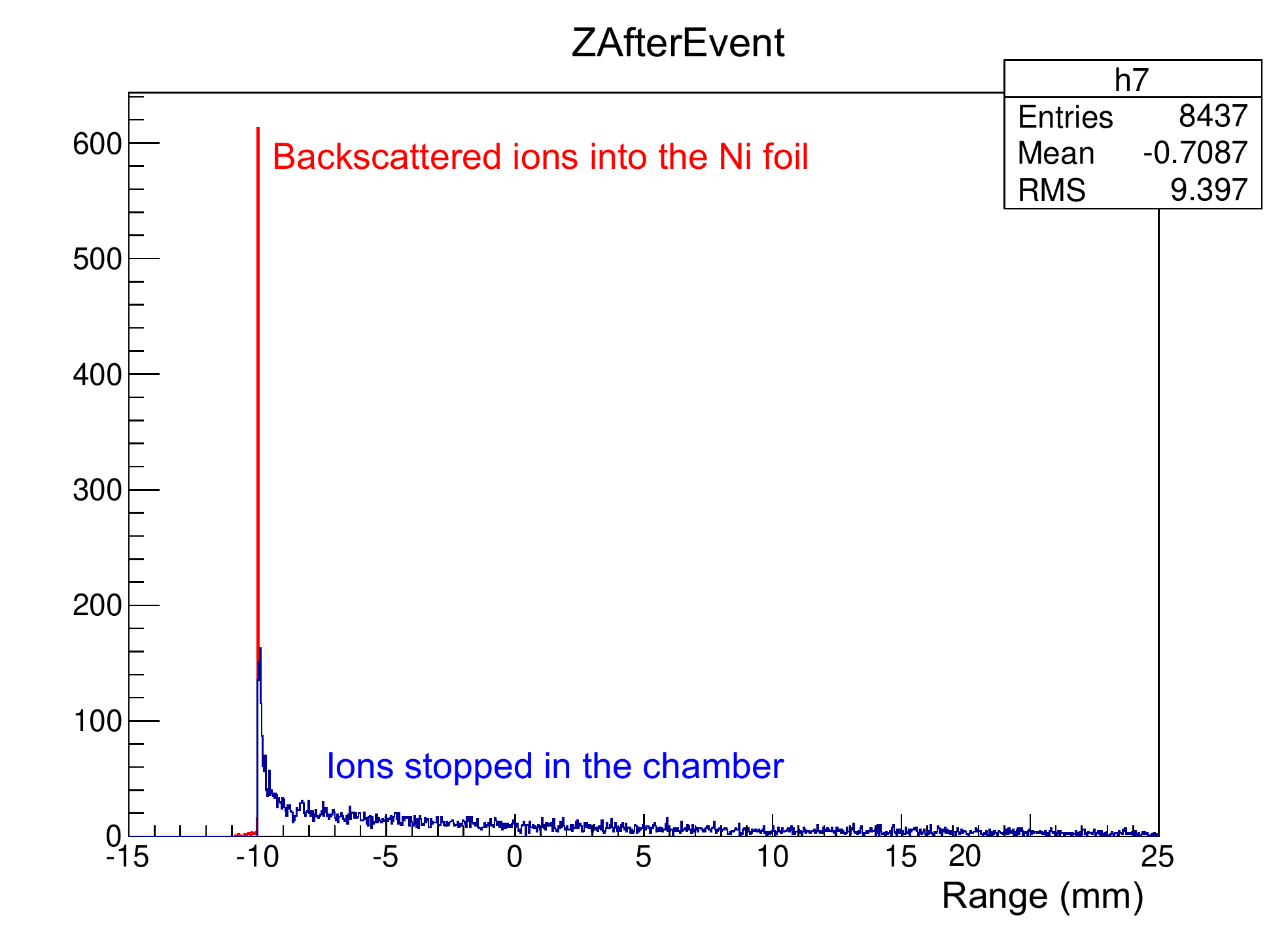}   
    \includegraphics[width=0.495\textwidth]{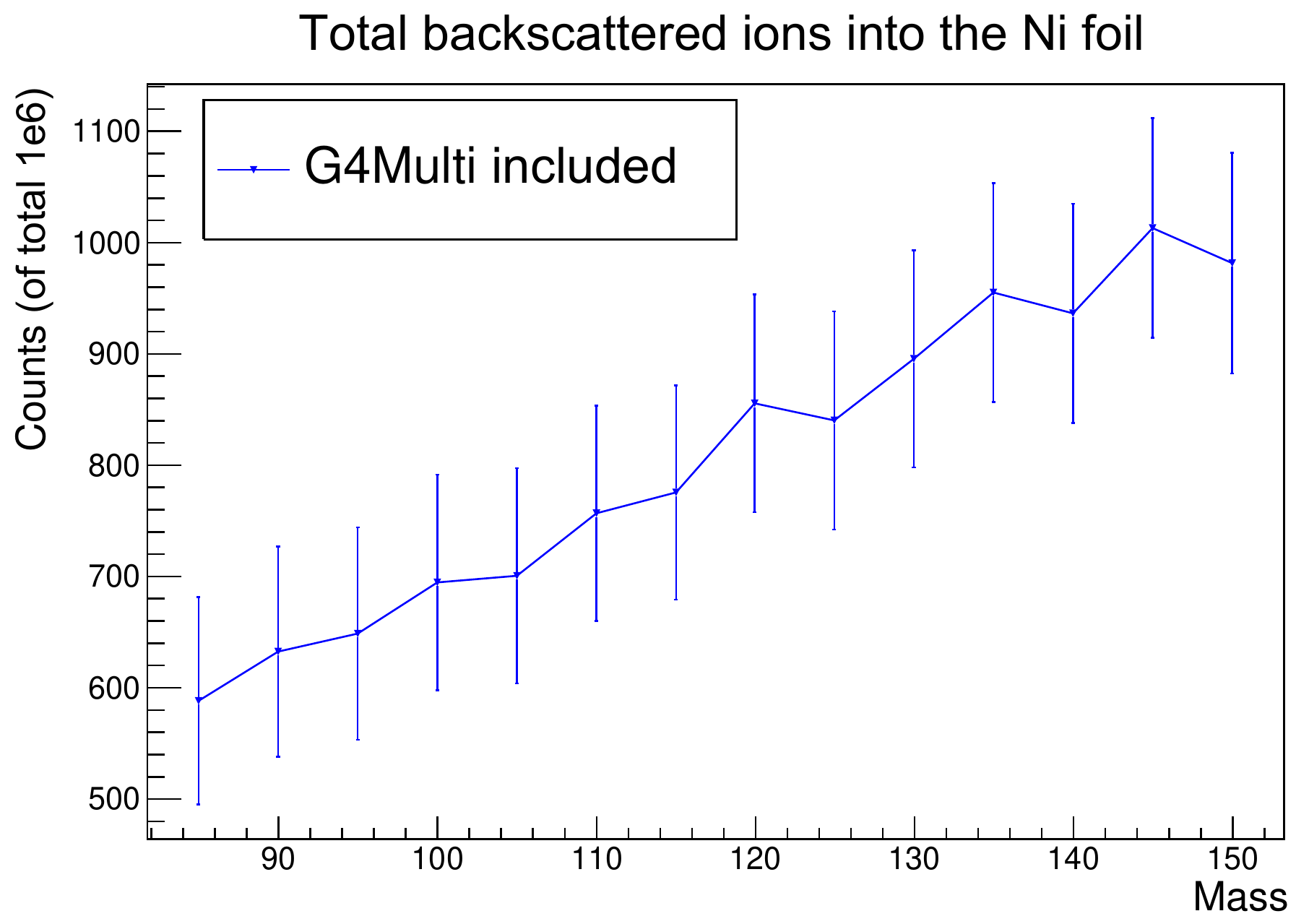}  	
\caption[Ion backscattering.]{a) Ions stopped in the Ni foil (red) and backscattered ions (blue). b) The backscattered ions compared to the ions passing the Ni foil. c) The number of backscattered ions is in the order of 10\% of the total stopped ions and is moreover mass dependent.}
\label{Fig33}
\end{figure}

During the simulations an anomaly was observed. A significant amount of the ions passing through the Ni foil would backscatter after entering the He gas. This effect is improbable, considering the mass ratio between the heavy fission fragments and the light He atoms. Figure~\ref{Fig33}a shows the ions stopped in the Ni foil (in red). At the end of the Ni foil, the backscattered ions can be seen (in blue). The backscattered ions fulfil the cut conditions in sect. \ref{sectsimulations}, i.e. they have entered the He gas, but then scattered back into the Ni foil. Figure~\ref{Fig33}b shows the ions that pass the Ni foil (in blue) and the backscattered ions (the red peak at -10\,mm).

This seemingly strange effect is in addition mass dependent as shown in Fig.~\ref{Fig33}c. The number of backscattered ions reaches up to 1000 ions out of $10^6$, which amounts up to 10\% of the stopped ions. Heavy fragments are twice as prone to backscatter compared to light ones. This large artefact is non-negligible and had to be addressed. 

The \class{G4hMultipleScattering} process (using the model \class{G4IonCoulombScatteringModel}) responsible for the elastic scattering in Geant4 is probably not suited for low energy heavy ions. This was verified by comparing the transport of low energy ions ($A = 127$, $E = 10$\,keV) in a pure He gas in Geant4 and in SRIM. No sign of backscattering was found in the SRIM results, contrary to the Geant4 results where about 40\% of the heavy ions were backscattered by the He gas i.e. 40\% of the ions emitted in the $Z$-direction ended up with negative $Z$-coordinate. Consequently, Geant4 was run in two configurations enabling different processes: 

\begin{description}
\item[Configuration A:] \mbox{\class{G4ionIonisation} + \class{G4NuclearStopping} + \class{G4hMultipleScattering}}
\item[Configuration B:] \mbox{\class{G4ionIonisation} + \class{G4NuclearStopping}}
\end{description}

When removing the \class{G4hMultipleScattering} process (conf. B) the backscattering problem seemed to be solved. Note that the energy loss due to elastic scattering is included in both configurations.

\begin{figure}[]
\centering
    \includegraphics[width=0.95\textwidth]{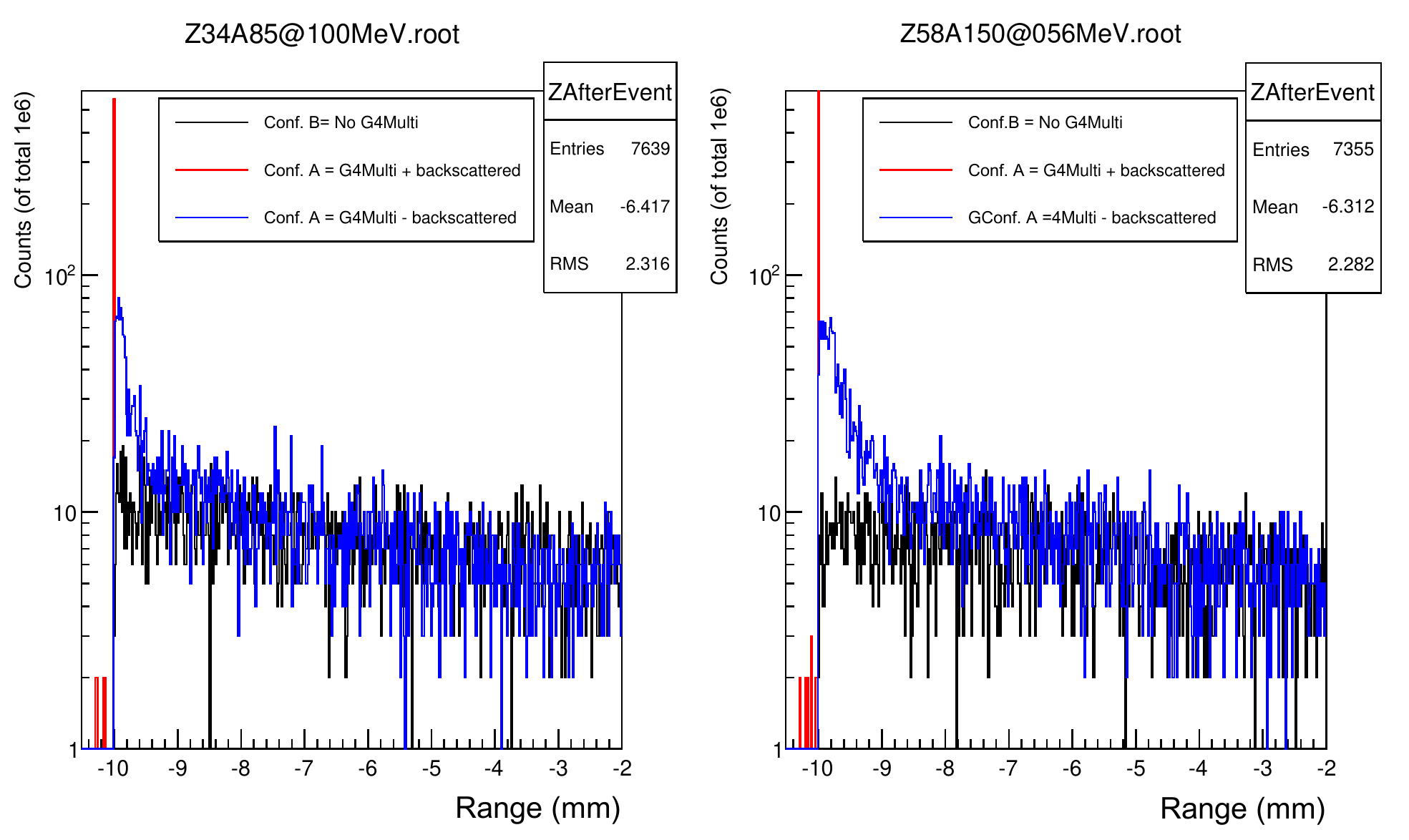}      	
\caption[Geant4 simulations with/without \class{G4hMultiplescattering}.]{Three different cases shown from the Geant4 simulations. The black curve shows the results without \class{G4hMultiplescattering}. The red and blue curves have enabled the \class{G4hMultiplescattering} class, one including the backscattered ions (red) and one excluding them (blue).}
\label{Fig34}
\end{figure}

Figure~\ref{Fig34} shows the ion range for three different cases. The first case (black curve) shows the ion range for conf. B. In the other two cases multiple scattering was enabled (conf. A). The red curve shows a large peak at the end of the Ni foil ($Z = -10$\,mm) due to the backscattered ions. The blue curve excludes the backscattered ions but shows that more ions are stopped right after the Ni foil as compared to the run without \class{G4hMultiplescattering}. The discrepancy is mainly found in the first mm after the Ni foil. As will be seen in sect. \ref{sectsrim}, SRIM confirms the ion range distribution of conf. B.

\begin{figure}[]
\centering
    \includegraphics[width=0.95\textwidth]{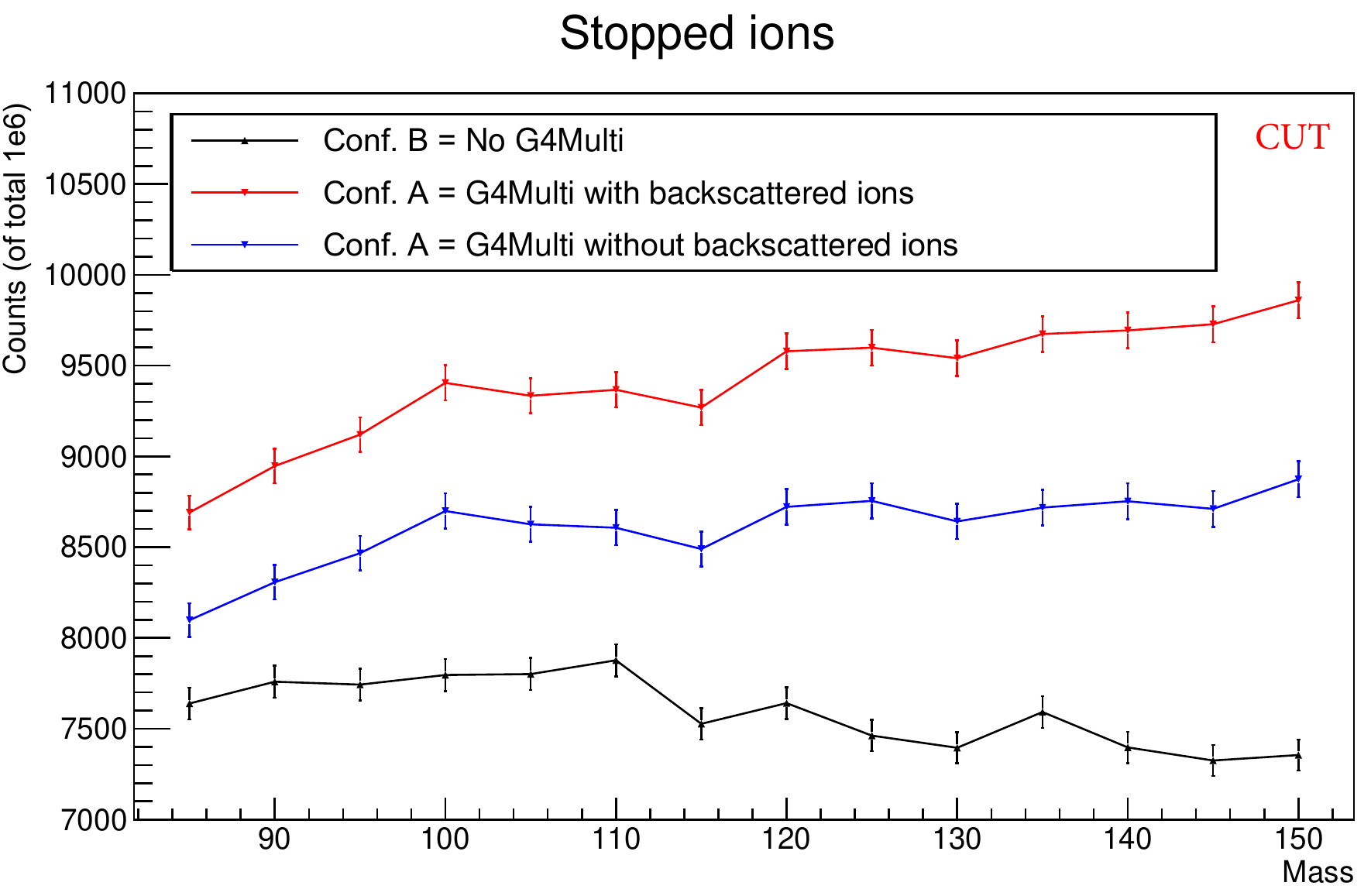}      	
\caption[The mass-dependence when disabling \class{G4hMultiplescattering}]{The mass-dependence of the stopped ions for the three cases in Geant4. By removing the \class{G4hMultiplescattering} class, the stopping efficiency reduces a little and a slight negative trend as a function of mass emerges.}
\label{Fig35}
\end{figure}

The mass-dependence of the ion stopping efficiency is presented in Fig.~\ref{Fig35} for the three cases. Despite the significant differences found in the range, the number of stopped ions does not vary much as a function of mass for all cases. The amount of stopped ions vary between 7 and 10\%. The relative trends go from slightly positive to slightly negative when removing \class{G4hMultiplescattering}. 

In the next section we show that the Geant4 calculations for all 14 isotopes agree better with the SRIM calculations under the B configuration. Hence this is another reason for disabling \class{G4hMultiplescattering} in all simulations in this report. All results presented in sections~\ref{sectIntro} - ~\ref{sectUTarg} are based on the B configuration.

\begin{figure}[]
\centering
    \includegraphics[width=0.32\textwidth]{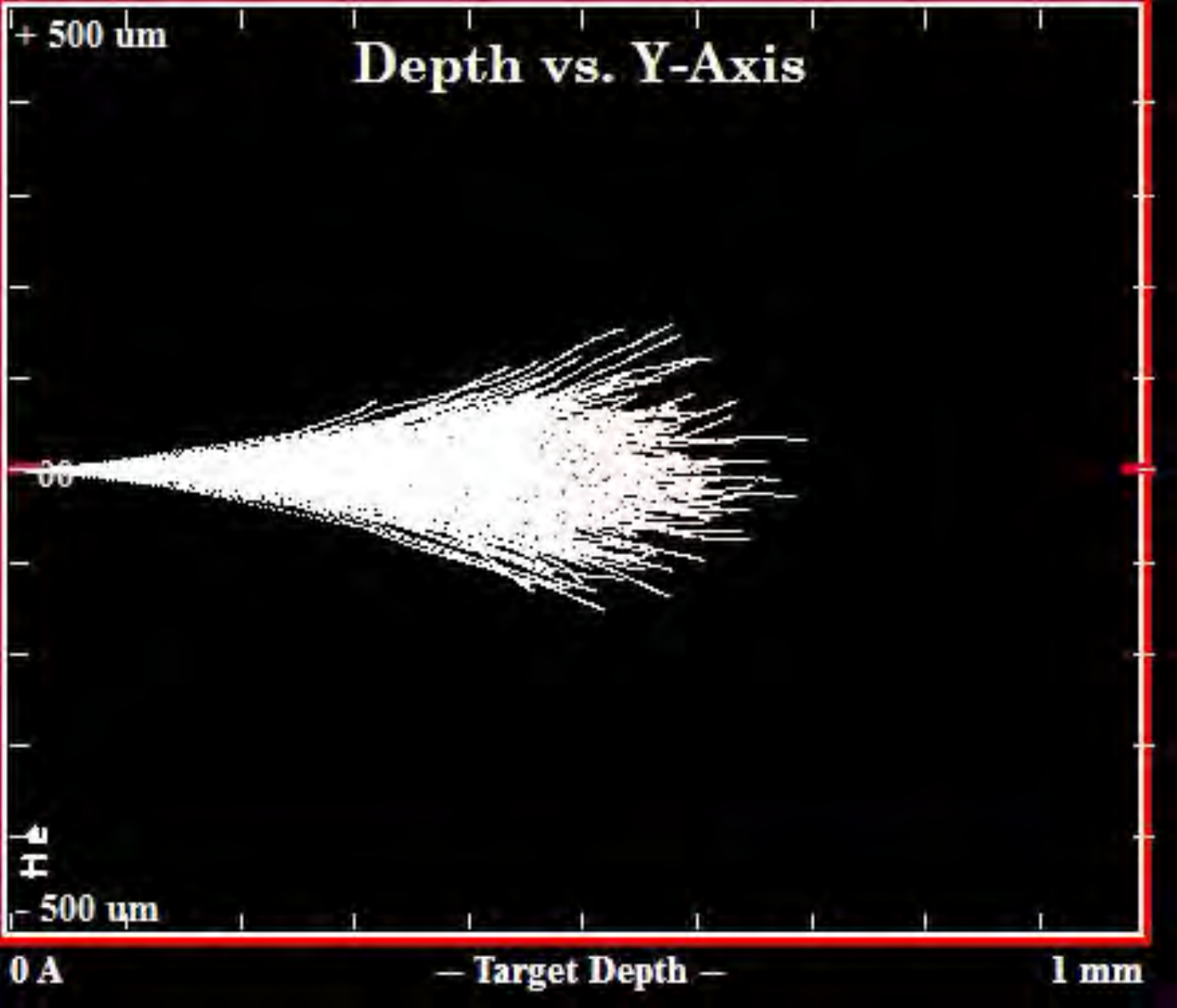}      	
    \includegraphics[width=0.32\textwidth]{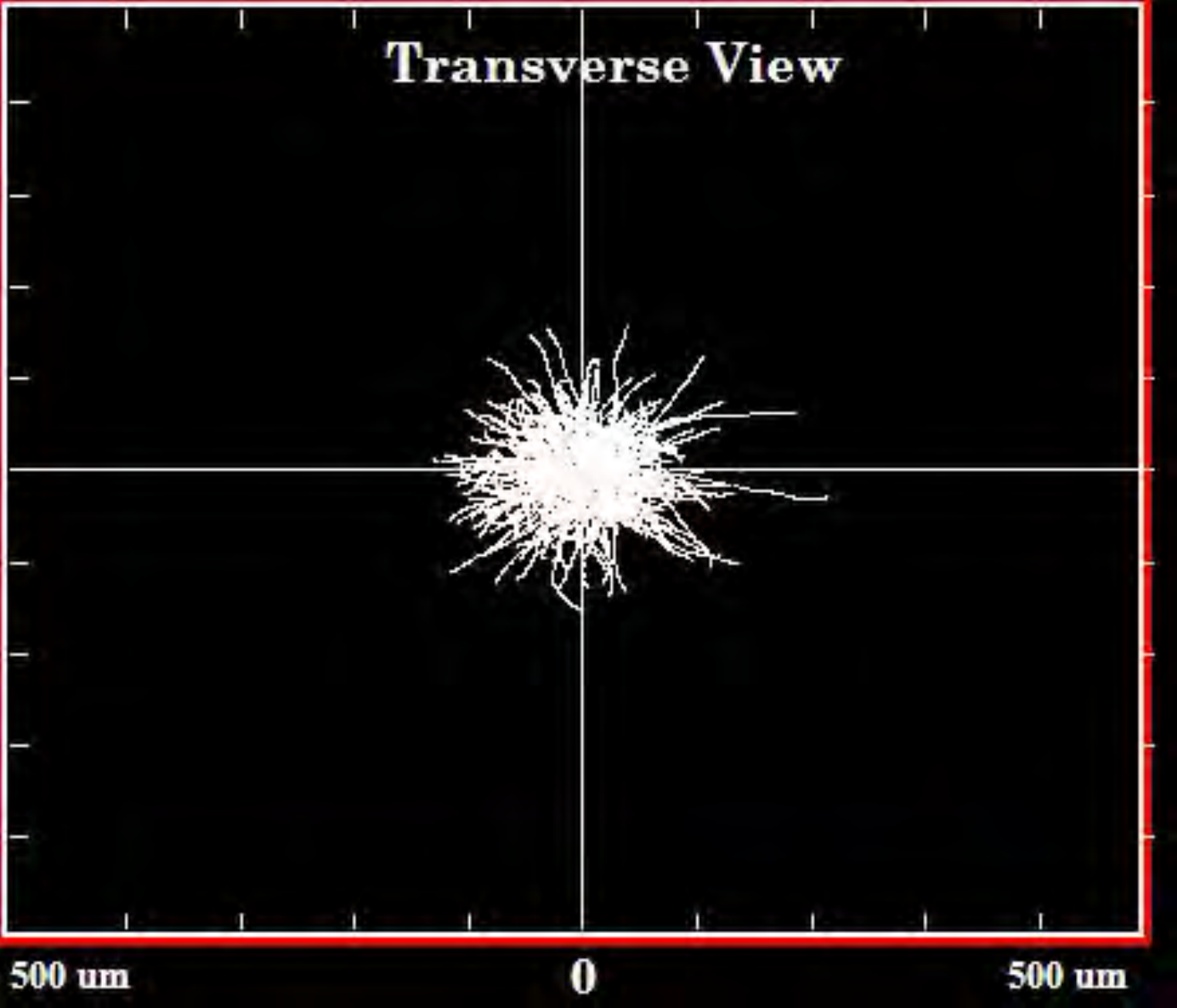}      	    
    \includegraphics[width=0.32\textwidth]{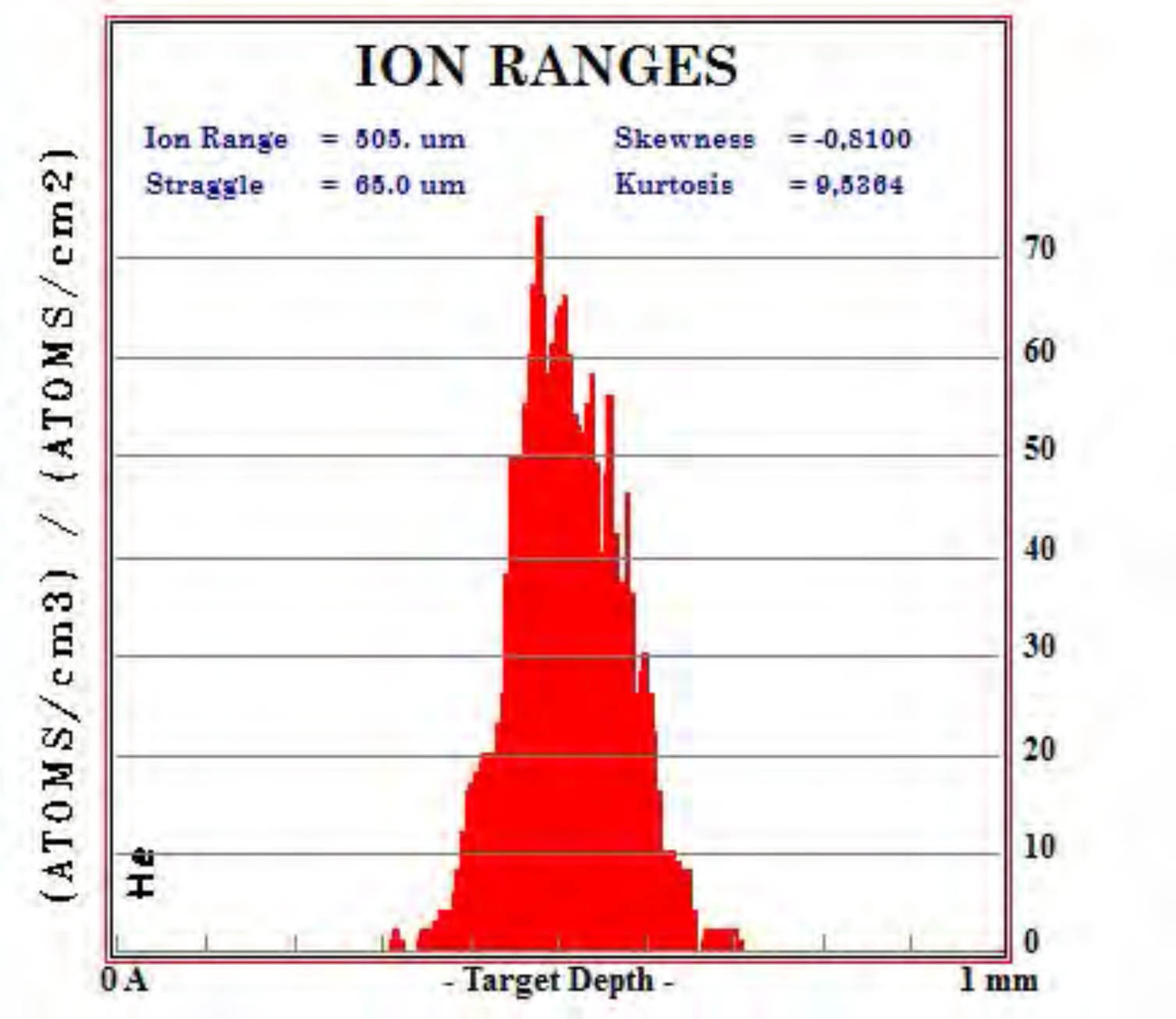}      	
\caption[The stopping range of 10\,keV Kr ions in 200\,torr He, using the SRIM code.]{The stopping range of 10\,keV Kr ions in 200\,torr He, using the SRIM code. The spread in the ion position is in the order of 200\,$\mu$m which is the approximate error introduced in the Geant4 simulations by selecting configuration B (see sect. \ref{sectMultiplescattering}).}
\label{Fig36}
\end{figure}

The systematic uncertainty introduced by disabling the elastic scattering channel can be estimated via the SRIM calculations. Figure~\ref{Fig36} shows the spread of 10\,keV Kr ions in He gas. The ion spread in $X,Y,Z$ is due to the angular straggling associated with elastic scattering. The difference in ion position introduced by omitting the angular change due to elastic scattering  is of the order of 0.2\,mm. This is believed to be below the overall systematic uncertainty of the simulations. 

\subsection{Step length}
The step length is a crucial parameter. It was set to 0.5\,nm inside the U target and the Ni foil. Larger step lengths resulted in less accurate results in the lower energy region. In order to limit the step length, the  \class{G4StepLimiter} class was added to the physics list. For the He gas in the stopping chamber the step length was set to 1\,$\mu$m due to the lower density. 

\clearpage
\section{\label{sectsrim}Comparison to SRIM calculations}

In sect. \ref{sectMultiplescattering} the angular part of the modelling of the elastic scattering process in Geant4 was identified as the reason for a strange back-scattering phenomenon. A first case study utilizing the SRIM package was discussed and confirmed the results for the B configuration listed in sect. \ref{sectMultiplescattering}. The SRIM package was further used in a wider scale to study the mass-dependent trends and to compare with the Geant4 calculations.

Due to very long calculation times, limited statistics were collected ($3\times10^5$). A random selection of starting position in the U target was implemented in SRIM as well as a random emission angle. However, due to the limited geometry design in SRIM, the U target was not tilted. In order to be compared with Geant4, the tilt was also removed in one of the comparison runs of Geant4. As it turned out, this has a minor impact on the number of collected ions. The Geant4 results in this comparison are without activating the \class{G4hMultipleScattering} process (conf. B).

\begin{figure}[H]
\centering
    \includegraphics[width=0.95\textwidth]{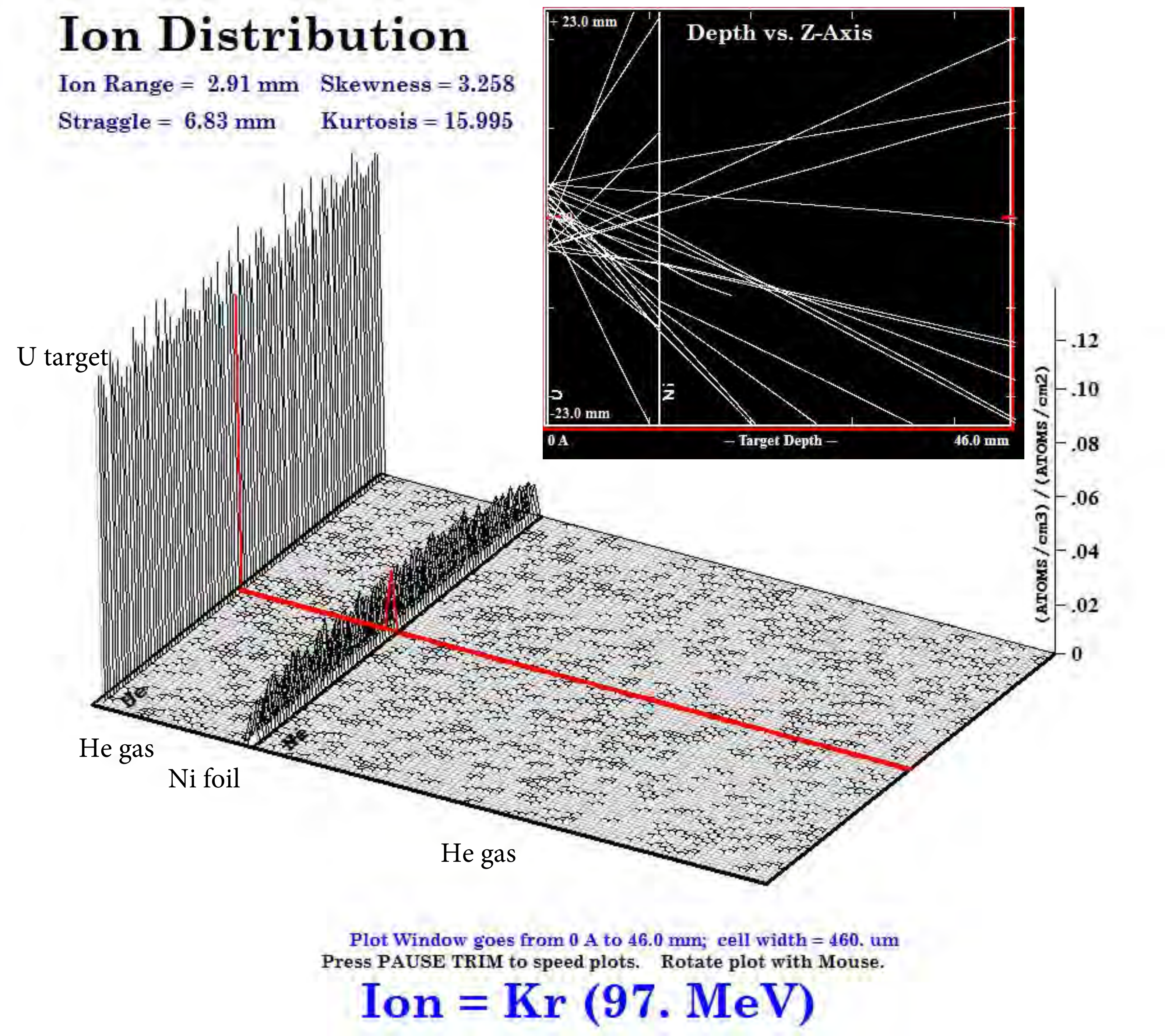} 	
\caption[The stopped ions in the SRIM calculations for Kr at 97\,MeV]{The stopped ions in the SRIM calculations for Kr at 97\,MeV. The inset shows some typical ion tracks.}
\label{Fig37}
\end{figure}

Figure~\ref{Fig37} shows the stopped ions for 97\,MeV Kr in SRIM. Most ions stop in the U target and in the Ni foil and only a small fraction stops in the stopping chamber, as found earlier with Geant4. The backscattering effect was not found in the SRIM results and agrees better with the Geant4 run without the \class{G4hMultiplescattering} implemented. 

\begin{figure}[b!]
\centering
    \includegraphics[width=0.98\textwidth]{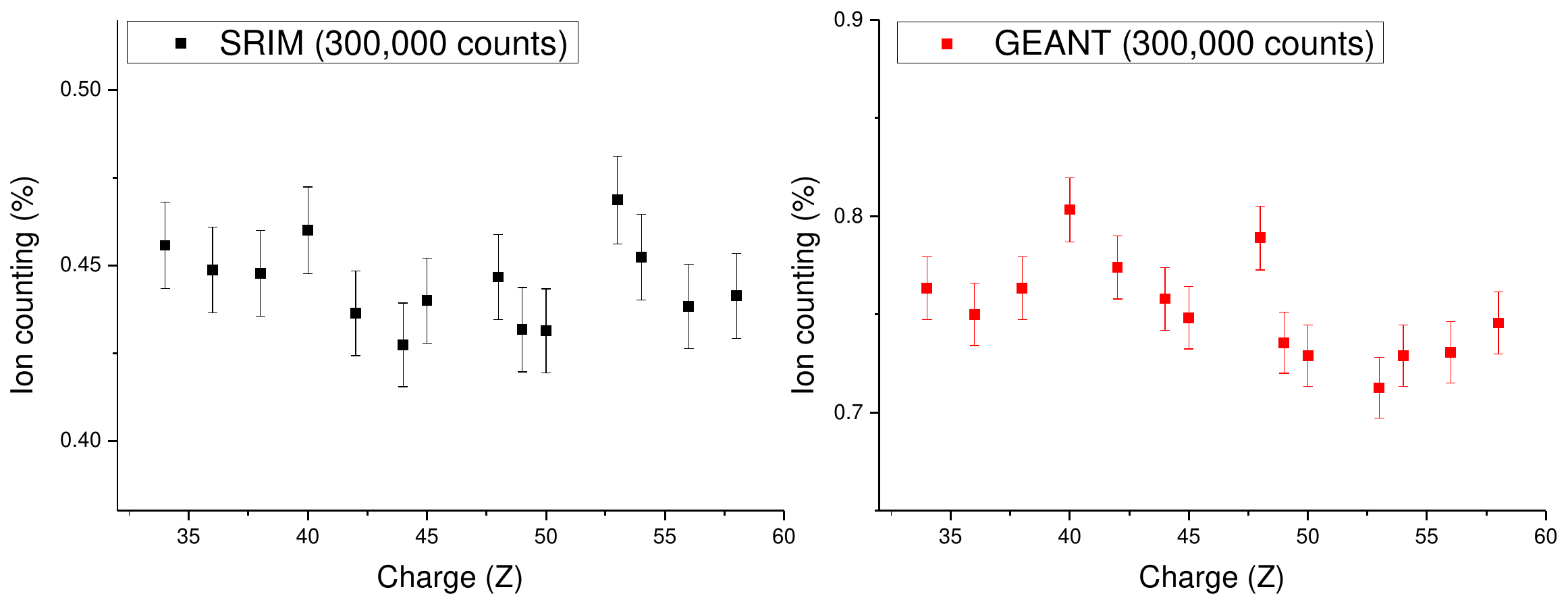}  	
\caption[The number of stopped ions from Geant4 and SRIM]{The fraction of stopped ions from both simulation codes, relative to all ions emitted in 2$\pi$. The agreement is fair as both codes give the same trend. Both simulations were performed without the 7$^\circ$ tilt in the U target.}
\label{Fig38}
\end{figure}

\begin{figure}
\centering
    \includegraphics[width=0.8\textwidth]{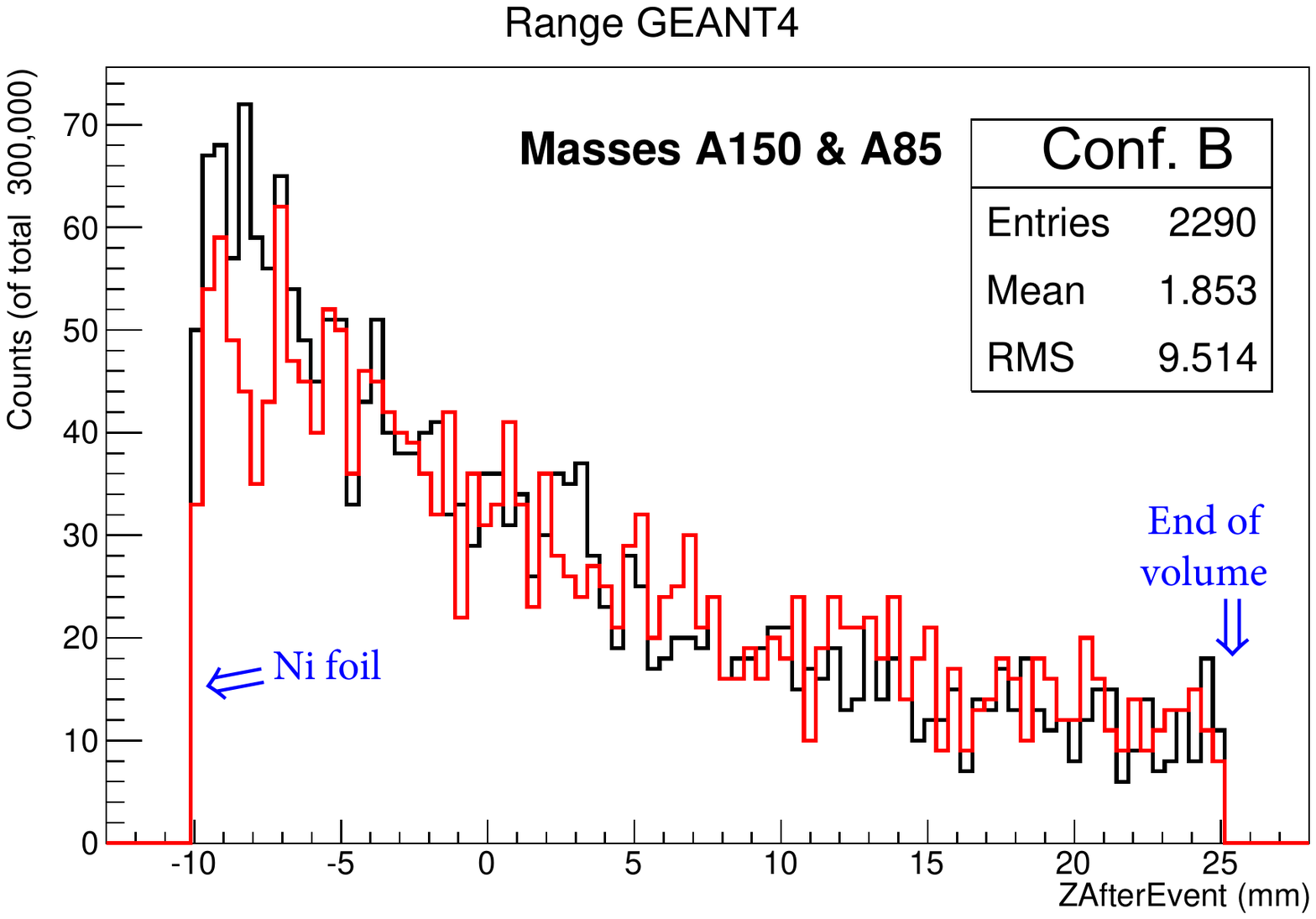}
    \includegraphics[width=0.8\textwidth]{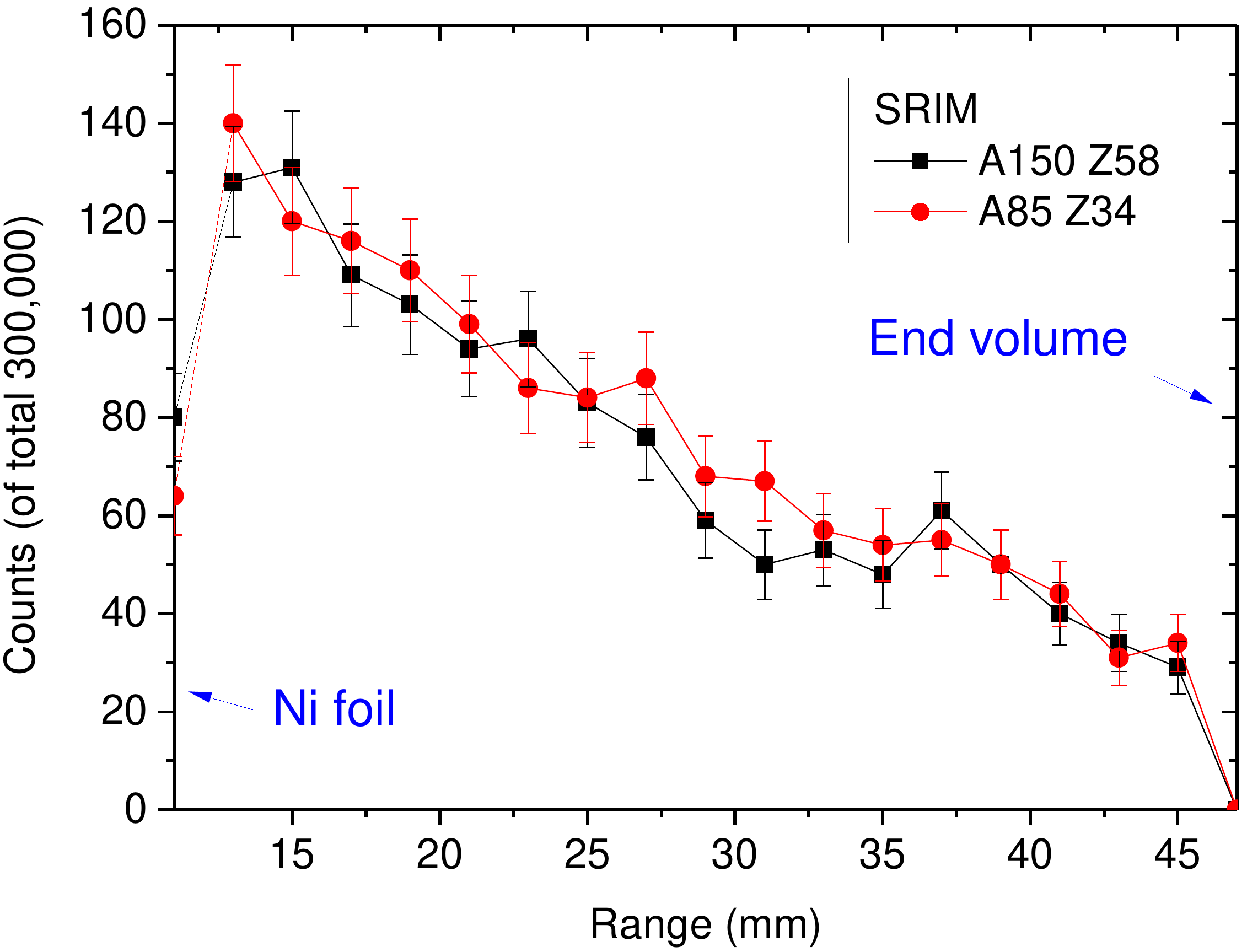}   	
\caption[Ion range after the Ni foil using Geant4 and SRIM]{The ion range after the Ni foil. The results from Geant4 (configuration B) (upper) and SRIM (lower) are in good agreement.}
\label{Fig39}
\end{figure}

The number of stopped ions as a function of fragment mass is presented in Fig.~\ref{Fig38}. The trend agrees well between the two codes. The stopping efficiency varies slightly in favour for lighter fragments. In absolute terms, SRIM gives about 0.45\% stopped ions in the gas whereas Geant4 gives about 0.75\% (fraction of ions emitted in 2$\pi$). Although the difference is significant in relative terms, it is a fairly small absolute difference considering the different physical processes taken into account by the two codes. The ions have lost their full energy and the stopping powers of Geant4 and SRIM are not identical. Moreover, the relative changes among the simulated ions are in good agreement. Figure~\ref{Fig39} shows the range of stopped ions after the Ni foil. The results are in  good agreement and show that most ions stop in the first few mm's after the Ni foil. The comparison to SRIM confirms the use of configuration B with Geant4 (sect.~\ref{sectMultiplescattering}). 

\clearpage
\section{Conclusions and summary}

Geant4 has been used to simulate the ion counting efficiency in the fission ion guide at the IGISOL facility. The SRIM code was also used as a comparison. The following issues have been treated: 
\begin{itemize}

\item \emph{\textbf{Mass dependence of the stopping efficiency}}: simulations of 14 different masses, with mean kinetic energies derived from GEF, showed that the mass dependence is relatively small despite the large variation in the fission fragments mass, charge and energy. This is thanks to the target which is thick enough to allow all ions to lose enough energy to be stopped in the chamber. Still, small variations were observed, where the heaviest fragments were collected about 9\% less efficient than the lightest. 

\item \emph{\textbf{Energy dependence}}: 5 different energies were simulated for each chosen fission fragment mass showing a staggering effect as a function of energy. However the variation in the stopping efficiency due to energy is of less importance than those due to mass and no clear tendencies were observed. Again due to the thickness of the U target, the full ion energy spectrum can be collected.

\item \emph{\textbf{Place of origin}}: The ions that are stopped in the chamber originate from around the center of the target (for low emission angles) up to the edge of the target towards to the stopping chamber (for higher emission angles). This showed a rather significant dependence as a function of mass.

\item \emph{\textbf{Uranium target thickness}}: The Uranium target of 15 mg/cm$^2$ was found to be thick enough. It is important to have enough thickness to ensure that the full ion energy spectrum can be obtained. A target with half the thickness proved to cause a significantly larger mass dependence of the stopping efficiency. 

\item \emph{\textbf{Ni foil thickness}}: Different Ni foil thicknesses were simulated. It was found that the mass dependence of the stopping efficiency could be reduced by optimizing the Ni foil thickness. Based on the simulations, the 1\,mg/cm$^2$ is well suited for this geometry design. The heavy fragments stop more efficiently in the gas compared to the light ones, as the Ni thickness increases. The optimum case would be a Ni foil thickness between 0.5 and 1\,mg/cm$^2$ to reduce the negative trend of the mass-dependence. It was also noted that the ions need to have kinetic energies below 2-3\,MeV after foil in order to be stopped in the chamber. This is in good agreement with previous IGISOL simulations \cite{sonoda}.

\item \emph{\textbf{Gas pressure}}: The nominal pressure of 200\,torr was compared to a higher pressure of 400\,torr. The stopping efficiency was doubled by increasing the pressure, however unfortunately leading to an increased mass-dependence. By choosing a lower Ni thickness one may minimize the mass-dependence and yet obtain higher statistics.

\item \emph{\textbf{Ion charge}}: The ionic charge state of the stopped ions was considered. However, due to the use of the effective ion charge model, the ion charge is a function of its kinetic energy. Therefore, all ions at rest end up in the 1+ state and hence no investigation of charge state distribution was possible.

\end{itemize}

\section{Outlook}

Further investigations are needed on the ion charge distribution with different methods, as we were unable to do this using Geant4. In addition different geometries will be studied, relevant for the neutron induced fission at the IGISOL facility.

\section{Acknowledgements}

We would like to acknowledge our colleagues at the IGISOL facility, University of Jyv\"askyl\"a, for their support in preparing this report and for their pioneering work in developing the ISOL-technique over the last 30 years.


\end{document}